\documentclass[twocolumn,showpacs,preprintnumbers,amsmath,amssymb]{revtex4}

\usepackage{graphics}
\usepackage{dcolumn}
\usepackage{bm}

\input epsf
\usepackage{amsmath,amssymb,epsfig,placeins,subfigure,wrapfig,indentfirst}

\begin{document}

\title{Physics of the interior of a spherical, charged black hole with
  a scalar field}   

\author{Jakob Hansen}
\affiliation{Niels Bohr Institute, Blegdamsvej 17, DK-2100 Copenhagen,
  Denmark} 
\author{Alexei Khokhlov} 
\affiliation{Department of Astronomy and Astrophysics, The
    University of Chicago, 5640 Ellis 
    Avenue, Chicago, IL 60637, USA}

\author{Igor Novikov}
\affiliation{Niels Bohr Institute, Blegdamsvej 17, DK-2100 Copenhagen,
  Denmark} 
\affiliation{Department of Astronomy and Astrophysics, The
    University of Chicago, 5640 Ellis 
    Avenue, Chicago, IL 60637, USA}
\affiliation{Astro Space Center of P.N. Lebedev Physical
  Institute, Profsoyouznaja 83/32, Moscow 
     118710, Russia}

\date{\today}

\begin{abstract}
We analyse the physics of nonlinear gravitational processes inside a 
spherical charged black hole perturbed by a self-gravitating massless 
scalar field. For this purpose we created an appropriate numerical
code. Throughout the paper, in addition to investigation of the
properties of the mathematical singularities where some curvature
scalars are equal to infinity, we analyse the properties of the
physical singularities where the Kretschmann curvature scalar is equal
to the planckian value. Using a homogeneous approximation we analyse the
properties of the spacetime near a spacelike singularity in
spacetimes influenced by different matter contents namely a scalar
field, pressureless dust and matter with ultrarelativistic isotropic
pressure. We also carry out full 
nonlinear analyses of the scalar field and geometry of spacetime
inside black holes by means of an appropriate numerical code with
adaptive mesh refinement capabilities. We use this code to 
investigate the nonlinear effects of gravitational focusing, mass
inflation, matter squeeze, and these effects dependence on the initial
boundary conditions. It is demonstrated that the position of the
physical  singularity inside a black hole is quite different from the
positions  of the mathematical singularities. In the case of the
existence of a strong outgoing flux of the scalar field inside  a
black hole it is possible to have the existence of two null
singularities and one central $r=0$ singularity simultaneously. 
\end{abstract}


\maketitle

\section{\label{sec:1}Introduction}
The problems of the internal structure of black holes are a real great
challenge and has been the subject of very active
analytical and numerical researches during the last decades
\cite{Goldwirth87, Poisson90, Ori91, Ori92,  Gnedin93, Bonanno94, Brady95,
  Droz96, Burko97b, Burko97c,  Burko98, Burko98c,
  Burko99, Burko02, Burko02b, Oren03, Hod97, Hod98, Hod98b,
  Ori99, Ori99b, Berger02, Hamilton04a, Hamilton04b}. There has been a great    
progress in these researches in the last few years and we now know 
many important properties of the realistic black hole's interior,
but some details and crucial problems are still the subject of much
debate.

Many important results have been obtained under simplifying
assumptions. One of the most widely used test toy models is a
spherical, charged, non-rotating black hole, nonlinearly perturbed by a
 minimally coupled and self-gravitating massless, uncharged, scalar
 field. While this toy model is not very realistic, it share many
 properties, including causal structure, with the more
realistic rotating black holes (e.g. \cite{Burko97b} page 5 and
\cite{Dafermos04}) which is why it is believed that insights into this
model may give us important understandings about rotating black
holes. 

The purpose of this paper is to continue the analysis of the physical
processes in the interior of black holes in the framework of this
toy model.

Inside a black hole the main sights 
are the singularity. A number of rigorous theorems (see references in 
\cite{Frolov98}) imply that singularities in the structure of
spacetime 
develops inside black holes. Unfortunately these theorems tell us
practically nothing about the locations and the nature of the
singularities. It has been found that in principle two types of
singularities can  
exist inside black holes corresponding to the toy model: A strong
spacelike singularity and a weak 
null singularity (instead of the inner horizon of a
Reissner-Nordstr{\" o}m black hole). Probably both types of 
singularities can exist simultaneously in the same black hole
and probably it is possible to have cases where only the strong
spacelike singularity exists. 
In the works \cite{Gurtug02} and references therein, some physical  
and geometrical properties of the singularities have been
investigated. Numerical simulations of the fully nonlinear
evolution of the scalar field and the geometry inside the spherical
charged black hole has been carried out in \cite{Burko97, Burko97c,
  Burko02, Burko02b}. 

Near the strong spacelike singularity, one can use a homogeneous
approximation in 
which it is supposed that temporal gradients are much
greater then the spatial gradients. Hence it may be assumed that all
processes near the singularity depend on the time coordinate
only. This uniform 
approximation has been used in \cite{Burko97b} (page 212) and
\cite{Burko98,Burko98b} to gain important new knowledge 
about the processes near the singularity. We will use the same
homogeneous approximation to extend these analyses and
clarify some fundamental physical processes near spacelike
singularities under the 
influence of three different matter contents, namely for the case of
pressureless dust, a massless scalar field and matter with ultrarelativistic
isotropic pressure. This investigation is done by means 
of a suitable numerical code which we develop for this purpose.

Subsequently we will study the nonlinear processes in large regions
inside the toy model 
black hole, not just limited to the homogeneous approximation near the
singularity. This will be done by using a stable and second order accurate,
numerical code with adaptive mesh refinement capabilities. We will
perturb the black hole with initial infalling 
scalar fields of different forms and strengths to further investigate
the behaviour of the singularities and physical processes near
them. We will also investigate the influence of outgoing scalar fluxes
on the interior regions and the singularity. Such outgoing fluxes
will unavoidably appear as a result of the scattering of ingoing
scalar field flux by the curvature of the spacetime and will also be emitted
from the surface of a star collapsing to a black hole.

Today it is widely believed that in the
singularity of 
a realistic black hole, the curvature of the spacetime tends to
infinity. Close to the singularity, where the curvature approaches the 
Planck value ($\left( \frac{\hbar G}{c^3}\right)^2 \approx 1.5\cdot
10^{131}$ cm$^{-4}$ \cite{MTW73}), classical
General Relativity is not applicable. There is not yet a final version
of the quantum theory of gravity, thus any extension of the
discussion of physics in this region would be highly speculative and
we will consider these regions as singularities from the
classical point of view throughout the paper.  

The paper is organised as follows; In section \ref{sec:2} we present 
our model of the spherically symmetric, charged black hole. In
section \ref{sec:5} we dicuss the mass function and some important
nonlinear effects which are fundamental for understanding the physical
processes inside black holes. In section \ref{sec:6} we use a 
homogeneous approximation to analyse the (spacelike) singularity for
three different matter contents: dust with zero
pressure, a massless scalar field and matter with relativistic
isotropic pressure.  In
section \ref{sec:7} we analyse the full nonlinear equations of the
model from section \ref{sec:2} using a numerical approach. Finally we
summarize our conclusions in section \ref{sec:8}. Details of the
numerical code used to obtain the results in section \ref{sec:7} and
analysis of it 
are given in appendices \ref{sec:3} and \ref{sec:4}. \\

\section{\label{sec:2}The model}
We wish to study the geometry inside a spherically symmetric 
black hole with a fixed electrical charge $q$ (i.e. Reissner-Nordstr{\"
  o}m metric), which is nonlinearly perturbed by a selfgravitating,
minimally coupled, massless scalar field. While astrophysical black
holes are more likely to be described by 
the Kerr metric, it is believed that this toy model captures the
essential physics, since the causal and horizon structures of the
Reissner-Nordstr{\" o}m and Kerr black holes are known to be very
similar \cite{Burko97b} (page 5). However, it is much simpler to make a
numerical model of the toy model since this can be simulated in a
two-dimensional spacetime. In constructing the toy model, we follow
here the approach of Burko and Ori \cite{Burko97, Burko02, Burko02b}
who have done similar investigations. \\

\subsection{Field equations}
In spherical symmetry, the general line element in double null-coordinates can
be written as:
\begin{equation}
  \label{eq:lineelement}
  ds^2 = -2e^{2\sigma (u,v)}du dv + r^2(u,v) d\Omega^2
\end{equation}
where $d\Omega^2=d\theta^2+\sin^2(\theta)d\phi^2$ is the line element
on the unit two-sphere and $r$ is 
a function of the null coordinates $u$ and $v$ (in- and outgoing respectively). 

With this metric the non-zero components of the Einstein tensor are:
\begin{subequations}
\label{eq:einstein-tensor}
\begin{eqnarray} 
G_{uu} &=& \frac{4\, r_{,u}\,\sigma_{,u}-2\, r_{,uu}}{r}\\
G_{vv} &=& \frac{4\, r_{,v}\,\sigma_{,v}-2\, r_{,vv}}{r}\\
G_{uv} &=& \frac{e^{2\sigma} +2\, r_{,v}\, r_{,u}+2\, r\, r_{,uv}}{r^2} \\
G_{\theta\theta} &=& -2\, e^{-2\sigma} r \left( r_{,uv}+r \,\sigma_{,uv}\right)\\
G_{\phi\phi} &=& -2\, e^{-2\sigma}\, r\, \sin^2 (\theta)\left( r_{,uv}+r\, \sigma_{,uv}\right)
\end{eqnarray} 
\end{subequations}

The energy-momentum tensor can be written as a sum of contributions from electromagnetic and
scalar fields:
\begin{equation}
T_{\mu\nu}=T^s_{\mu\nu} +T^{em}_{\mu\nu}
\end{equation} 

The energy-momentum tensor of a massless scalar field $\Phi$ is \cite{MTW73}:
\begin{equation}
\label{eq:energy-momentum-tensor}
  T^s_{\mu\nu}= \frac{1}{4\pi}\left( \Phi_{,\mu}\Phi_{,\nu}
  -\frac{1}{2}g_{\mu\nu}g^{\alpha\beta}\Phi_{,\alpha}\Phi_{,\beta}\right)
\end{equation}
whose non-zero components for the metric \eqref{eq:lineelement} are:
\begin{subequations}
\label{eq:energy-momentum-tensor-a}
\begin{eqnarray} 
T^s_{uu} &=& \frac{1}{4\pi}\Phi_{,u}^2\\
T^s_{vv} &=& \frac{1}{4\pi}\Phi_{,v}^2\\
T^s_{\theta\theta} &=& \frac{1}{4\pi}r^2\,e^{-2\sigma}\Phi_{,u}\Phi_{,v}\\
T^s_{\phi\phi} &=& \frac{1}{4\pi}r^2\, \sin^2(\theta )\, e^{-2\sigma}\Phi_{,u}\Phi_{,v}
\end{eqnarray} 
\end{subequations}

The energy-momentum tensor of an electric field in spherical symmetry
and null coordinates is \cite{MTW73}:
\begin{equation}
\label{eq:energy-momentum-tensor2}
T^{em}_{\mu\nu}=F_{\mu\alpha}F^{\alpha}_{\nu}+\frac{1}{4}g_{\mu\nu}F_{\mu\nu}F^{\mu\nu}
\end{equation}
whose non-zero components for the metric \eqref{eq:lineelement} are:
\begin{subequations}
\label{eq:energy-momentum-tensor-c}
\begin{eqnarray} 
T^{em}_{uv} &=& \frac{q^2}{8\,\pi\, r^4} e^{2\sigma}\\
T^{em}_{\theta\theta} &=& \frac{q^2}{8\,\pi\, r^4}r^2 \\
T^{em}_{\phi\phi} &=& \frac{q^2}{8\,\pi\, r^4}r^2 \sin^2(\theta )
\end{eqnarray} 
\end{subequations}

From the Einstein and energy-momentum tensors we can write up the
Einstein equations, $G_{\mu\nu}=8\,\pi\, T_{\mu\nu}$ (with $c=1, G=1$),
governing the spacetime. The $u-u$, $v-v$,
$u-v$ and $\theta-\theta$ components of the Einstein equations respectively are:

\begin{eqnarray}
& &  r_{,uu} - 2\, r_{,u}\,\sigma_{,u} + r\, \left(\Phi_{,u} \right)^2 =0  \label{eq:constraint1}\\
& &  r_{,vv} - 2\, r_{,v}\sigma_{,v} + r\left(\Phi_{,v} \right)^2 =0  \label{eq:constraint2}\\
& &  r_{,uv} +\frac{r_{,v} r_{,u}}{r}+\frac{e^{2\sigma}}{2r} \left( 1 - \frac{q^2}{r^2}\right) =0  \label{eq:evolve1}\\
& &  \sigma_{,uv} - \frac{r_{,v} r_{,u}}{r^2}- \frac{e^{2\sigma}}{2r^2} \left( 1 -
    2\frac{q^2}{r^2}\right) + \Phi_{,u}\Phi_{,v}  = 0   \label{eq:evolve2}
\end{eqnarray}

Lastly, the scalar field must satisfy the Gordon-Klein equation (note
that Gordon-Klein is a consequence of the Einstein equations
for the scalar field \cite{MTW73}), $\nabla^\mu\nabla_\mu\Phi = 0$,
which in the metric \eqref{eq:lineelement} becomes: 
\begin{equation}
  \label{eq:evolve3}
  \Phi_{,uv}  +
  \frac{1}{r}\left( r_{,v}\Phi_{,u}+r_{,u}\Phi_{,v}\right)= 0
\end{equation}

Equations \eqref{eq:evolve1} - \eqref{eq:evolve3} are evolution equations
which are supplemented by the two constraint equations
\eqref{eq:constraint1} and \eqref{eq:constraint2}. It is noted that
none of these equations depends on the scalar field $\Phi$ itself, but
only on the derivatives of $\Phi$, i.e. the derivative of the scalar
field is a physical quantity, while the absolute value of the scalar
field itself is not. Specifically we note the $T_{uu}=(\Phi_{,u})^2/4\pi$
and  $T_{vv}=(\Phi_{,v})^2/4\pi$ components of the energy-momentum
tensor which are part of the constraint equations. Physically
$T_{uu}$ and $T_{vv}$ represents the flux of the scalar field through
a surface of constant $v$ and $u$ respectively. These fluxes will play
an important role in our interpretation of the numerical results in
section \ref{sec:7}. \\

\subsection{Initial value problem}
We wish to numerically evolve the  unknown functions $r(u,v), 
\sigma (u,v)$ and $\Phi (u,v)$ throughout 
some computational domain. We do this by following the approach of
\cite{Burko97, Burko97c, Burko02b} to numerically integrate the
three evolution equations \eqref{eq:evolve1} -
\eqref{eq:evolve3}. These equations form a well-posed initial value
problem in which we can specify initial 
values of the unknowns on two initial null segments, namely an outgoing
($u=u_0=$ constant) and an ingoing ($v=v_0=$ constant) segment. We impose
the constraint equation  \eqref{eq:constraint1}
and \eqref{eq:constraint2} on the initial segments.
Consistency of the evolving fields with the constraint 
equations is then ensured via the contracted Bianchi identities
\cite{Burko97}, but we use the constraint equations throughout the
domain of integration to check the accuracy of the numerical
simulation.
 
On the initial null segments, the constraint equations reduces the
number of unknowns by one on $v=v_0$ and $u=u_0$ respectively. The remaining two 
unknowns expresses only one degree of physical freedom, while the
other unknown expresses the gauge freedom associated with the
transformation $u\rightarrow \tilde{u}(u), v\rightarrow \tilde{v}(v)$,
(the line element \eqref{eq:lineelement} and
eqs. \eqref{eq:constraint1}-\eqref{eq:evolve3} are invariant to such a
transformation). We choose a standard gauge in which $r$ is linear in $v$ and $u$ on the
initial null segments. Specifically we choose:
\begin{equation}
  \label{eq:2_10}
  r(u_0,v)=v, \ \ \ \ \ \  r(u,v_0)=r_0+u\, r_{u0}
\end{equation}
We also choose that the outgoing segment should run along $u_0=0$.

We can now use \eqref{eq:constraint1} and \eqref{eq:constraint2} to find:
\begin{subequations}
  \label{eq:2_11}
  \begin{equation}
    \sigma_{,v}(u_0,v) = \frac{1}{2r_{,v}}\left(
    r_{,vv}+r\left(\Phi_{,v}\right)^2\right)= 
    \frac{v}{2}\left(\Phi_{,v}\right)^2
  \end{equation}
  \begin{equation}
    \sigma_{,u}(u,v_0) = \frac{1}{2r_{,u}}\left(
    r_{,uu}+r\left(\Phi_{,u}\right)^2\right) = 
    \frac{r_0+u\ r_{u0}}{2r_{u0}}\left(\Phi_{,v}\right)^2
  \end{equation}
\end{subequations}
which can can be readily integrated to find $\sigma(u,v)$ on the initial  
null segments if $\Phi$ and the constants $r_0, r_{u0}$ and $\sigma
(u_0,v_0)$ are specified on these.  

Following \cite{Burko97, Burko02, Burko02b} we choose $\sigma
(u_0,v_0)=-\frac{\ln (2)}{2}$ and $r_0=5$. The parameter $r_{u0}$, can 
be related to the initial mass and charge of the black hole via the mass
function (the mass function is further discussed in section
\ref{sec:5}) which in the metric \eqref{eq:lineelement} has the form:
\begin{equation}
  \label{eq:2_12}
  m(u,v) = \frac{r}{2}\left( 1+\frac{q^2}{r^2}+4\frac{r_{,u} r_{,v}}{2e^{2\sigma}}\right)
\end{equation}
which in our choice of gauge, at the point of intersection of the
initial null segments, take the form:
\begin{equation}
  \label{eq:2_13}
  m_0 =m(u_0,v_0)= \frac{r_0}{2}\left( 1+\frac{q^2}{r_0^2}+4r_{u0}\right)
\end{equation}
hence $r_{u0}$ can be determined by $r_0$, $m_0$ and $q$ as:
\begin{equation}
  \label{eq:2_14}
  r_{u0}=\frac{1}{4}\left( \frac{2}{r_0}\left(m_0-\frac{q^2}{2r_0}\right)-1\right).
\end{equation}

Hence, by specifying a distribution of the scalar field $\Phi$ on
the initial null segments, choosing a gauge and initial charge and
mass of the black hole we can specifiy complete initial conditions
on the initial null segnemts. Using a numerical code (described in
appendix \ref{sec:3}) we can then use the evolution equations,
eqs. \eqref{eq:evolve1} - \eqref{eq:evolve3} to evolve the unknown
functions throughout the computational domain.

\section{\label{sec:5}Nonlinear effects; internal mass function}
We will in the next sections consider the evolution of the scalar
field together with the geometry of the interior of a black hole. This
evolution is highly nonlinear. One of the main parameters of this
evolution is the mass function which represents the total effective
mass in a sphere of radius $r(u,v)$ \cite{Poisson90, Frolov98,
  Poisson89a}. We give here different expressions for the mass
function, which emphasize its different characteristics. 

In the metric:
\begin{eqnarray}
  ds^2 &&= g_{tt} dt^2 + g_{rr}dr^2+r^2 d\Omega^2\nonumber \\
  d\Omega^2 &&= d\theta^2+\sin^2\theta d\phi^2  \label{eq:5.1}
\end{eqnarray}
the mass function can be written in the following forms:
\begin{equation}
  \label{eq:5.2}
  m=\frac{r}{2}\left( 1+\frac{q^2}{r^2}-g_{rr}^{-1}\right)
\end{equation}
or
\begin{equation}
  \label{eq:5.3}
  m=4\pi \int_{r_1}^{r_2}T_t^t r^2 dr + m_0
\end{equation}

In the metric
\begin{equation}
  \label{eq:5.4}
  ds^2 = -\alpha^2 dudv + r^2 d\Omega^2
\end{equation}
it has the form \cite{Burko97,Oren03}:
\begin{equation}
  \label{eq:5.5}
  m=\frac{r}{2}\left( 1+\frac{q^2}{r^2}+4\frac{r_{,u}r_{,v}}{\alpha^2}\right)
\end{equation}
or (for the scalar field, $\Phi$) \cite{Oren03}:
\begin{equation}
  \label{eq:5.6}
  m_{,uv}=2\frac{r^3}{\alpha^2}\Phi_{,u}^2\Phi_{,v}^2-r\left(1-\frac{2m}{r}+\frac{q^2}{r^2}\right)\Phi_{,u}\Phi_{,v}
\end{equation}
There are two important physical processes which can lead to a nonlinear change of the
mass parameter:

\begin{itemize}
\item[\textbf{1.}] The mass $m$ inside a sphere can change because
  of the work of pressure forces on the surface of the sphere. A
  clear manifestation of this squeeze effect is the change of the mass of a
  spherical volume in a homogeneous model of the Universe filled
  with relativistic gas (see \cite{Zeldovich83a}, page 13). In section
  \ref{sec:6} we will consider another example, namely for the case of
  the imitation of the interior of a black hole. For the description
  of this process, it is most appropriate to use the form
  \eqref{eq:5.3} for the mass function. Remember that inside the event
  horizon, $r$ is a time-like coordinate. 
\item[\textbf{2.}] Mass inflation \cite{Poisson90}. This process inside the
  black hole exists if near the Cauchy horizon (CH) there are
  simultaneous ingoing
  and outgoing fluxes of a massless field (for example scalar
  field). Actually the existence of the outgoing flux together with
  the ingoing is inevitable because of backscattering of part of
  the ingoing flux by the spacetime curvature. 
  The simplest exact
  model of the mass inflation process inside of a black hole has been
  constructed by Ori \cite{Ori91}.  For the description of
  this process it is most useful to use formula
\eqref{eq:5.6} from which it can be seen that evolution of $m$ with
  both $u$ and $v$ 
is possible only if there are both $\Phi_{,u}^2$ and
$\Phi_{,v}^2$ fluxes simultaneously.
\end{itemize}
 One can often observe the simultaneous manifestation of both these
 processes.

Another important nonlinear effect is the focusing effect by the
gravity of beams of opposite fluxes of radiation. A particular
manifestation of the focusing effect is the contraction of the CH under the
gravity of transverse irradiation by the outgoing radiation. 
Eventually the CH singularity shrinks down to a point-like size and meets
a central (probably spacelike) singularity $r=0$. It should be
mentioned that it is incorrect to say that the CH singularity is
transformed into a $r=0$ spacelike singularity, because the formation of
the $r=0$ singularity is causally absolutely independent from the
formation of the CH singularity.

We want to mention that it is possible, in principle, to have the situation when the mass
function depends on only one null coordinate, say $v$, while it does
not depend on the other $u$ coordinate. This situation is described by
a charged Vaidya solution \cite{Bonnor70}. In this solution there is
an effect of a linear change of $m$, because of an ingoing lightlike
radial flux of energy into the black hole (without any scattering of
this radiation by a curvature of the spacetime). Of course this
effect is compatible with formula \eqref{eq:5.6}.

\section{\label{sec:6}Homogeneous approximation}
In the close vicinity of the spacelike singularity of a black hole all
processes, as a rule, have high temporal gradients, much higher than
the spatial gradients 
along the singularity, and the processes depend on a very restricted
space region. So for clarification of some physical processes one can
use a homogeneous approximation and assume that all processes depend
on the time coordinate only. This approach has been proposed by
Burko \cite{Burko97b, Burko98b} and we will use it at the
beginning of our analysis to clarify some main properties of the
singularity before coming to the full analysis of the spherical model
in section \ref{sec:7}.

\subsubsection{Leading order analysis}
The general homogeneous, spherically symmetric line element has the
form:
\begin{eqnarray}
  ds^2 &&= g_{tt}(r) dt^2 + g_{rr}(r)dr^2+r^2 d\Omega^2\nonumber\\
  d\Omega^2 &&= d\theta^2+\sin^2\theta d\phi^2  \label{eq:6.1}
\end{eqnarray}
Inside a black hole in the region between the event horizon and the
Cauchy horizon (or the spacelike singularity) $r$ is timelike and $t$
is spacelike. To describe the contraction of the CH, we should thus consider the
variation of the time coordinate $r$ from bigger to smaller values.
The $r-r$, $t-t$ and $\theta -\theta$ components of the Einstein
equations (with $c=1, G=1$) are then given by:
\begin{eqnarray}
\frac{g_{tt} - g_{rr}\,g_{tt} + r\,g_{tt}'}{r^2\,g_{rr}\,g_{tt}} &=&
8\pi \left( T_r^r + E_r^r\right)  \label{eq:6.2}\\
\frac{g_{rr} - {g_{rr}}^2 - r\,g_{rr}'}{r^2\,{g_{rr}}^2} &=&8\pi \left(
  T_t^t + E_t^t\right)  \label{eq:6.3} 
\end{eqnarray}
\begin{eqnarray} 
    &\frac{1}{4r\,{g_{rr}}^2\,  
     {g_{tt}}^2}\{ g_{tt} [ 
        2g_{rr}( g_{tt}' + r\,g_{tt}'' )-(
     r\,g_{rr}'\,g_{tt}' ) ] & \nonumber\\
      &-2{g_{tt}}^2\,g_{rr}' -
        r\,g_{rr}\,{g_{tt}'}^2\} 
=8\pi (    T_\theta^\theta + E_\theta^\theta )&
        \label{eq:6.4}   
\end{eqnarray}
where the primes denotes differentiation with respect to $r$ (the
$\phi -\phi$ component of the Einstein equations again yields 
equation \eqref{eq:6.4}). The tensor $E$ represents here contribution
from a free electric field corresponding to a charge $q$, which we
will assume to be constant:
\begin{equation}
  \label{eq:6.4b}
  E_r^r=E_t^t=-E_\theta^\theta =-\frac{q^2}{8\pi r^4},
\end{equation}
while the tensor $T$ represent contributions from other matter
contents.

To clarify the meaning of different processes we will consider three
different physical matter contents (in addition to the electric field) with
different equations of state. Namely, we will consider: 
\begin{itemize}
\item[{\bf A)}] Dust
\item[{\bf B)}] A massless scalar field
\item[{\bf C)}] Ultrarelativistic gas 
\end{itemize}
From the Einstein equations one can find the following expressions for
the non-zero components of $T$ for these matter contents:\\

{\bf A) Dust (with pressure $P=0$):}
\begin{equation}
 \label{eq:6.5}
      T_r^r =-\epsilon = \epsilon_0 \left( \frac{g_{tt,
    init}}{g_{tt}}\right)^{\frac{1}{2}}\left(\frac{r_{init}}{r}\right)^2 
\end{equation}
where $\epsilon_0, g_{tt,init}$ and $r_{init}$ are constants.\\

{\bf B) Massless scalar field \cite{Burko97b}:} 
\begin{subequations}
 \label{eq:6.7}
\begin{eqnarray}
      T_r^r&&=-\epsilon\\
      T_t^t&&=\epsilon\\
      T_\theta^\theta&&= \frac{\epsilon}{g_{rr}}\\
      \epsilon &&= \epsilon_0 \left( \frac{g_{tt,
    init}}{g_{tt}}\right)\left(\frac{r_{init}}{r}\right)^{4}=\frac{-1}{8\pi g_{rr}}\left(\frac{d\Phi}{dr}\right)^2\\ 
     \epsilon_0&&=\frac{d^2}{8\pi}\frac{1}{g_{tt,init}r_{init}^4}   
\end{eqnarray}
\end{subequations}

where $d$ is a constant which were used in \cite{Burko97b}.\\

{\bf C) Ultrarelativistic gas (with isotropic pressure $P=\frac{\epsilon}{3}$, $\epsilon
  $ being matter density):}  
\begin{subequations}
 \label{eq:6.6}
\begin{eqnarray}
      T_r^r &&=-\epsilon\\
      T_t^t &&= T_\theta^\theta=\frac{\epsilon}{3}\\
    \epsilon &&= \epsilon_0 \left( \frac{g_{tt,
    init}}{g_{tt}}\right)^{\frac{2}{3}}\left(\frac{r_{init}}{r}\right)^{\frac{8}{3}}
\end{eqnarray}
\end{subequations}\\

Substitution of \eqref{eq:6.5}-\eqref{eq:6.6} into \eqref{eq:6.2} and
\eqref{eq:6.3} enables us to find the unknown functions $g_{rr}=g_{rr}(r)$
and $g_{tt}=g_{tt}(r)$ and hence solve the problem for each of the
three cases. Equation
\eqref{eq:6.4} can be used as a control of the calculations.

Formally metric \eqref{eq:6.1} corresponds to a metric of a special
class of the ``homogeneous cosmological models'' considered by
 Zeldovich and Novikov \cite{Zeldovich83a} (page 535), Grishchuk
 \cite{Grishchuk70} and others. 

\begin{figure*}
\subfigure[{\label{fig:6_2a}}]{\includegraphics[width=0.495\textwidth]{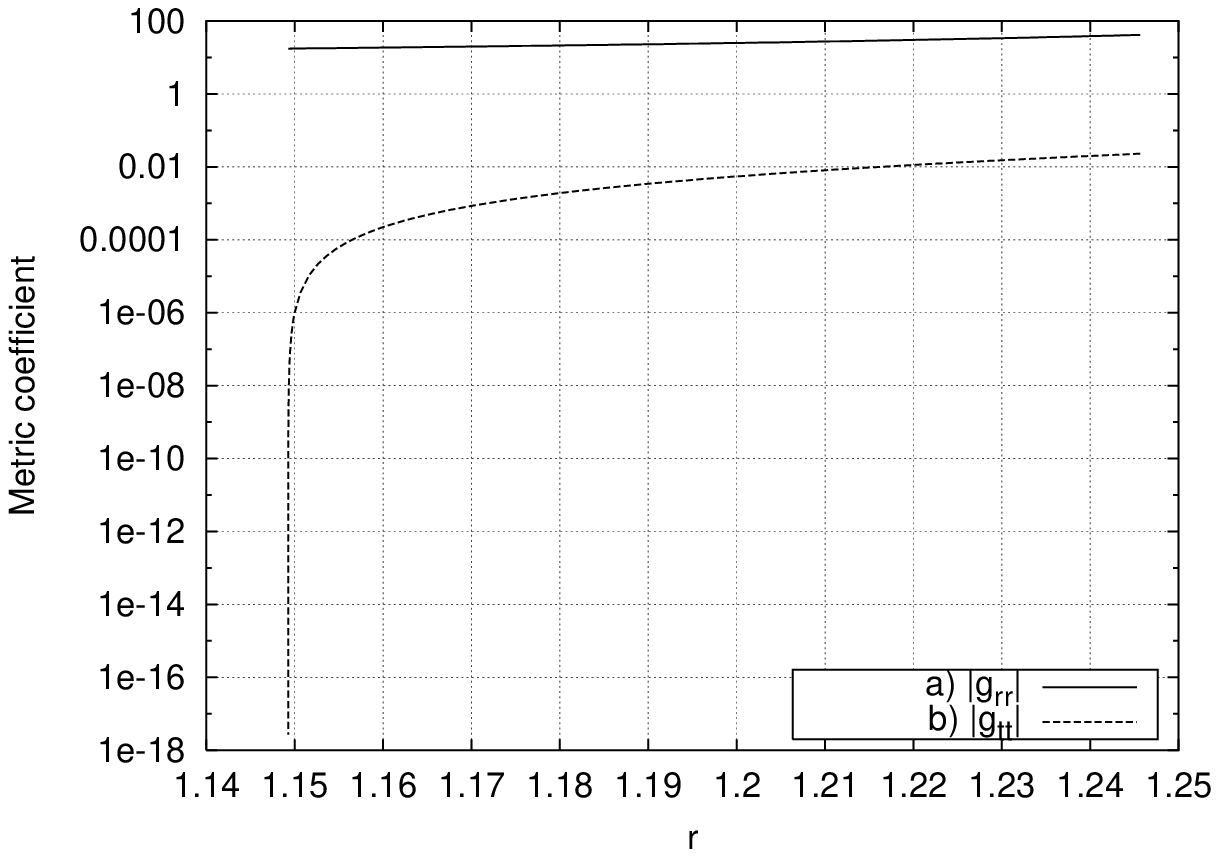}}
\subfigure[{\label{fig:6_2b}}]{\includegraphics[width=0.495\textwidth]{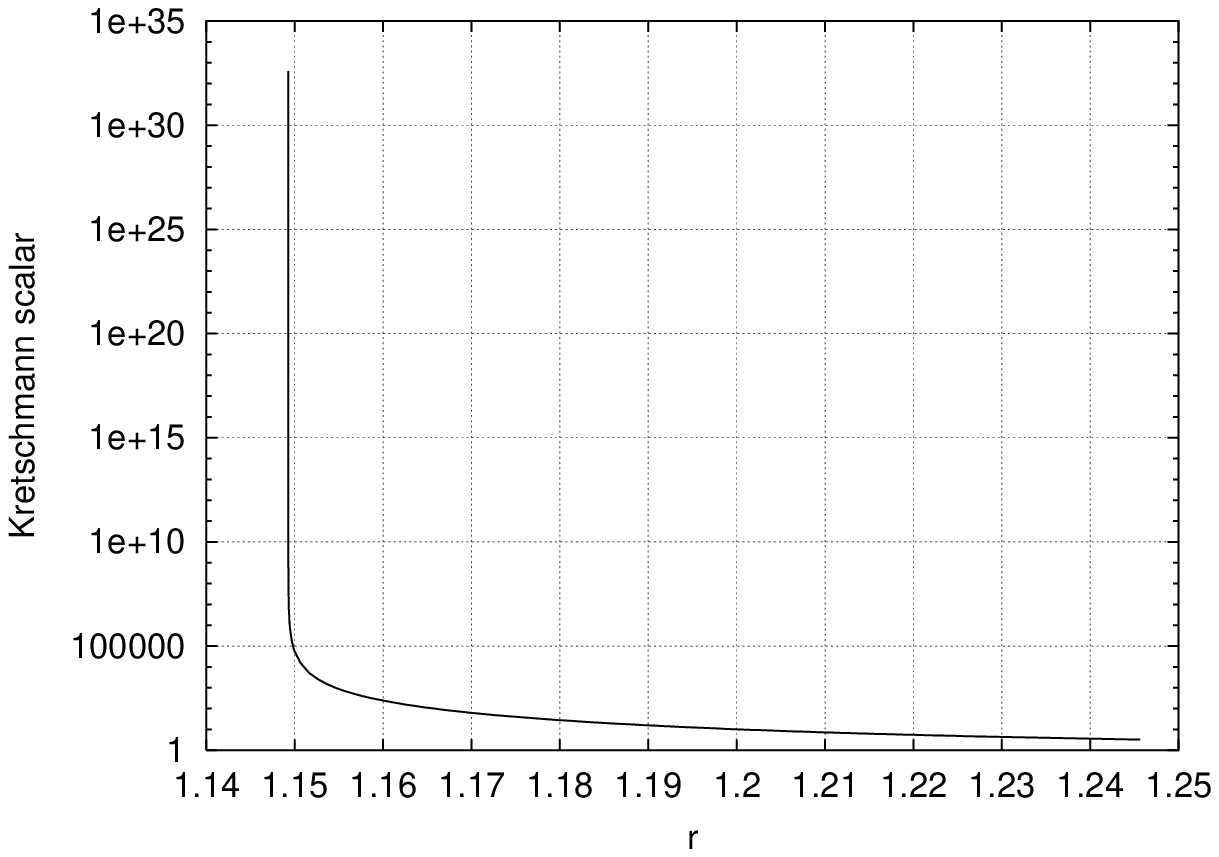}}
\caption{{Metric coefficients (plot a) and Kretschmann
    scalar (plot b) versus $r$ for the case of dust with $\epsilon_0$= 0.03\label{fig:6.2} }}
\end{figure*}
\subsection{Dust, $P=0$}
Let us start the discussion with the simplest case, namely the case of
dust with pressure $P=0$.

In this case it is possible to have a singularity at $r=r_{sing}\neq 0$ with
$r_{CH}< r_{sing}<r_{EH}$, where $r_{CH}$ and $r_{EH}$ are the positions
of the Cauchy Horizon and the Event Horizon in the absence of dust.

Let us consider the leading order terms in a series expansion for
the metric functions and leading order terms in the Einstein
equations, near the singularity. Close to the singularity, where
$g_{tt}\rightarrow 0$, a leading order expansion of
eqs. \eqref{eq:6.2}+\eqref{eq:6.5} gives us:
\begin{equation}
  \label{eq:6.8}
  \frac{dg_{tt}}{dx}\frac{1}{g_{tt}\, r_{sing}}= - \frac{8\pi
  \left( g_{rr}\right)_{sing}\,\epsilon_0} {\left(\frac{g_{tt}}{
  g_{tt,init}}\right)^{1/2}\left(\frac{r_{sing}}{r_{init}}\right)^{2}}  
\end{equation}
where $(g_{rr})_{sing}=g_{rr}(r_{sing})$ and $g_{tt,init}=g_{tt,init}(r_{init})$
and where we assume that $g_{tt}=A x^\alpha$ and $x=r-r_{sing}$. Also
$A$, $\alpha$ are constants and $(g_{rr})_{sing}=g_{rr}(r_{sing})$
is the value of $g_{rr}$ at the singularity
$r=r_{sing}$. 

From \eqref{eq:6.8} one find:
\begin{equation}
  \label{eq:6.10}
  \alpha = 2
\end{equation}
which leads in turn to (remember that $(g_{rr})_{sing}$ is negative for
$r_{CH}<r<r_{EH}$):
\begin{equation}
  \label{eq:6.11}
  r_{sing}=-\frac{4\pi\, (g_{rr})_{sing}\,\epsilon_0\, r_{init}^2\, \sqrt{g_{tt,init}}}{\sqrt{A}}
\end{equation}
Using the proper time $\tau : d\tau =
\sqrt{|g_{rr}|}dr$ we have for the vicinity of the singularity:
\begin{eqnarray}
  & &g_{tt}\propto \tau^2,\nonumber\\ & &r\propto \tau^0=const.,\nonumber\\& &\tau =0 \mbox{ at the
  singularity}  \label{eq:6.12} 
\end{eqnarray}
For the Kretschmann scalar $K\equiv R_{iklm}R^{iklm}$ we have \cite{Burko97b}:
\begin{equation}
  \label{eq:6.13}
  K=\frac{12}{\tau^4}
\end{equation}
Thus this spacelike singularity does not correspond to $r=0$

\subsubsection{Numerical analysis}
To understand the behaviours of the model \eqref{eq:6.1} as functions
of the parameters of the model we perform use a simple numerical code
to numerically solve
\eqref{eq:6.2}+\eqref{eq:6.3} substituting \eqref{eq:6.5} for the
stress-energy tensor. 
According to the remark in the introduction we consider the region
with $K=K_{planck}$ as a physical singularity and will
consider only the region with $r>r_c$, where $r_c$ corresponds to the
critical value of $r$ at which the Kretschmann scalar is equal to the
planckian value. We will take
this restriction into account in all our subsequent analyses.

We note that for the case of dust $P=0$ there are no nonlinear effects causing
an increase of the mass function $m$. This is seen from eq. \eqref{eq:5.3} because $T^t_t = P =
0$. 

To analyse the change of $r_c$ with variation of the matter
contents we numerically integrate
eqs.\eqref{eq:6.2}, \eqref{eq:6.3}, \eqref{eq:6.5}. As initial values
we use $r_{init}=0.95\cdot 
r_{EH}\approx 1.25$ and set $g_{tt,init}$ and $g_{rr,init}$ equal to
their values at $r_{init}$ for the
Reissner-Nordstr{\" o}m solution (with initial mass $m_0=1$ and charge
$q=0.95$) and vary the initial matter density
$\epsilon_0$. 

Figure \ref{fig:6_2a} shows an example of the variation of the metric
functions with $r$ for the case $\epsilon_0 = 0.03$. It is seen that
$g_{tt}\rightarrow 0$ as $r\rightarrow 
r_c\approx 1.149$. As $g_{tt} (r)\rightarrow 0$, the density and curvature 
increases rapidly, which can easily be understood from 
eq. \eqref{eq:6.5}. This is indicated in fig. \ref{fig:6_2b} which
shows the variation of $K(r)$ for the same case. The line in this
figure does not visibly reach $K=K_{planck}\approx 1.5\cdot
10^{131}$, however this is solely 
due to limitations in numerical resolution because of the catastrophic
blowup of $K(r)$ as indicated by the vertical line in the figure.

\begin{figure*}
\includegraphics[width=0.495\textwidth]{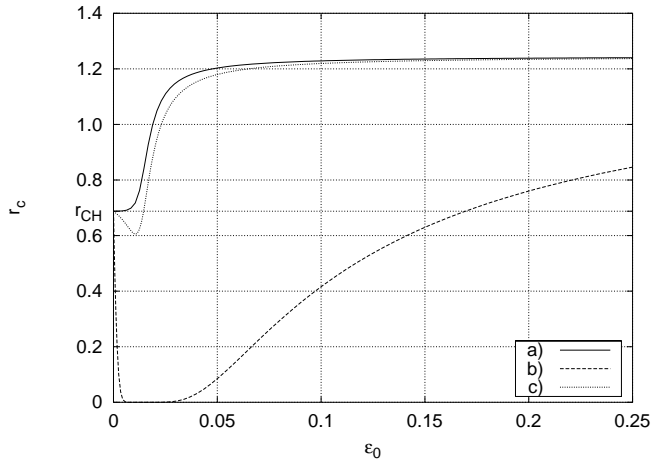}
\caption{Critical value $r_c$ as a function of $\epsilon_0$ for
  a) Dust case, b) Scalar case and c) Ultrarelativistic gas. \label{fig:6.1}}
\end{figure*}

The dependence of $r_c$ on the initial matter density $\epsilon_0$ can
be seen in fig. \ref{fig:6.1}. Near the mathematical singularity,
$r_{sing}$, the scalar $K$ increases very
rapidly with decreasing $r$, so 
approximately $r_{sing}\approx r_c$ (physical singularity). 
As one can see from the figure, for the case of dust, $r_c$ decreases with
decreasing $\epsilon_0$ until $r_c \rightarrow
r_{CH}$ at $\epsilon_0\rightarrow 0$. This behaviour is easily understood: for smaller matter
contents it takes a longer time to compress the dust to the critical density at
$r_c$. On the other hand, in the Reissner-Nordstr{\" o}m solution without
additional matter, the volume of the uniform reference frame
\eqref{eq:6.1} tends to zero when $r\rightarrow r_{CH}$ (because
$g_{tt}\rightarrow 0$). So when $\epsilon_0\rightarrow 0$ and the
behaviours of the solutions are close to the Reissner-Nordstr{\" o}m
solution, the matter density of dust must tend to infinity when the volume
tends to zero at $r\rightarrow r_{CH}$.
Note that this spacelike singularity $r=r_{sing}$ is
not a central singularity $r=0$. The physical singularity, where
$K=K_{planck}$, practically coincide with the mathematical one, where $K=\infty$.

Finally we note that the laws \eqref{eq:6.12}, \eqref{eq:6.13} has been
confirmed by numerical calculations.

\subsection{Massless scalar field}
The case of a scalar field has been analysed by Burko in
\cite{Burko97b}. Here we extend his analysis. 

This case differs drastically from the case of dust. In this case a mathematical
singularity $r_{sing}$ can exist only at $r_{sing}=0$. In the vicinity
of this singularity the solution \eqref{eq:6.2}, \eqref{eq:6.3},
\eqref{eq:6.7} can be written in the first approximation as follows:
\begin{subequations}
  \label{eq:6.14}
\begin{eqnarray} 
    g_{tt}&=&2mCr^\beta\\
    g_{rr}&=&-(\beta +2)\frac{1}{2m}r^{\beta+2}\\
    \Phi &=& \sqrt{\beta +1}\ln r, 
\end{eqnarray} 
\end{subequations}
where $m,C$ and $\beta$ are constants. We note that we use constants
$m$ and $C$ which are different from 
Burko's constants. Our constants have direct physical meanings: $m$ is
the black hole mass, $C$ is a gauge parameter, related to the
possibility of changing the scale of measurement of the $t$ space
coordinate. Also we have $d^2 = \frac{(\beta +1)}{(\beta +2)}4m^2
C$ where $d$ is a constant used by Burko in \cite{Burko97b}. 

The exponent $\beta$ depends on the amplitude of the scalar field. As
Burko demonstrated $\beta > 0$ if $q\neq 0$. So in the vicinity of
the singularity the value of $\left( \frac{d\Phi}{dr}\right)^2$, which
is the only term in the equations \eqref{eq:6.2},
\eqref{eq:6.3},\eqref{eq:6.7}, which 
determines the strength of the scalar field, can not be smaller then
$\frac{1}{r^2}$ (unless it is equal to zero identically). To
understand the behaviour of the singularity in this case let us note
the following;
\begin{figure*}
\subfigure[Line a) $\epsilon_0=0.0001$, b) $\epsilon_0=0.001$,c)
           $\epsilon_0=0.0025$ and d) $\epsilon_0=0.005$
\label{fig:6.3a}]{\includegraphics[width=0.495\textwidth]{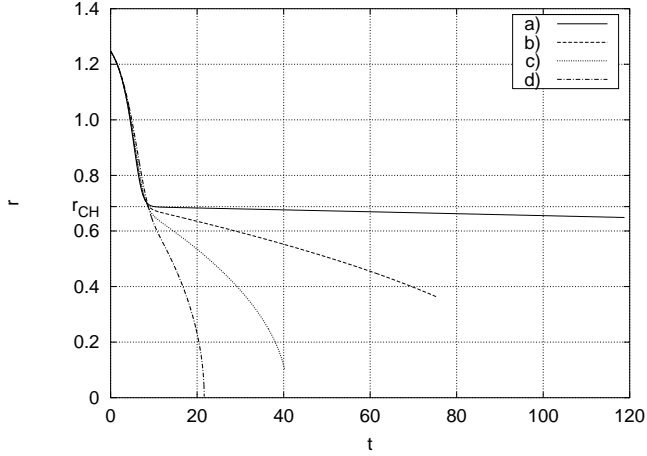}}
\subfigure[Line a) $\epsilon_0=0.010$, b) $\epsilon_0=0.025$, c)
           $\epsilon_0=0.050$, d) $\epsilon_0=0.250$
\label{fig:6.3b}]{\includegraphics[width=0.495\textwidth]{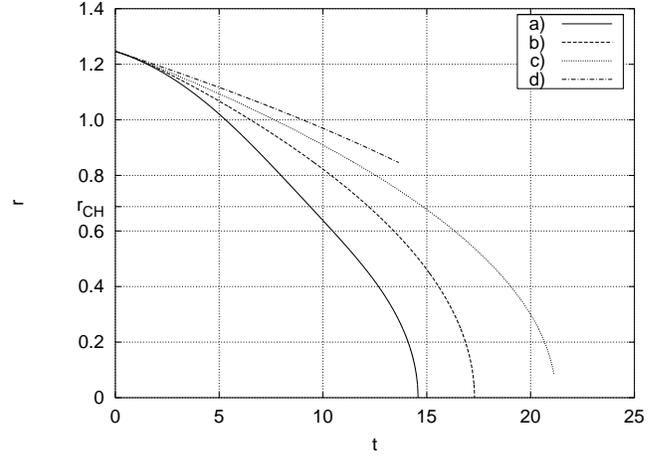}}
\caption{{ $r$ versus $t$ for the scalar case for various
           $\epsilon_0$. \label{fig:6.3}}} 
\end{figure*}

\begin{figure*}
\subfigure[Line a) $\epsilon_0=0.0001$, b) $\epsilon_0=0.001$,
c) $\epsilon_0=0.0025$ and line d) $\epsilon_0=0.005$
\label{fig:6.4a}]{\includegraphics[width=0.495\textwidth]{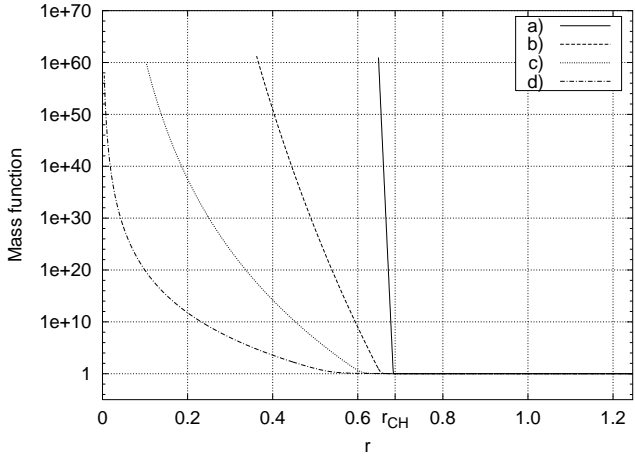}}
\subfigure[Line a) $\epsilon_0=0.010$, b) $\epsilon_0=0.025$,
c) $\epsilon_0=0.050$ and line d) $\epsilon_0=0.250$
\label{fig:6.4b}]{\includegraphics[width=0.495\textwidth]{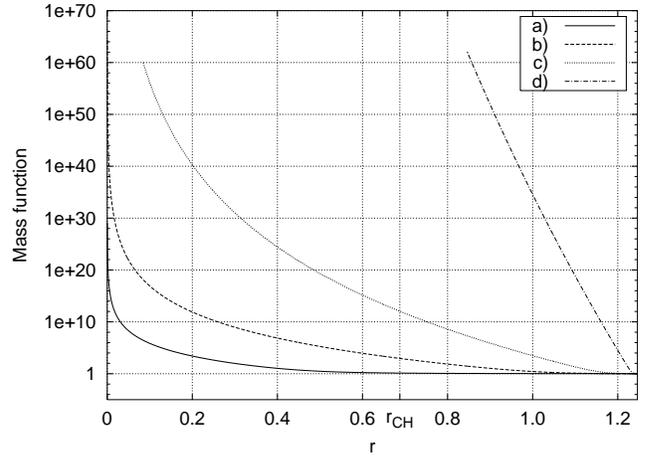}}
\caption{{Mass function versus $r$ for the scalar case for
    various $\epsilon_0$.\label{fig:6.4}}}
\end{figure*}

\begin{figure*}
\subfigure[$\epsilon_0=0.0001$\label{fig:6.5a}]{\includegraphics[width=0.495\textwidth]{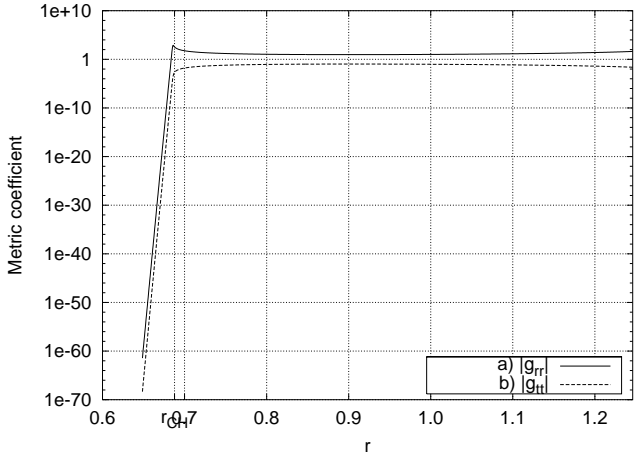}}
\subfigure[$\epsilon_0=0.01$\label{fig:6.5b}]{\includegraphics[width=0.495\textwidth]{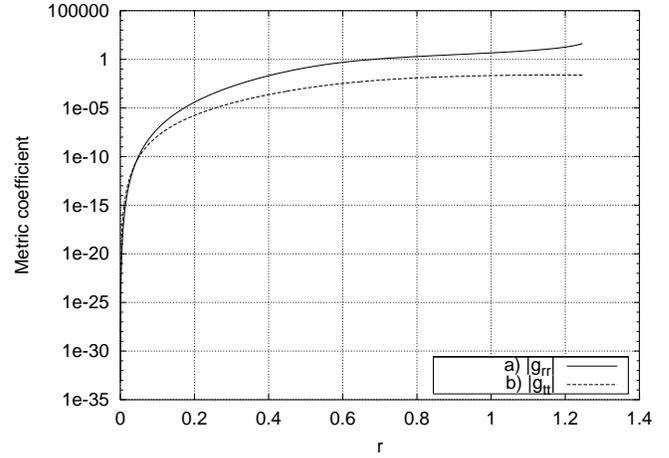}}
\subfigure[$\epsilon_0=0.05$\label{fig:6.5c}]{\includegraphics[width=0.495\textwidth]{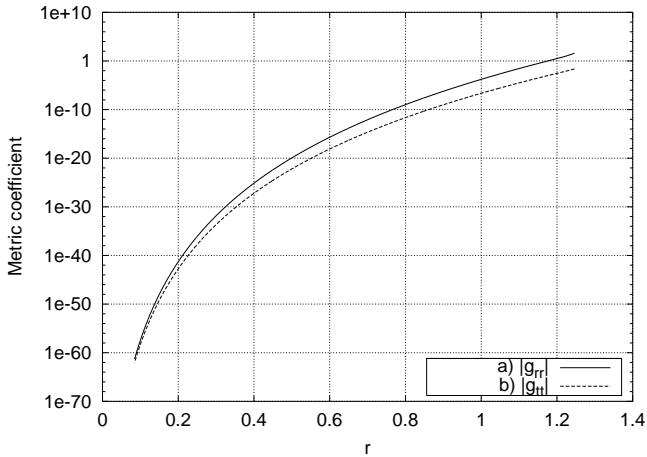}}
\subfigure[Kretschmann scalar for a) $\epsilon_0=0.0001$, b)
$\epsilon_0=0.01$ and c) $\epsilon_0=0.05$,\label{fig:6.5d}]{\includegraphics[width=0.495\textwidth]{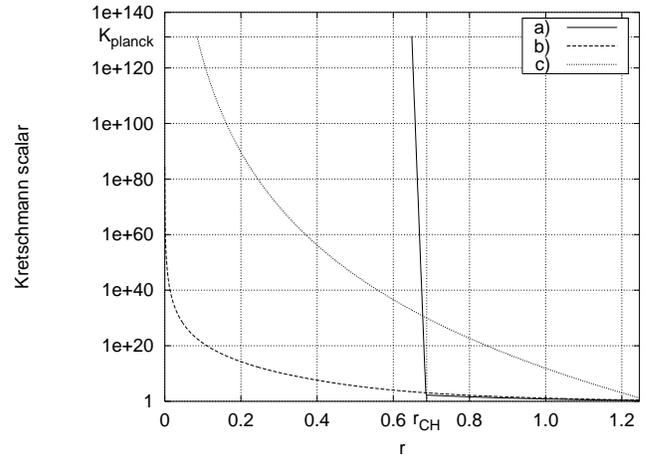}}
\caption{{Metric functions (a-c) and Kretschmann
    scalar (d) versus $r$ for the
    scalar case for various $\epsilon_0$.\label{fig:6.5}}}
\end{figure*}

\begin{figure*}
\subfigure[\label{fig:6.6a}]{\includegraphics[width=0.495\textwidth]{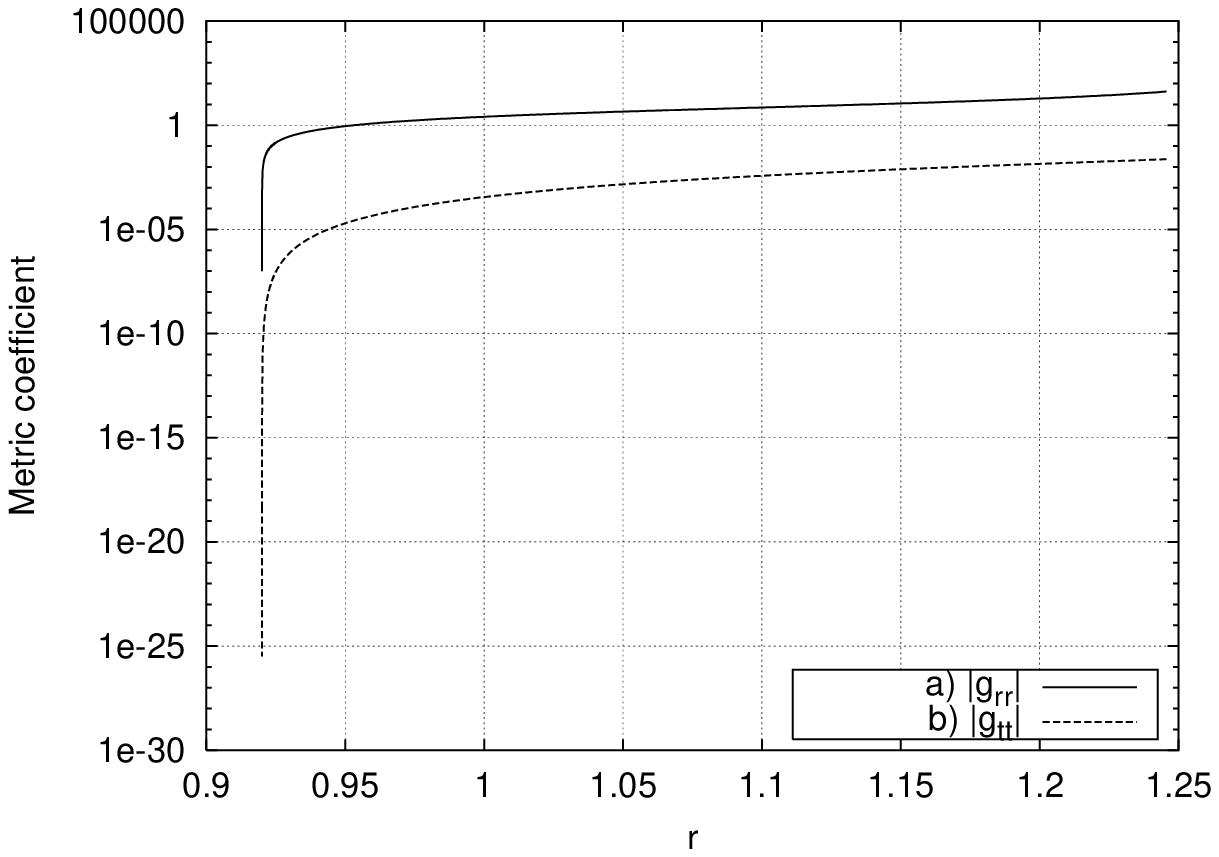}}  
\subfigure[\label{fig:6.6b}]{\includegraphics[width=0.495\textwidth]{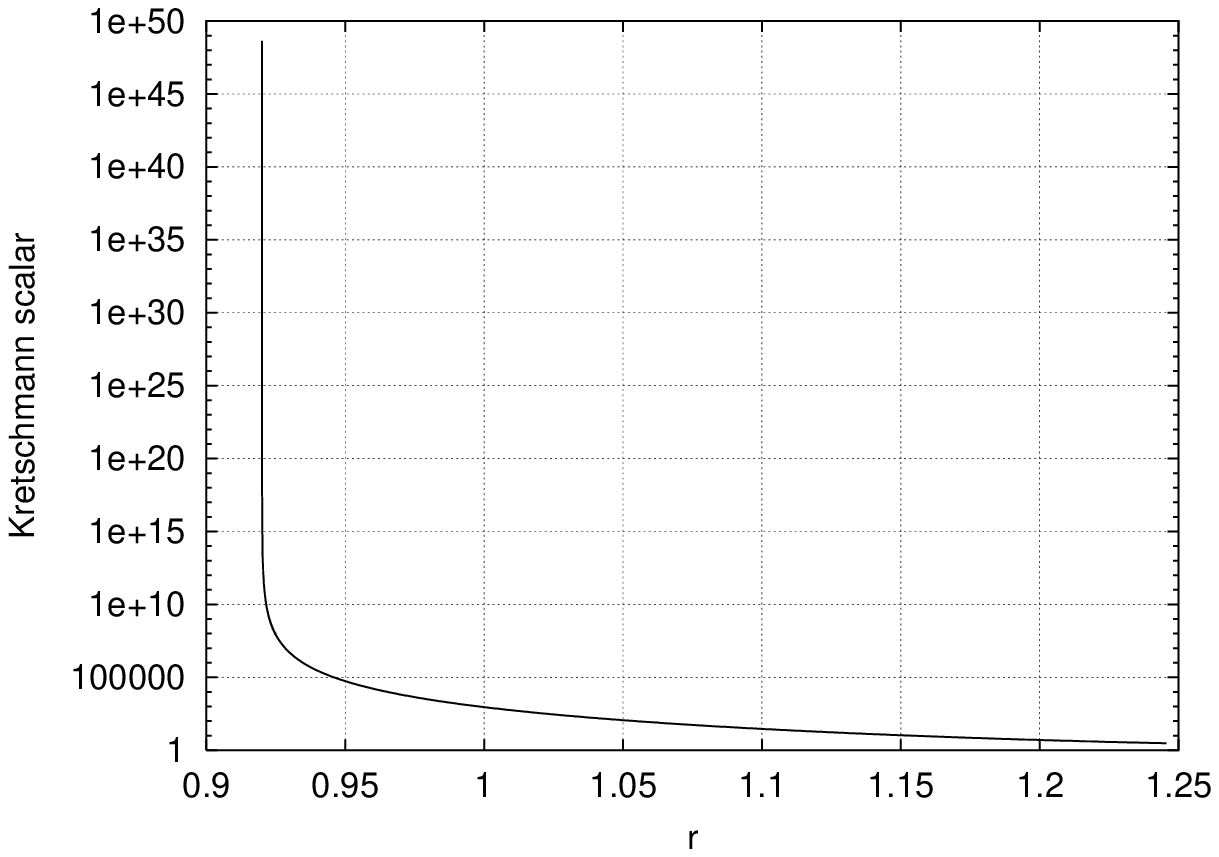}}
\caption{{Metric functions (plot a) and Kretschmann
    scalar (plot b) versus $r$ for the case of
    ultrarelativistic gas for $\epsilon_0=0.02$.\label{fig:6.6}}}
\end{figure*}

In the metric \eqref{eq:5.4} the scalar field can be
represented as a sum of two equal fluxes moving in opposite
directions along the (spacelike) $t$-axis with the fundamental velocity
$c$. Indeed, let us suppose that in $u,v$ coordinates there are
everywhere and always two equal, opposite directed, fluxes along these
coordinates, hence 
$\frac{d\Phi}{du}=\frac{d\Phi}{dv}$ which depend on $r=u+v$, but not
on $t=u-v$. Then there is a coordinate transformation: 
\begin{equation}
  \label{eq:6.15}
  u=R-t, v=R+t
\end{equation}
which corresponds to a transformation to the metric \eqref{eq:6.1} but
with another time coordinate $R : dR =
\sqrt{\frac{-g_{rr}}{g_{tt}}}dr$. The transformation \eqref{eq:6.15}
corresponds to a transformation of the tensor of the scalar field:
\begin{subequations}
  \label{eq:6.16}
  \begin{eqnarray}
    T_{rr}&=&T_{tt}=\tilde{T}_{uu} + \tilde{T}_{vv}\\
    T_{\theta\theta} &=&\tilde{T}_{\theta\theta}\\ T_{\phi\phi}&=&\tilde{T}_{\phi\phi}\\
    \mbox{All other } T_{ik}&=&0 
  \end{eqnarray}
\end{subequations}
Applying transformation \eqref{eq:6.15} to \eqref{eq:6.7}, the new
energy-momentum tensor depends on the 
timelike coordinate $r$ but not
on $t$. The existence of two opposite fluxes near the Cauchy Horizon
should lead to two nonlinear effects: mass inflation and shrinking of the
CH down to $r=0$. The uniformity and equality of the two fluxes lead to
the situation where both effects manifest themself simultaneously and
there are not any gradients in space. To see these effects we
perform the numerical integration of the system
\eqref{eq:6.2},\eqref{eq:6.3},\eqref{eq:6.7}. Here, as for the case of dust,
we start the computation from $r_{init}=0.95\cdot r_{EH}\approx 1.25$, put the
initial values of $g_{tt}$ and $g_{rr}$ equal to their values for the zero
matter content Reissner-Nordstr{\" o}m solution (with $q=0.95$,
$m=1.0$) at $r_{init}$ and vary the characteristic of the initial
amplitude of the scalar field, $\epsilon_0$.

In fig. \ref{fig:6.3} one can see the
propagation ($r$ vs. $t$)
of the ingoing signal with the velocity $c$ ($c=1$) in models with
different $\epsilon_0$. 
Fig. \ref{fig:6.4} shows the mass function as a
function of $r$ for the
same choices of $\epsilon_0$ and
fig. \ref{fig:6.5} presents examples of the
evolution of the metric functions, $g_{tt}$ and $g_{rr}$,
and $K$ in the models with different $\epsilon_0$. Also we refer to
fig. \ref{fig:6.1}, (line b) depicting $r_{c}$ as
a function of $\epsilon_0$.

From these figures it is clearly seen that in models with very
small $\epsilon_0$ (e.g. $\epsilon_0=0.0001$) there is a
manifestation of a mass inflation at $r$ 
close to the CH. First of all we see that the light signal propagates
along $r\approx r_{CH}$ during a long period (line ``a'' in
fig. \ref{fig:6.3a}). This is a nessesary condition for mass 
inflation to occur. 
Secondly, we see more directly that the
massfunction, which was small at large $r$, starts to  
manifest mass inflation at $r$ close to $r_{CH}$
(line ``a'' in fig. \ref{fig:6.4a}). The metric functions $g_{tt}$ and
$g_{rr}$ behaves 
like the case of the pure Reissner-Nordstr{\" o}m solution at larger
$r$, but in the vicinity of $r_{CH}$ they start to collapse
(fig. \ref{fig:6.5a}). We also see that $K$
demonstrates a sudden sharp increases at $r$ close to $r_{CH}$
(fig. \ref{fig:6.5d}, line ``a''), and it reaches the $K_{planck}$
value at $r_c$ 
close to $r_{CH}$ before the shrinkage of the CH manifests itself strongly
(fig. \ref{fig:6.5d}). Thus here
we have the \textit{physical} singularity at $r$ close to
$r_{CH}\neq 0$.

At larger $\epsilon_0$, the term
associated with scalar matter in the Einstein equations 
starts to dominate over a term which represents the electric charge much earlier,
hence the manifestation of the electric field (which is responsible for the
origin of the CH) is not so essential in this case. Functions $g_{tt}$ and $g_{rr}$
differ from the case of the Reissner-Nordstr{\" o}m at $r$ essentially
larger than $r_{CH}$ (see figs. \ref{fig:6.5b},\ref{fig:6.5c} and
compare with fig.\ref{fig:6.5a} which essentially behaves like the
Reissner-Nordstr{\" o}m solution for $r>r_{CH}$ ). For the cases $\epsilon_0\ge
0.001$, the light signal  propagates at $r$ close to 
$r_{CH}$ for a very short period of time (line ``b,c,d'' on
Fig. \ref{fig:6.3a}). For $\epsilon_0 \ge 0.01$ the light signal does
not feel the presence of $r_{CH}$ at all (see fig. \ref{fig:6.3b}). For
$0.01\le \epsilon_0\le0.03$, we observe the shrinkage of the model to
$r$ close to $r=0$ before $K$ reaches $K_{planck}$ (see
line ``b''  on fig. \ref{fig:6.1}). So for these values of $\epsilon_0$ the physical
singularity is at $r$ close to $r=0$. 

At big values of
$\epsilon_0$ (for example $\epsilon_0 \ge 0.050$) there is not any
manifestation of the effects near $r=r_{CH}$ because in this case the light 
signal does not propagate long enough along $r\approx r_{CH}$ for mass
inflation to occur. The mass function nevertheless increases
impetously with decreasing $r$ due to the compression of the model and
$K$ reaches the critical value $K_{planck}$  at rather big $r$ (see
figs. \ref{fig:6.1}, \ref{fig:6.5c} and \ref{fig:6.5d}).

\subsection{Ultrarelativistic gas, $P=\frac{\epsilon}{3}$}
The case $P=\frac{\epsilon}{3}$ is in some sense intermediate between
the cases of pressureless dust and scalar field as it is seen in
fig. \ref{fig:6.1}. In fig. \ref{fig:6.6} one can
see the contraction of the model and 
corresponding increase of the Kretschmann scalar $K$. There is not any
manifestation of the mass inflation near $r\approx r_{CH}$, but only
the nonlinear effect of the increase of the mass function because of
the matter squeeze.

\section{\label{sec:7}Physics of the interior}
\begin{figure*}
\subfigure[Lines of constant $\log_{10} \left(
  T_{vv}\right)$. Lines are from
$\log_{10} \left( T_{vv}\right)=-10.05$ to $\log_{10} \left(
  T_{vv}\right)=-0.85$ in $\Delta \log_{10} \left( T_{vv}\right)=0.20$
  intervals. Thick dotted line marks 
$\log_{10} \left( T_{vv}\right)=-5.45$. Fully drawn thick line marks
apparent horizon. \label{fig:7.1a}]{\includegraphics[width=0.495\textwidth]{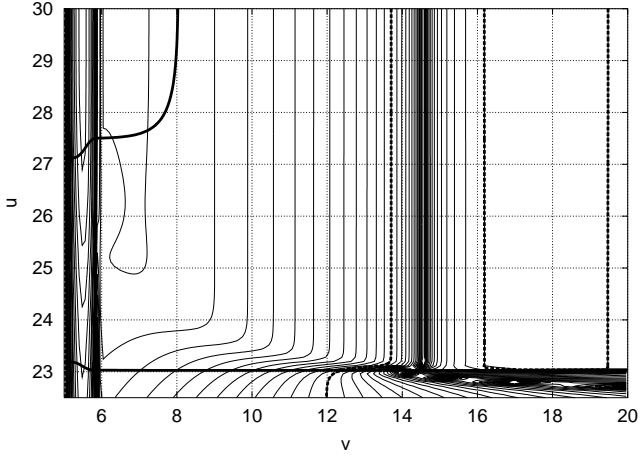}}  
\subfigure[$\log_{10} \left( T_{vv}\right)$ along $u=26.00$.
\label{fig:7.1c}]{\includegraphics[width=0.495\textwidth]{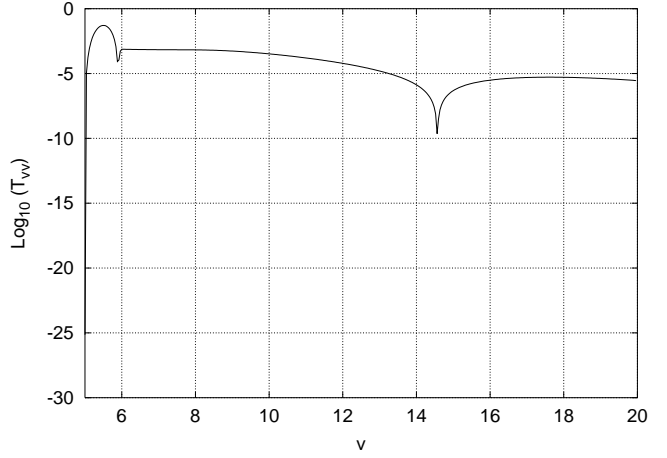}}
\subfigure[Lines of constant $\log_{10} \left(
  T_{uu}\right)$. Lines are from
$\log_{10} \left( T_{uu}\right)=-10.0$ to $\log_{10} \left(
  T_{uu}\right)=-2.00$ in  $\Delta \log_{10} \left(
  T_{uu}\right)=0.25$ intervals. Thick dotted line marks 
$\log_{10} \left( T_{uu}\right)=-3.25$. Fully drawn thick line marks
apparent horizon. 
\label{fig:7.1d}]{\includegraphics[width=0.495\textwidth]{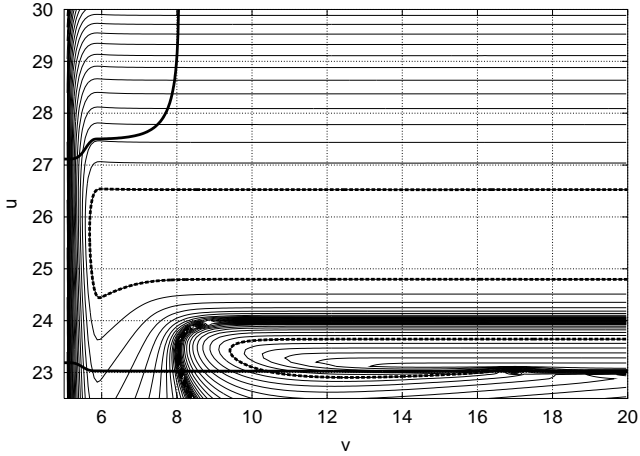}} 
\subfigure[$\log_{10} \left( T_{uu}\right)$ along
$v=10.00$. \label{fig:7.1e}]{\includegraphics[width=0.495\textwidth]{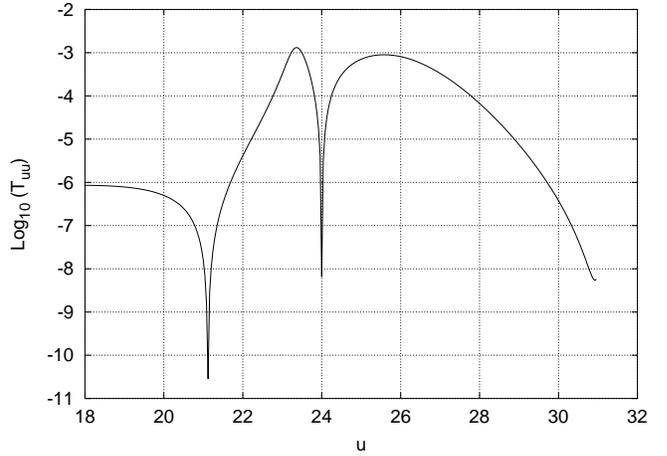}} 
\caption{{\label{fig:7.1} $T_{vv}$ and $T_{uu}$
    for the simple compact pulse, case: $\Delta = 1.0$, $A=0.05$.}} 
\end{figure*} 
 In this section we analyse the fully nonlinear processes inside a
 spherical, charged black hole with a scalar field, as described in
 section \ref{sec:2}, using results from numerical simulations. Our
 numerical code is described in appendix \ref{sec:3} and tested in
 appendix 
 \ref{sec:4}. As mentioned in the introduction, some parts of this
 problem have been discussed in works  \cite{Burko97c, Burko02,
 Burko02b}. In this section we extend these analyses and reveal new
 aspects of the  problem. In subsection \ref{subsec:7.1} we
 investigate a simple compact pulse. In subsection
 \ref{subsec:doublepulse} we investigate a  somewhat more complicated
 compact pulse and in subsection \ref{subsec:7.3} we investigate the
 influence of the $T_{uu}$ flux on the singularities.

 To perform this analysis we specify different boundary conditions
 along some initial $u=u_0=0.00$ and $v=v_0=5.00$ to imitate some physical fluxes
 of energy into the charged black hole, perform numerical simulations
 and analyse the results. Throughout this section, the black hole,
 prior to influence from scalar pulses, has initial mass $m_0=1.00$ and
 charge $q=0.95$. Also, our domain of integration is from $5.0<v<20.0$ and
 $0.0<u<30.0$. For all simulations the gauge is chosen as described in
 section \ref{sec:2}.  

\subsection{\label{subsec:7.1}Simple compact pulse}
We start from the simplest case when the flux of the scalar field
into the charged black hole is
specified along initial $u=u_0$ outside of the black hole in the
following way: 
\begin{equation}
  \label{eq:7.1}
  \Phi_{,v} (u_0, v)= A \sin^2 \left(\pi\frac{v-v_0}{v_1-v_0}\right)
\end{equation}
where $v_0$ and $v_1$ marks the beginning and end of the ingoing
scalar pulse, respectively (i.e. we set the beginning of the pulse
equal to the beginning of our computational domain) and $A$ measures
the amplitude of the pulse. This can readily be integrated to give: 
\begin{equation}
  \label{eq:7.1.1}
    \Phi (u_0, v)= \frac{A}{4\pi} \left( 2\pi\left( v-v_0\right)
    -\left( v_1-v_0\right)\sin\left(2\pi\frac{v-v_0}{v_1-v_0} \right)   \right)
\end{equation}
After the pulse, at $v>v_1$, the flux through $u=u_0$ is set equal
to zero, i.e. $\Phi_{,v}  (u_0,v)=0$. The flux of the scalar field
through initial ingoing segment $v=v_0$ is set equal to zero:
$\Phi_{,u} (u,v_0)=0$. This
means that there is no flux of energy from the surface of a
collapsing charged star into the computational domain. 

Note that we formulate the initial condition directly
for the flux $T_{vv}=(\Phi_{,v})^2/4\pi$ of the scalar field through the surface
$u=u_0$, rather than for $\Phi$ itself since $T_{vv}$ has the direct
physical meaning. Also remember from section \ref{sec:2}, that once the
flux through the two initial surfaces has been chosen, all other
initial conditions are determined by our choice of gauge and the
constraint equations.

In our computations we vary the width of the signal $\Delta
=(v_1-v_0)$, and its amplitude $A$, in a broad range. In
fig. \ref{fig:7.1} is seen a typical example of the evolution of the scalar
field $\Phi$ for the case of $\Delta = 1.00, A=0.05$.
Fig. \ref{fig:7.1a} and \ref{fig:7.1c} represents the evolution of the flux 
$T_{vv}$ of the scalar energy into the black
hole. Fig. \ref{fig:7.1d} and 
\ref{fig:7.1e} shows the $T_{uu}$ flux which arises as a 
result of $T_{vv}$ being scattered by the spacetime curvature.
In fig \ref{fig:7.1a}-\ref{fig:7.1c} the initial pulse (between
$5.0<v<6.0$) and subsequent tails with resonances are clearly seen.
In different regions, $T_{uu}$ and $T_{vv}$ are converted
into one another due to curvature and resonances. In some
regions, $T_{vv}$ is locally greater than $T_{uu}$, it is especially
noted that the highest local flux is $T_{vv}$ inside the pulse (between
$5.0<v<6.0$, fig. \ref{fig:7.1c}). 

We will now consider some direct effects related to these fluxes.

\subsubsection{Focusing effects}
\begin{figure*}
\subfigure[IAH and mass function. Line a) is the IAH (left and bottom
axis). Lines b)-d) represents the mass function along $u=27.338, u=27.533$ and
$u=27.884$ respectively (right and bottom
axis).\label{fig:7.2a}]{\includegraphics[width=0.495\textwidth]{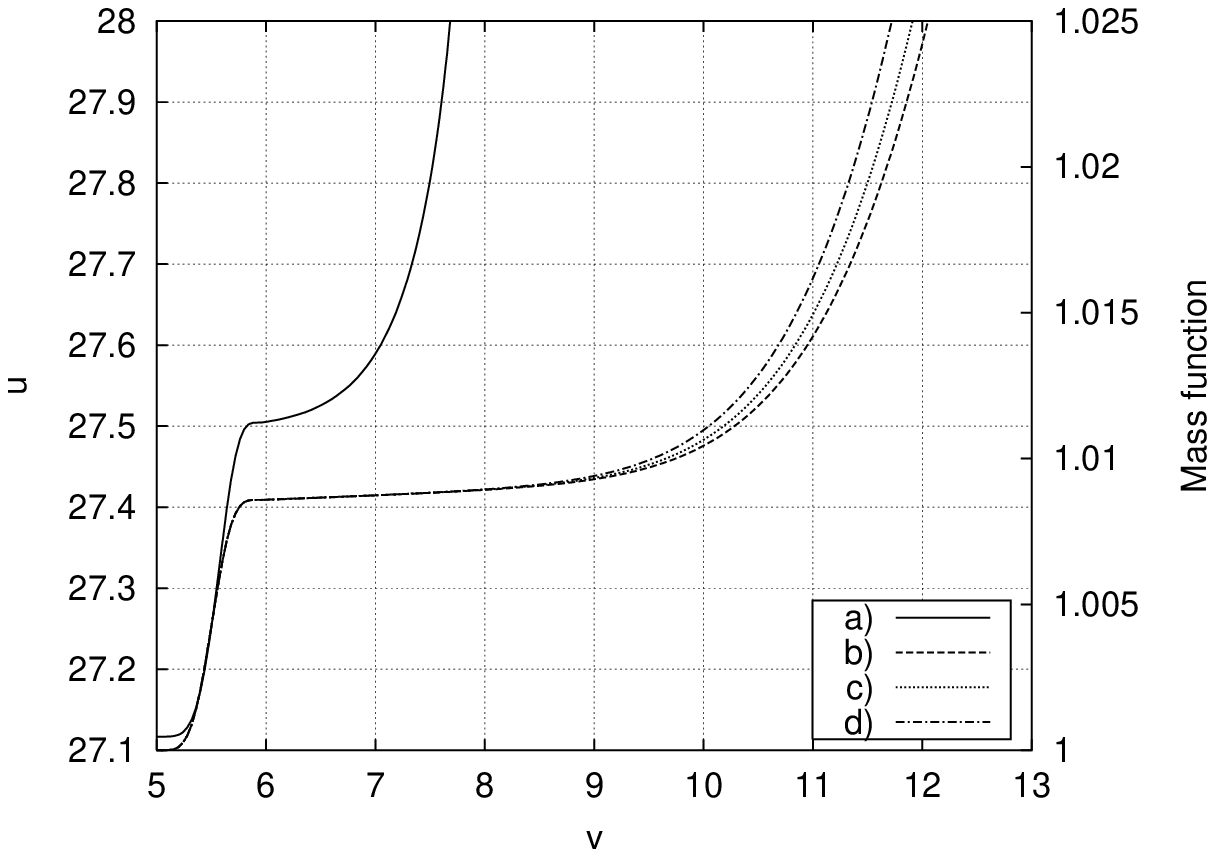}}  
\subfigure[Mass function along lines of constant u. Separation between
lines is $\Delta u=0.40$, bottom  line is along $u=24.00$, top line is along
$u=30.00$, thick dotted line is along
$u=26.00$.\label{fig:7.2b}]{\includegraphics[width=0.495\textwidth]{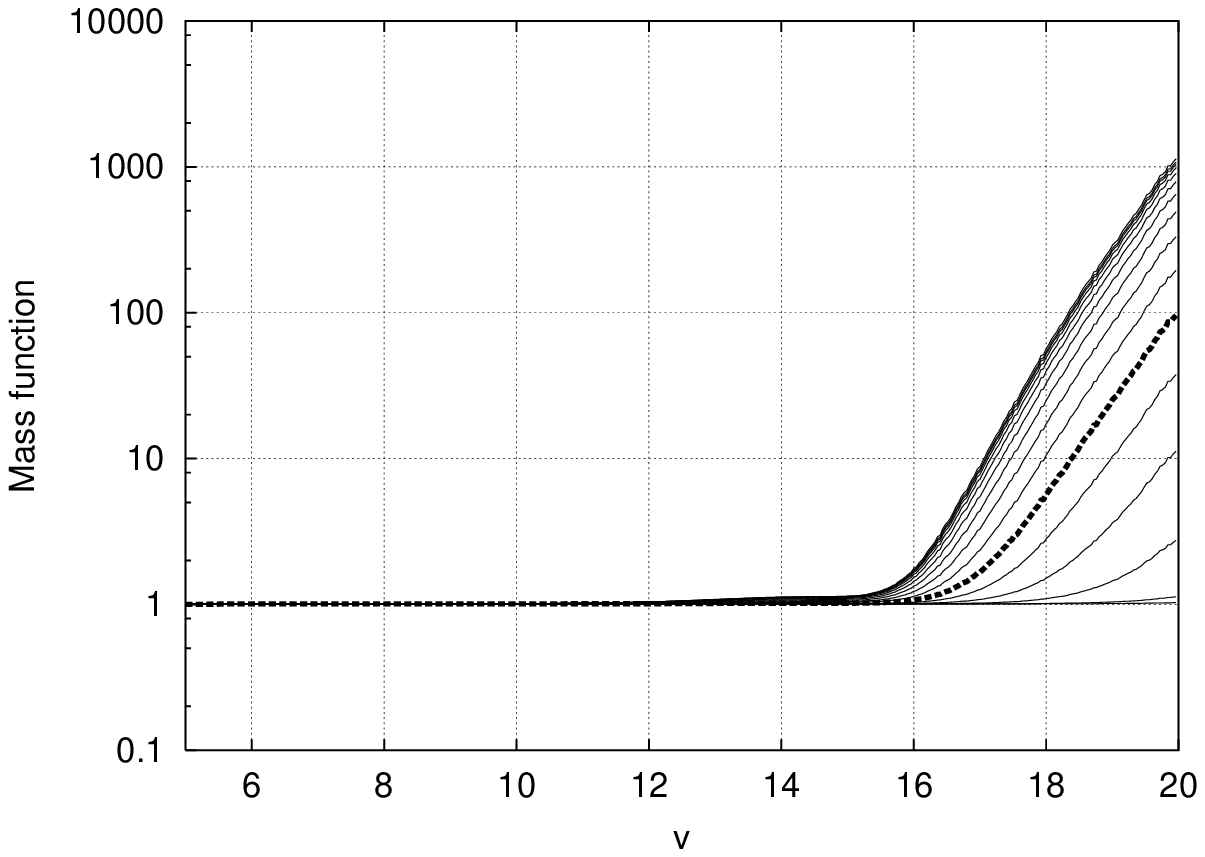}} 
\caption{{Illustrations of the mass function for
    the simple compact pulse, case: $\Delta =1.0$, $A=0.05$.\label{fig:7.2}}}
\end{figure*}
\begin{figure*}
\subfigure[Lines are from $r=0.680$ (bottom line) to $r=0.660$ (top left line) in
$\Delta r=0.001$ increments. Thick dotted line is
$r=0.669$.\label{fig:7.3b}]{\includegraphics[width=0.495\textwidth]{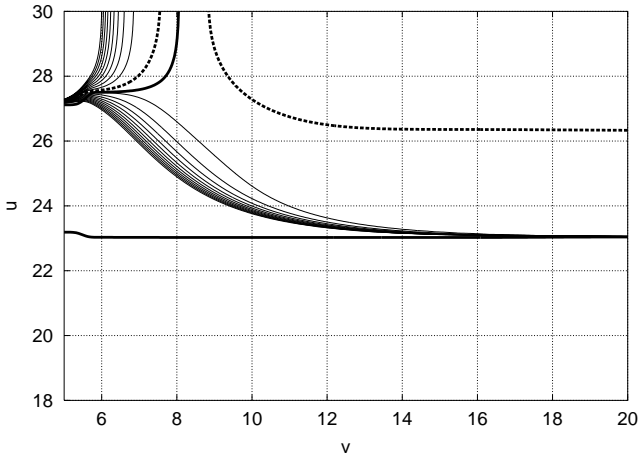}} 
\subfigure[Lines are from $r=0.68710$ to $r=0.68665$ in $\Delta
r=5\cdot 10^{-5}$ increments. Thick dotted line is
$r=0.68690$.\label{fig:7.4b}]{\includegraphics[width=0.495\textwidth]{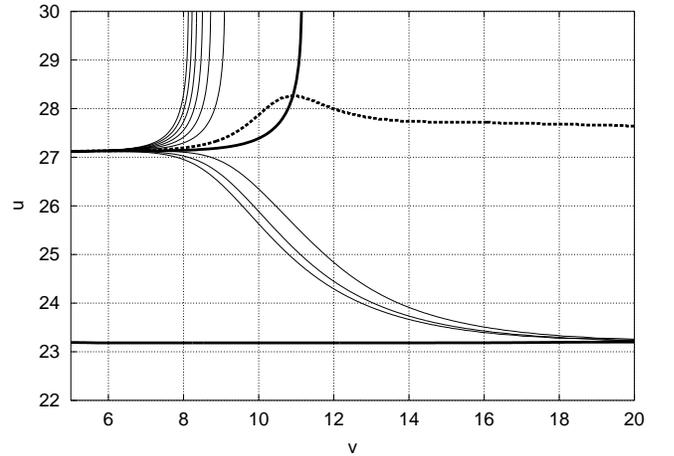}}
\caption{{ 
    Lines of constant $r$ for the simple compact pulse, cases (a): $\Delta =1.0$, $A=0.05$ 
    and (b): $\Delta =1.0$, $A=0.01$. Fully drawn thick line marks
    apparent horizon.\label{fig:7.4}}} 
\end{figure*}
One of the first noticeable effect is that the initial
pulse $T_{vv}$ leads to an initial change of the outer apparent horizon (OAH)
 and inner apparent horizon (IAH) in the region within the pulse
 itself, e.g. $5<v<6$ for fig \ref{fig:7.1}. In fig. \ref{fig:7.2a},
 it can be seen that the mass 
 function, $m$ (eq. \eqref{eq:5.5}) near the IAH increases correspondingly
 as the IAH moves from $u\approx 27.12$ to $u\approx 27.51$. A similar
 change can be seen for the OAH in the same region (e.g. fig. \ref{fig:7.3b}). The
 increase of the mass function and the change of the apparent horizons
 in this region is the trivial effect of mass being pumped into the
 black hole by the $T_{vv}$-flux of the initial pulse.
\begin{figure*}
\subfigure[Lines of constant $\log_{10} \left(
  T_{uu}\right)$. Lines are from
$\log_{10} \left( T_{uu}\right)=-10.0$ to $\log_{10} \left(
  T_{uu}\right)=0.00$ in $\Delta \log_{10} \left( T_{uu}\right)=0.50$ intervals. Thick dotted line marks
$\log_{10} \left( T_{uu}\right)=-1.50$ (decreasing ``outwards''). Fully drawn thick line marks
apparent horizon. \label{fig:7.5a}]{\includegraphics[width=0.495\textwidth]{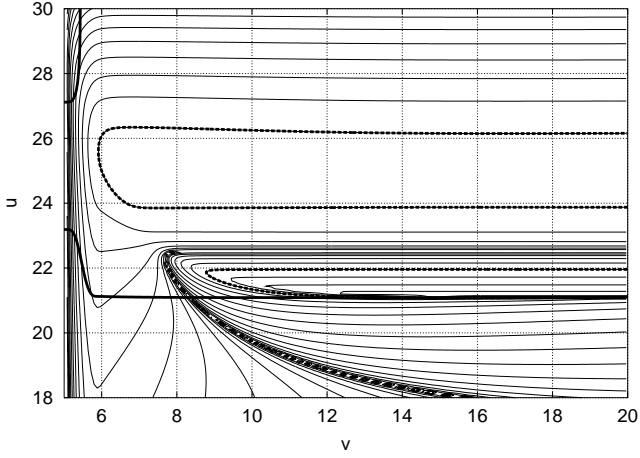}}
\subfigure[$\log_{10} \left(
  T_{uu}\right)$ along $v=15.00$ \label{fig:7.5b}]{\includegraphics[width=0.495\textwidth]{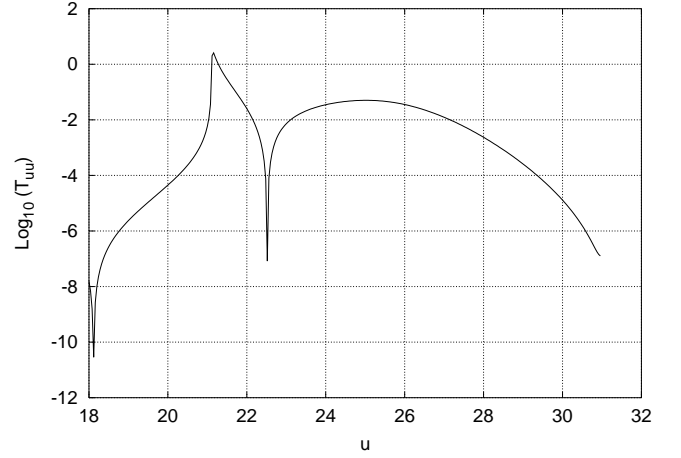}}
\subfigure[Lines of constant $r$. Lines are from $r=0.520$ (bottom left line) to $r=0.475$ (top right
  line) in intervals of $\Delta r = 0.005$. \label{fig:7.5c}]{\includegraphics[width=0.495\textwidth]{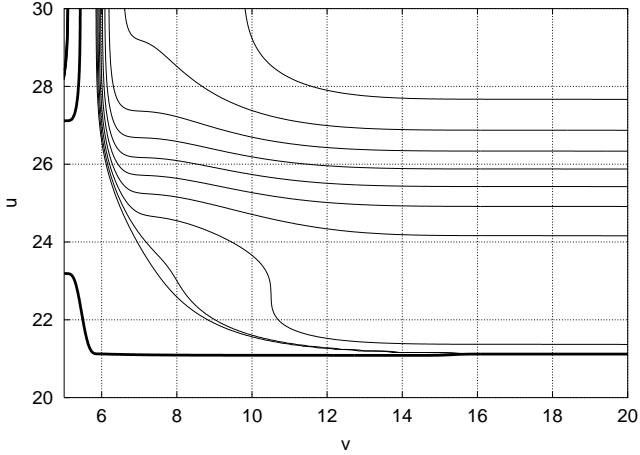}}
\subfigure[Lines of constant $\log_{10} \left(
  T_{\theta\theta}\right)$. Lines are from $\log_{10} \left(
  T_{\theta\theta}\right)=-2.0$ (bottom right line) to $\log_{10} \left(
  T_{\theta\theta}\right)=32.0$ (top right line) in intervals of $\Delta \log_{10} \left(
  T_{\theta\theta}\right)=2.0$. \label{fig:7.6}]{\includegraphics[width=0.495\textwidth]{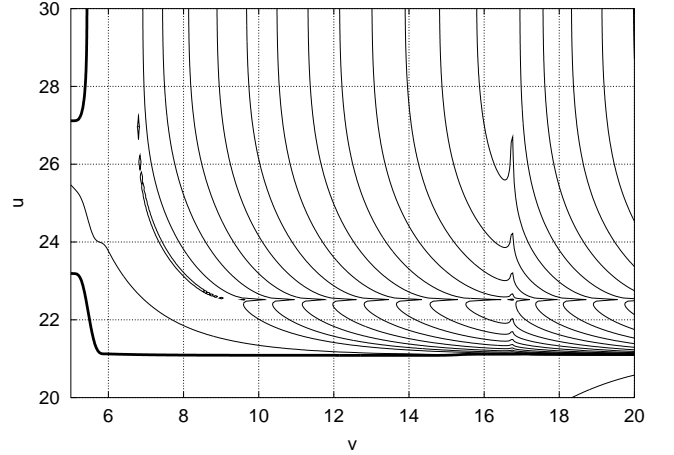}}
\caption{{\label{fig:7.5} Various contour plots
  for the simple compact pulse, case: $\Delta =1.0$, $A=0.20$ }}
\end{figure*}

The dramatic change of the IAH at $v\approx 7-8$, however, is related with
other nonlinear effects. We remember that worldlines of imaginary test
photons along $u=const$ and $v=const$ are under action of the gravity of the
radiation $T_{vv}$ and $T_{uu}$, which leads to a focusing effect.
For example, in the absence of scalar radiation, outgoing photons
along $u=const$ slightly above of the IAH will go to greater
$r$ as $v\rightarrow \infty$ in the  
Reissner-Nordstr{\" o}m solution. With the existence of
scalar radiation, a similar outgoing ray initially slightly above the
IAH will now, because of the focusing effect of the $T_{vv}$ and
$T_{uu}$ radiation, go to smaller $r$ and generate a maxima 
$\left(\frac{du}{dv}=0\right)_{r=const}$ which correspond to the
position of the IAH. This is seen in fig. \ref{fig:7.3b} and more clearly
in fig. \ref{fig:7.4b}. This effect leads to a drastic change of the
shape of the IAH.  It is seen by lines c) and d) in figure
\ref{fig:7.2a} that this change of the IAH in this region ($v\approx
7-8$) is not accompanied by any significant change of the mass
function. About the increase of the mass function at $v>8$ see below. 

In the case of smaller initial amplitude of the pulse $A$, the change
of the shape of IAH due to focusing starts later. For example, for the
case $\Delta = 1$, $A=0.01$ (see figure \ref{fig:7.4b}), the change
starts at $v\approx 10$. In this case the change of the 
OAH and IAH in the region $5<v<6$, related to the initial pulse of the
scalar field, is so small that it is invisible in the figure. 

In figs. \ref{fig:7.5a}-\ref{fig:7.5c} (case: $\Delta=1.0,A=0.2$) one can see
another manifestation of the focusing effect related with the change
of the flux
$T_{uu}$. The figures shows close correspondence between $T_{uu}$
and the rate of focusing of lines of constant $r$. Especially we note
that along the line $u\approx 22.5$ for $v>8$ there is a minimum of
$T_{uu}$ (seen by the collection of very closely spaced lines in
fig. \ref{fig:7.5a} and as the local minima in
fig. \ref{fig:7.5b}). Comparing with figure \ref{fig:7.5c}  we see that the lines of  
$r=const$ shows minimal focusing along this minima, compared to the
focusing at $u<22.5$ and $24<u<28$. At $u>28$ there is a decrease of $T_{uu}$ and
we see a corresponding decrease of the focusing effect.

\begin{figure*}
\subfigure[Mass function along lines of constant $u$. From $u=22.40$
(bottom line) to $u=30.00$ (top line) in $\Delta u = 0.40$
intervals. Thick dotted line is along
$u=26.00$.\label{fig:7.7a}]{\includegraphics[width=0.495\textwidth]{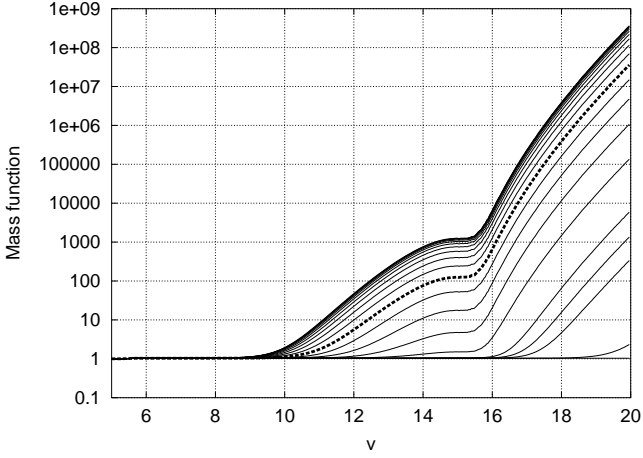}} 
\subfigure[Lines of constant $\log_{10} \left(
  T_{vv}\right)$. Lines are from
$\log_{10} \left( T_{vv}\right)=-10.0$ to $\log_{10} \left(
  T_{vv}\right)=-0.25$ in $\Delta log_{10} \left( T_{vv}\right)=0.25$
intervals. Thick dotted line marks 
$\log_{10} \left( T_{vv}\right)=-4.75$. Fully drawn thick line marks
apparent horizon. Closely spaced lines near $v\approx 15$ inside
apparent horizont marks local
minima.
\label{fig:7.7b}]{\includegraphics[width=0.495\textwidth]{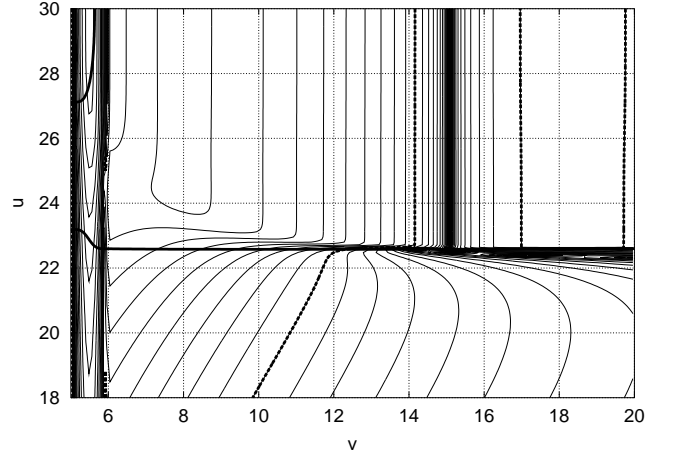}} 
\caption{{\label{fig:7.7}Plots for the simple compact pulse, case: $\Delta =1.0$, $A=0.10$.}}
\end{figure*}
\begin{figure*}
\subfigure[Mass function along
$v=20.00$.
\label{fig:7.8a}]{\includegraphics[width=0.495\textwidth]{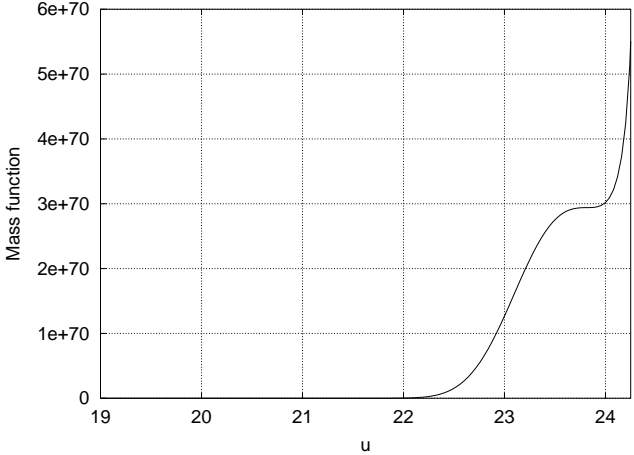}} 
\subfigure[$\log_{10} \left( T_{uu}\right)$ along
$v=20.00$.
\label{fig:7.8b}]{\includegraphics[width=0.495\textwidth]{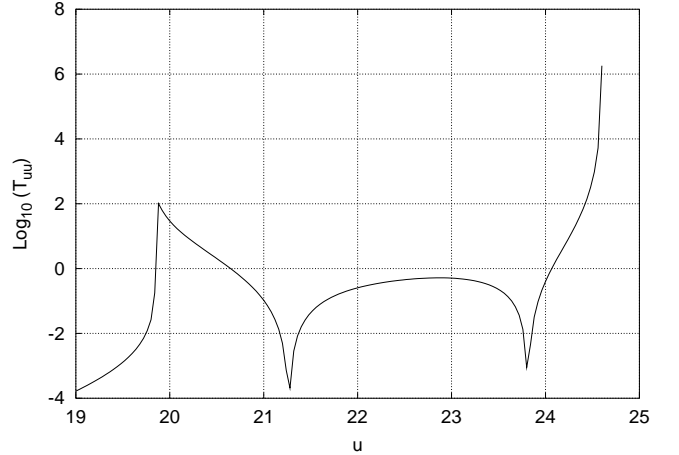}} 
\caption{{\label{fig:7.8} Plots for the simple compact pulse, case: $\Delta =2.0$, $A=0.20$.}}
\end{figure*}

We should also remember that in the dynamic equations (equations
\eqref{eq:evolve1}-\eqref{eq:evolve3}) and the expression for
$T_{\theta\theta}$, 
(eqs. \eqref{eq:energy-momentum-tensor-a} +
\eqref{eq:energy-momentum-tensor-c}) the 
scalar field appears only in the form of the product
$\Phi_{,u}\Phi_{,v}$ and the same nonlinear effects can be described in
terms of $T_{\theta\theta}$ component which is presented in
fig. \ref{fig:7.6}. For example, where $T_{uu}$ has its minima, a
corresponding effect is clearly visible in fig. \ref{fig:7.6}. 
We also note from fig. \ref{fig:7.5} that for this case 
the initial pulse is so strong that its flux changes the IAH inside
the pulse and there are no double turns as it was seen on
fig. \ref{fig:7.1}.

\subsubsection{Mass function}
\begin{figure*}
\subfigure[Mass function versus
$v$.\label{fig:7.2c}]{\includegraphics[width=0.495\textwidth]{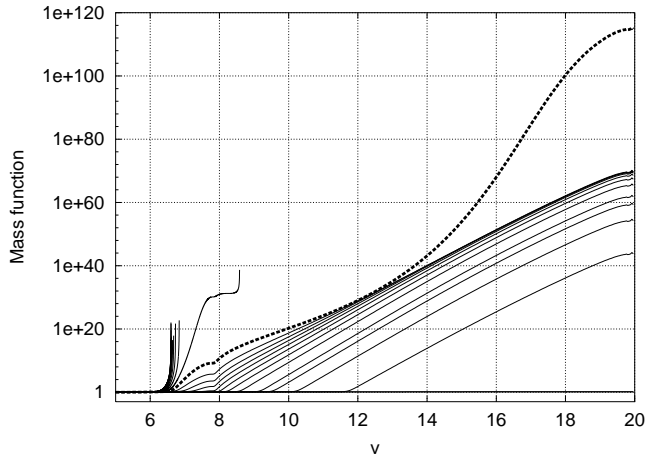}} 
\subfigure[Kretschmann scalar versus $v$. Dotted horizontal line marks
line of planckian curvature. \label{fig:7.12}]{\includegraphics[width=0.495\textwidth]{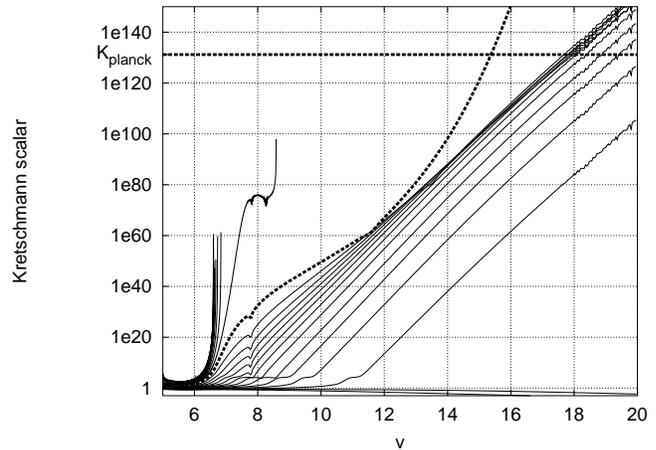}}  
\caption{\label{fig:7.12+7.2c}Mass function and Kretschmann scalar
    along lines of constant $u$ for the simple compact pulse, case: $\Delta =2.0, A=0.2$. Separation between
lines is $\Delta u=0.40$. Lines are from
    $u=19.40$ (lowest right line) to $u=29.80$ (near vertical line in
    lower left corner). Thick dotted line is along $u=24.60$.}
\end{figure*}
Let us come now to the discussion of the behaviour of the
mass function $m$. As we remember, the (modest) increase
of $m$ in the region $5<v<6$ in fig. \ref{fig:7.2a} is related to
the input of the energy in the initial pulse. The increase of $m$ seen in the region $8<V<15$
in fig. \ref{fig:7.2} is related partly with the compression-effect (see
section \ref{sec:5}), but still it is very difficult to separate this effect from the
beginning of the mass inflation. The essentially faster increase of
$m$ at $v>15$ (fig. \ref{fig:7.2b}) is the manifestiation of the mass
inflation when we come to the CH. 

Mass inflation depends mainly on the $T_{vv}$ flux
but also on the $T_{uu}$ flux and, as described in section \ref{sec:5}, the
co-existence of both fluxes
is essential for mass inflation to occur. An example of the dependence on $T_{vv}$ is seen in
fig. \ref{fig:7.7} ( case $\Delta =1.0, A=0.10$). It is seen that
where $T_{vv}$ has a minimum (the narrowly spaced lines at $v\approx 15$) the increase 
of $m$ almost stops. Fig. \ref{fig:7.8} demonstrates the dependence of
mass inflation on  $T_{uu}$ for the stronger pulse: $\Delta =2.0,
A=0.20$. This pulse is so strong that a $r=0$ singularity is formed in the
domain (this is further discussed in the next subsection). At the
minimum of $T_{uu}$ at $u\approx  
23.8$ the increase of $m$ also stops and at $u>24$ where $T_{uu}$ increases
rapidly, $m$ also has similar rapid increase. However, the line plotted
terminates at the $r=0$ singularity, thus the final rapid increase of
mass is a combination of the effects of compression and mass
inflation. 

The compression effect can be more clearly seen in
fig. \ref{fig:7.2c}, which shows
$m$, along lines of constant $u$, again for the case $\Delta =2.0,
A=0.2$. When one comes to the $r=0$ singularity, compression tends to
infinity and we see catastrophic infinite increase of $m$. This can be
seen by the near vertical lines in the lower left in the 
figure, which represents lines of high $u$. These lines  experience a
catastrophic infinite increase of mass as they approach $r=0$, as
indicated by these lines being near vertical.

The remaining right hand side lines which run in all the range
$5<v<20$, represents lines of constant $u$ which reach the CH and the
mass increase along those lines are due to the mass
inflation. The line marked by thick dashes represent a line
that comes close to the point where the $r=0$ and CH singularities
meet. The structure of this line is more complicated as it is
influenced by both processes.
 
Finally, in fig. \ref{fig:7.12} is seen the Kretschmann scalar for the
same lines.

\begin{figure*}
\subfigure[Case: $\Delta =2, A=0.010$. From $u=23.00$ (rightmost line) to
$u=29.80$ (leftmost line).\label{fig:7.9a}]{\includegraphics[width=0.495\textwidth]{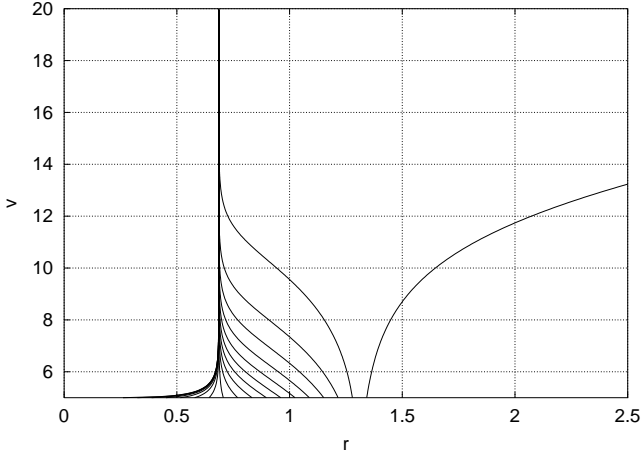}}
\subfigure[Case: $\Delta =2, A=0.125$. From $u=21.40$ (rightmost line) to
$u=29.80$ (leftmost line).\label{fig:7.9b}]{\includegraphics[width=0.495\textwidth]{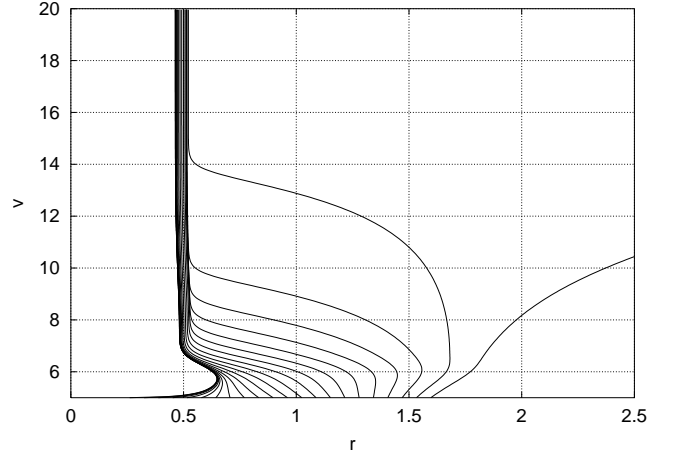}}
\subfigure[Case: $\Delta =2, A=0.180$. From $u=20.20$ (rightmost line) to
$u=29.80$ (leftmost line).\label{fig:7.9c}]{\includegraphics[width=0.495\textwidth]{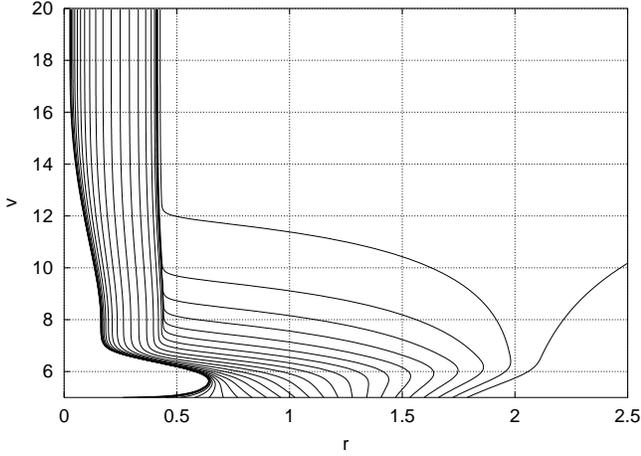}}
\subfigure[Case: $\Delta =2, A=0.200$. From $u=19.80$ (rightmost line) to
$u=29.80$ (bottom leftmost line).\label{fig:7.13}]{\includegraphics[width=0.495\textwidth]{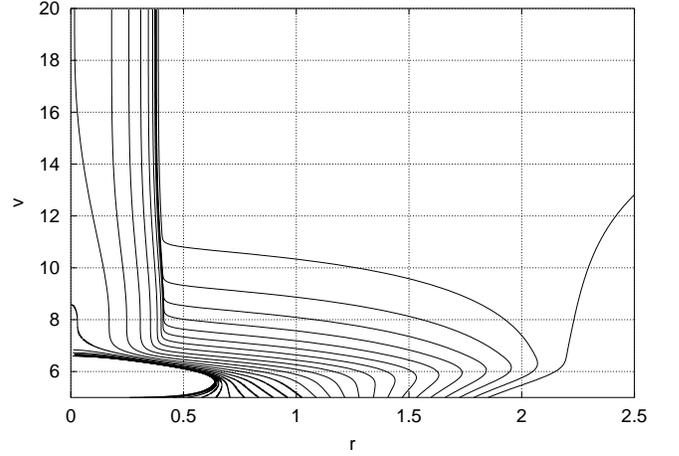}}
\caption{{\label{fig:7.9} $v$ versus $r$ along
    lines of constant $u$ for amplitudes for the simple compact pulse. Separation between
    lines of constant $u$ is $\Delta u = 0.40$.}}
\end{figure*}
\begin{figure*}
\subfigure[Case:$\Delta =2.0$, $A=0.18$. Lines are from 
    $r=0.45$ (bottom left line) to $r=0.03$ (top right line) in
    intervals of $\Delta r =0.01$\label{fig:7.10}]{\includegraphics[width=0.495\textwidth]{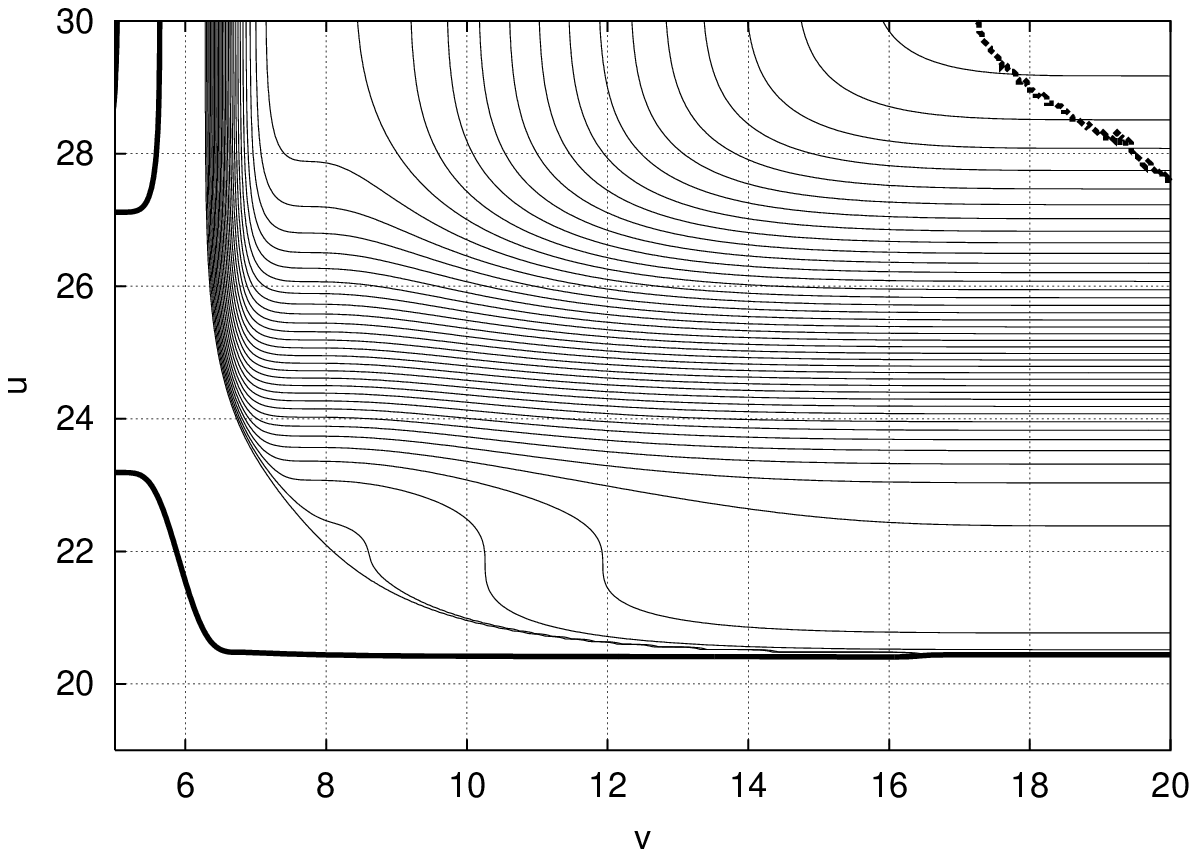}}
\subfigure[Case:$\Delta =2.0$, $A=0.20$. Lines are from 
    $r=0.42$ (bottom left line) to $r=0.20$ (top right line) in
    intervals of $\Delta r =0.01$. Top right thick line mark $r=0$
    singularity. \label{fig:7.11}]{\includegraphics[width=0.495\textwidth]{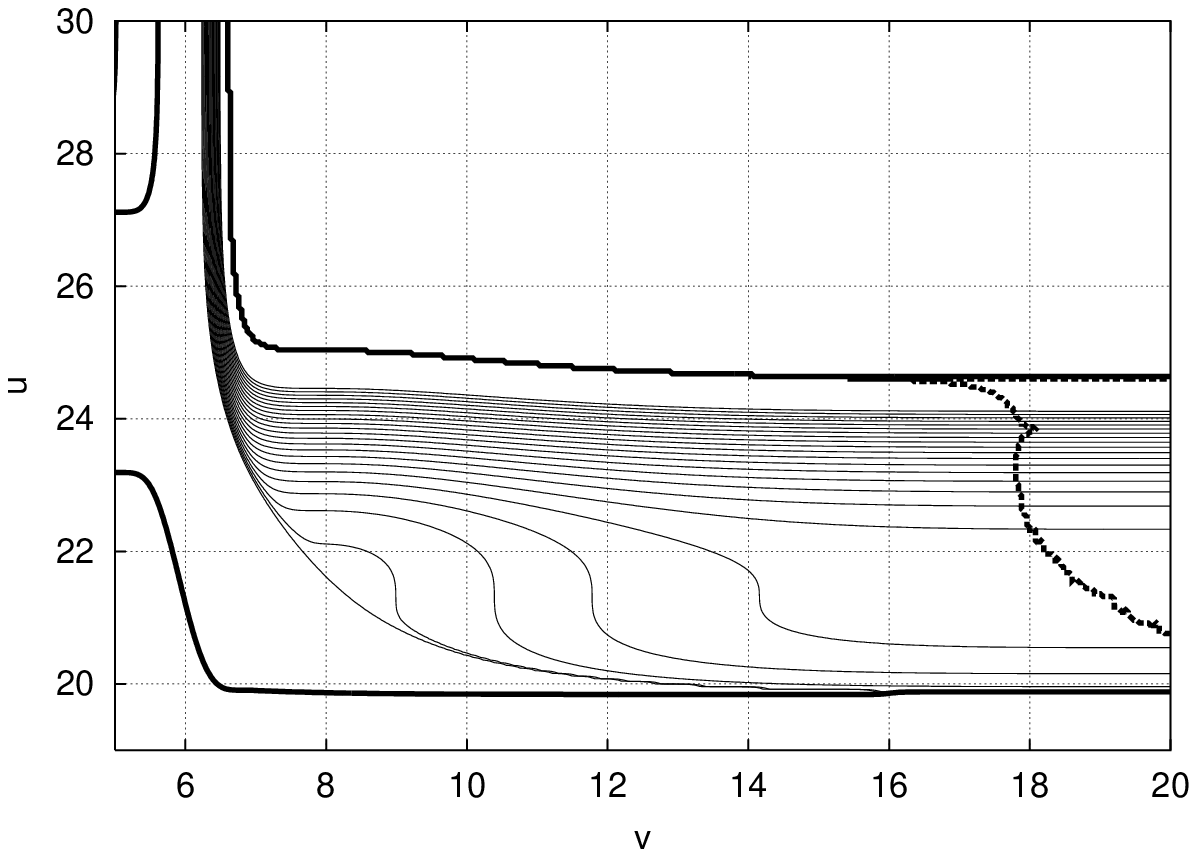}}
\caption{{\label{fig:7.10-11} Lines of constant $r$
    for two different simple compact pulses. Thick dotted lines in right part of the figures represents
    line of planckian curvature. Bottom fully drawn thick line marks OAH.}}
\end{figure*}

\subsubsection{\label{subsec:singularity}The singularity}
Now we will discuss the singularity. When the initial pulse is rather
weak we cannot see the manifestation of the spacelike singularity in
our computational domain. However, we can see the asymptotic approach
of $u=constant$ test photons to the CH singularity. In
fig. \ref{fig:7.9a} we see that for the case $\Delta =2.0, A=0.01$,
all our
test photons come asymptotically to the same value at $r\approx
0.69$, corresponding to the analytical value for $r$ at the CH for the
pure Reissner-Nordstr{\" o}m solution, i.e. the CH singularity itself
does not show any tendency to shrink down (within our computational
domain). 
\begin{figure*}
\subfigure[Lines of constant $\log_{10} \left(
  T_{vv}\right)$. Lines are from $\log_{10} \left(
  T_{vv}\right)=-10.0$ to $\log_{10} \left(
  T_{vv}\right)=2.0$ in intervals of $\Delta \log_{10} \left(
  T_{vv}\right)=0.50$.\label{fig:7.14a}]{\includegraphics[width=0.495\textwidth]{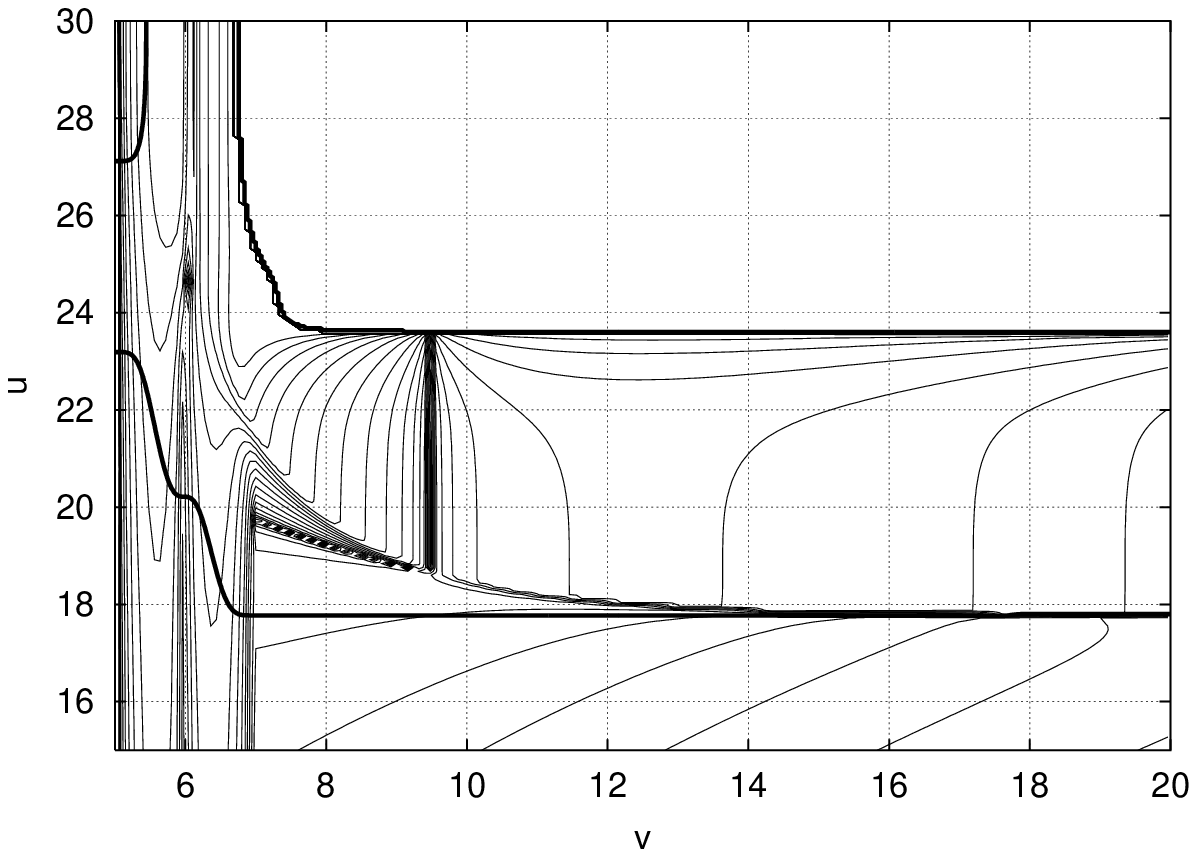}}
\subfigure[Lines of constant $r$. From $r=3.00$ (bottom line) to
$r=0.10$ (top right thin line) in $\Delta r =0.10$ intervals. Thick
dotted line marks $K=K_{planck}$. 
 \label{fig:7.14b}]{\includegraphics[width=0.495\textwidth]{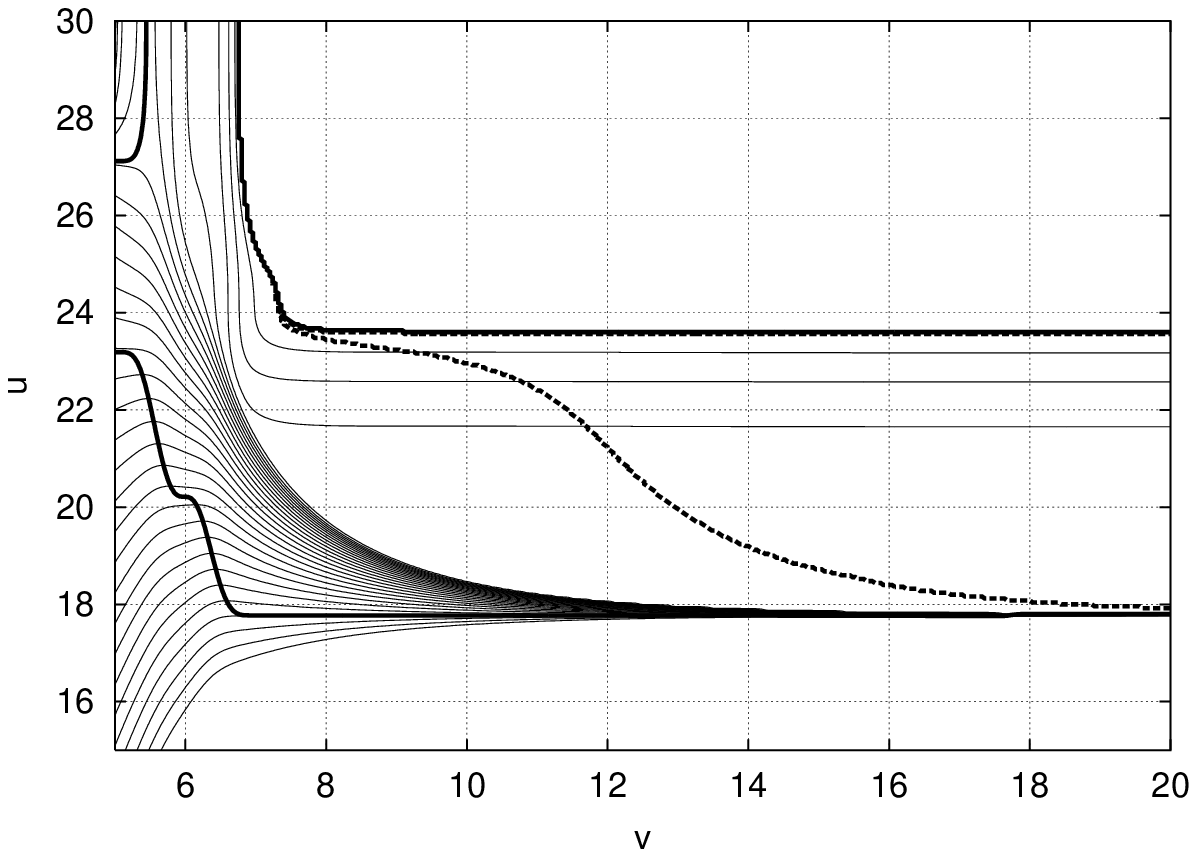}}
\caption{{\label{fig:7.14} Contour lines for
    double sine pulse of form of eq. \eqref{eq:7.2}, case $\Delta =2.0,
    A=0.25$. Thick bottom line is AH, thick upper line is $r=0$.}} 
\end{figure*}

With an increase of the amount of energy in the initial
pulse one can observe a nonlinear effect of shrinkage of the
CH-singularity under 
the action of the gravity of the irradiating flux $T_{uu}$ together
with the $T_{vv}$ flux. In
fig. \ref{fig:7.9b} (case $\Delta =2.0, A=0.125$) the test photons with
greater $u=constant$ comes asymptotically to smaller values of $r$ and
in fig. \ref{fig:7.9c} (case $\Delta =2.0, A=0.180$) the pulse is so
strong that the CH almost (but not quite) shrinks down to $r=0$. 

In fig. \ref{fig:7.13} (case $\Delta =2.0, A=0.200$) one can see both
the manifestation of the shrinkage of the CH singularity (photons with
higher $u$ come asymptotically to smaller $r$) and existence of the
$r=0$ singularity (photons with the highest $u$ come to $r=0$).

Figure \ref{fig:7.10-11} shows lines of constant $r$ and the position
of $K=K_{planck}$ (marked by the thick dotted line) for the two
strongest cases from fig. \ref{fig:7.9}. We remember that this line and
places with higher $K$ should be considered as a singularity from the 
point of view of classical physics. Thus, for both these cases the
physical singularity is placed at finite values of $v$ and is not
a null singularity. In fig. \ref{fig:7.11} we furthermore see the $r=0$
spacelike singularity inside of our computational domain. This
singularity can be considered as a result of mutual gravitational
focusing of $T_{vv}$ and $T_{uu}$ fluxes in the region between inner
and outer apparent horizons. At small $v<15$ the physical singularity
practically coincide with $r=0$, but for $v>17$ we see that the
spacetime structure of the physical singularity is quite different
from the structure of the mathematical singularity. The physical
singularity here depends mainly on the true CH-singularity, but its
position in the $u-v$ diagram is quite different from the position of
the true CH-singularity which is at $v=\infty$.

This can also be compared with fig. \ref{fig:7.12} from which we see
different behaviours of $K$ for the test photons for the strong case
of $\Delta =2.0, A=0.200$. The lines in the lower left hand corner, 
sharply increasing to near vertical, are the lines which come to $r=0$.
The thick dotted line is the line which comes to a point at the
singularity close to the meeting of the $r=0$ and CH
singularities. The remaining lines in the right hand side are the
lines which come to the CH-singularity.

\subsection{\label{subsec:doublepulse}Double sine pulse}
\begin{figure*}
\subfigure[Lines of constant $\log_{10} \left(  T_{uu}\right)$. Lines are from
$\log_{10} \left( T_{uu}\right)=-10.0$ to $\log_{10} \left(
  T_{uu}\right)=3.00$ in $\Delta \log_{10} \left( T_{uu}\right)=0.25$ intervals. Thick dotted line marks
$\log_{10} \left( T_{uu}\right)=1.50$, decreasing leftwards.\label{fig:7.16}]{\includegraphics[width=0.495\textwidth]{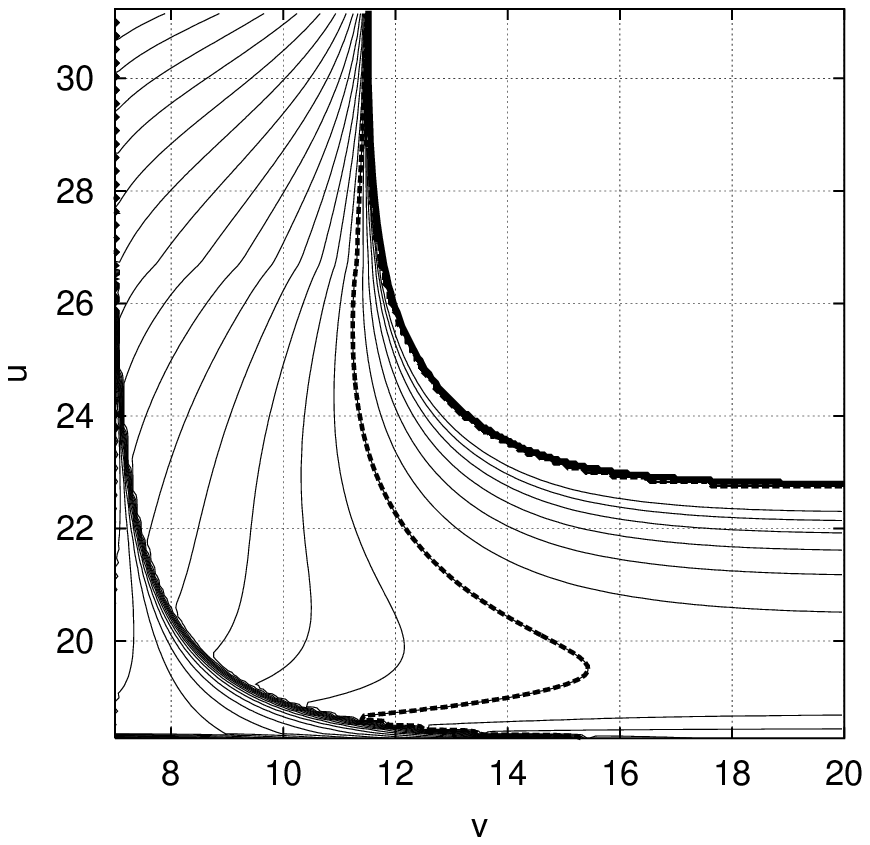}}
\subfigure[Lines of constant $\log_{10} \left(  T_{vv}\right)$. Lines are from
$\log_{10} \left( T_{vv}\right)=-10.0$ to $\log_{10} \left(
  T_{vv}\right)=3.00$ in $\Delta \log_{10} \left( T_{vv}\right)=0.25$ intervals. Thick dotted line marks
$\log_{10} \left( T_{vv}\right)=1.50$, decreasing downwards.\label{fig:7.17}]{\includegraphics[width=0.495\textwidth]{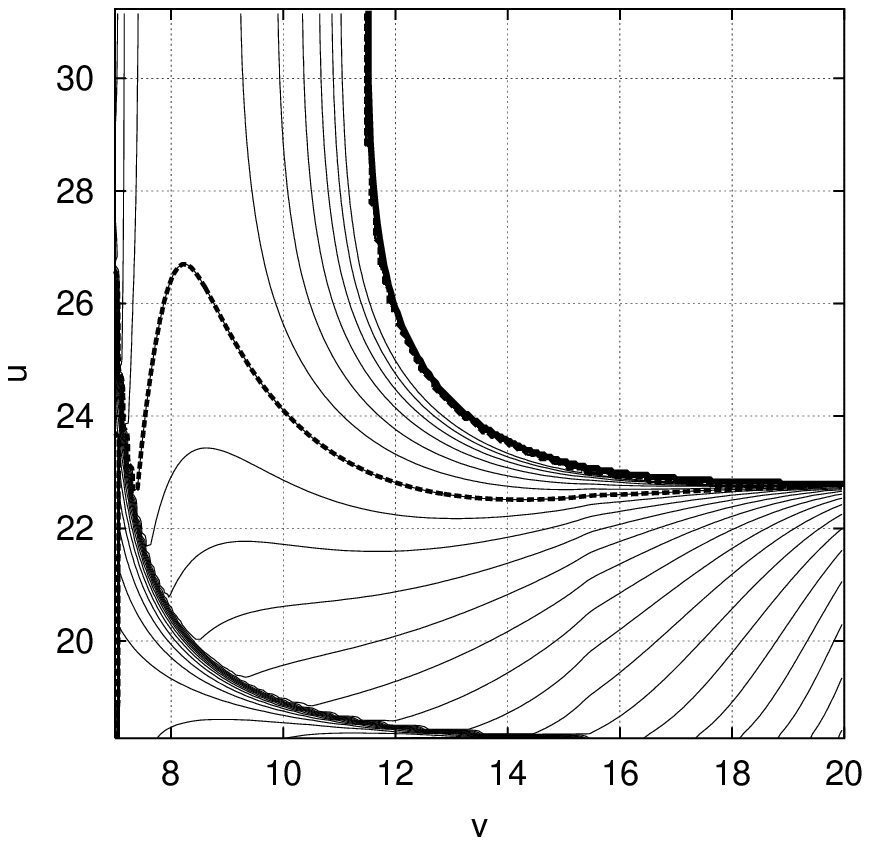}}
\caption{{\label{fig:7.15-17}Contours for the
  double-flux case based on the simple compact pulse: $\Delta =2.0$, $A=0.25$. Thick top right line marks $r=0$.}}
\end{figure*}
\begin{figure}
\includegraphics[width=0.495\textwidth]{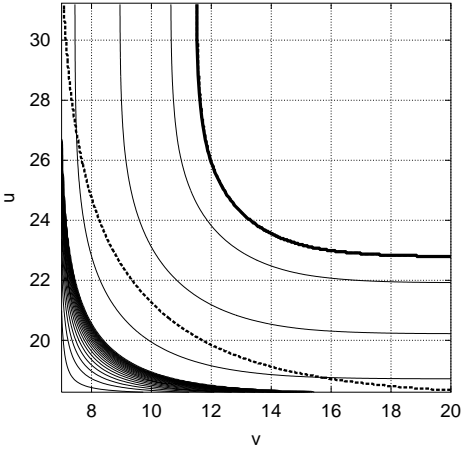}
\caption{{\label{fig:7.15}Lines of constant $r$ for the
  double-flux case based on the simple compact pulse: $\Delta =2.0$, $A=0.25$. Thick top right line marks
 $r=0$. Lines are from $r=2.50$ (bottom left line) to $r=0.10$ (top
 right  thin line) in intervals of $\Delta r = 0.10$. Thick dotted
 line marks  $K=K_{planck}$.  }}
\end{figure}
In this subsection we choose the ingoing flux to be of the form:
\begin{equation}
  \label{eq:7.2}
  \Phi_{,v}(u_0,v) =
  \sqrt{6}\, A\,\cos\left(\pi\frac{v-v_0}{v_1-v_0}\right)\sin \left(
  \pi\frac{v-v_0}{v_1-v_0}\right)^2 
\end{equation}
instead of equation \eqref{eq:7.1}. This readily integrates to give
the initial expression for $\Phi(u_0,v)$:
\begin{equation}
  \label{eq:7.3}
  \Phi (u_0,v) = \frac{\sqrt{\frac{2}{3}}\,
  A\,\left(v_1-v_0\right)\sin\left(\pi\frac{v-v_0}{v_1-v_0}\right)^3}{\pi} 
\end{equation}
where $v_0$ and $v_1$ as before, marks the beginning and end of the pulse
respectively and $A$ is the amplitude of the pulse. Also, as before we
set $\Phi_{,v}(u_0,v) = 0$ for $v>v_1$. The pulse is scaled in such a
way that for a given width $\Delta =v_1-v_0$ and amplitude $A$, the
integral of the initial flux, $\int_{v_0}^{v_1} T_{vv}\, dv$, is equal for
pulses of the form \eqref{eq:7.1} and \eqref{eq:7.2}.
Eq. \eqref{eq:7.2} has the shape of a double pulse, rather then
\eqref{eq:7.1} which is the shape of a single initial pulse. This
pulse is more complicated than \eqref{eq:7.1}, but is similar in 
shape to the pulse shapes used in some other papers
(e.g. \cite{Burko97,Burko98b}).  

We have performed a series
of computations based on eq. \eqref{eq:7.2} (all other initial conditions
equal to those in the previous subsection). The results of these
computations demonstrate a more complex picture of 
interplay between the scalar fluxes than in the case of subsection
\ref{subsec:7.1}. This is natural because of the more complex shape of
the initial pulse. But the main physics and principal properties of
the singularities are the same in the two cases. An example
illustrating the increased complexity in structure of the fluxes can
be seen in figure \ref{fig:7.14}, which illustrates the
$T_{vv}$ flux for the case:$\Delta =2.0,A=0.25$. The shapes of the
apparent horizons and the central singularity $r=0$ 
are now more complex as well as the distribution of the $T_{vv}$
field. Still the general characters of the $r=0$ and the
physical $K=K_{planck}$ singularities are the same.

\subsection{\label{subsec:7.3}The influence of the $T_{uu}$ flux} 
\begin{figure*}
\subfigure[Mass function along lines of constant $v$. Lines are from
$v=7.04$ (bottom right) to $v=19.04$ (top left) in $\Delta v=1.0$
intervals.
\label{fig:7.18}]{\includegraphics[width=0.495\textwidth]{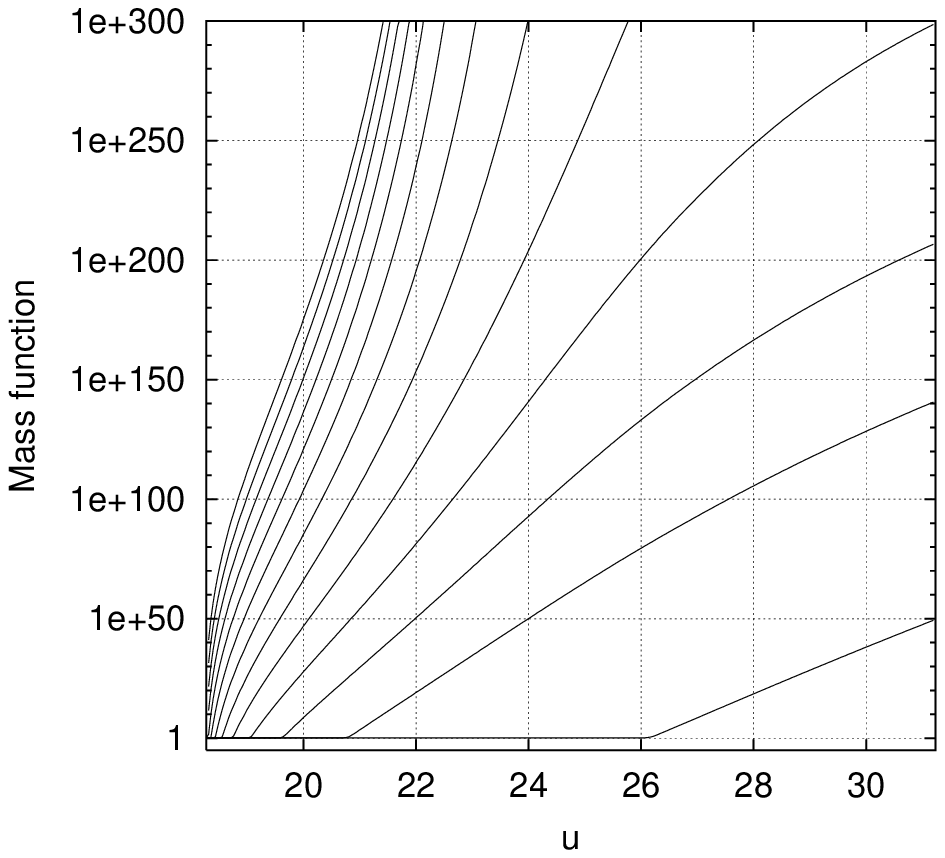}} 
\subfigure[Mass function along lines of constant $u$, corresponding to
the symmetrical equivalent of the lines in fig. \ref{fig:7.18}.
\label{fig:7.19}]{\includegraphics[width=0.495\textwidth]{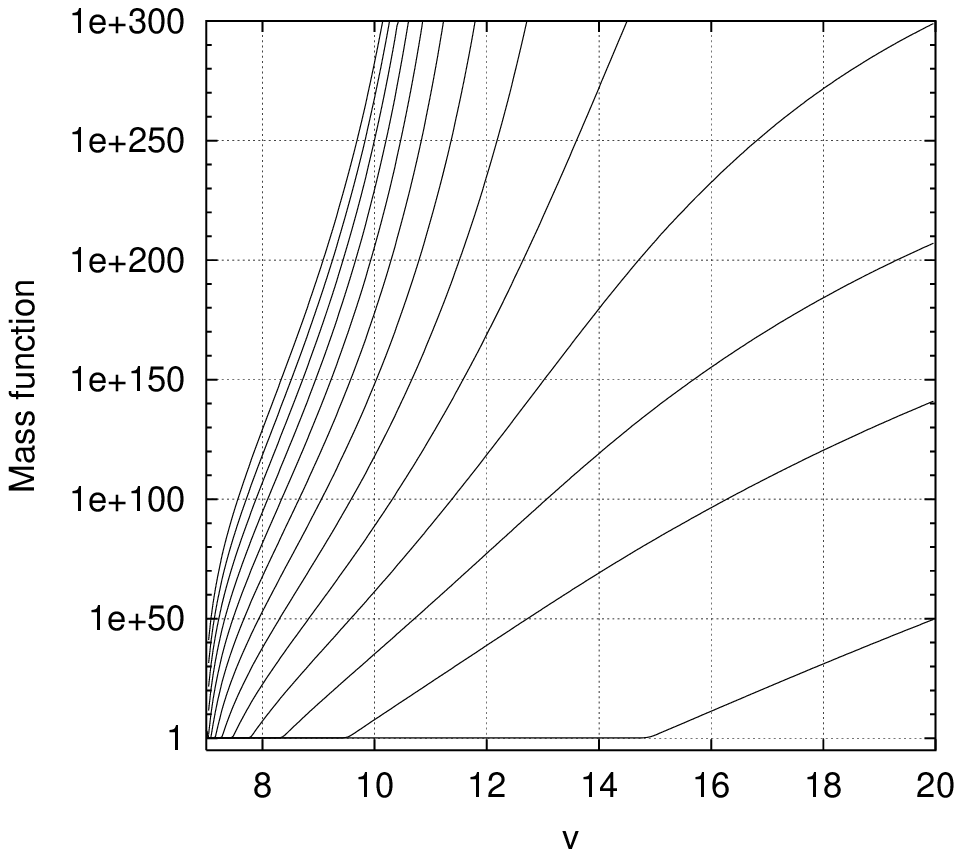}}
\caption{{\label{fig:7.18-19}Mass function for the double-flux case
    double-flux case based on the simple compact pulse: $\Delta =2.0$,
    $A=0.25$}} 
\subfigure[Kretschmann scalar along lines of constant $v$. Lines are from
$v=7.04$ (bottom right) to $v=19.04$ (top left) in $\Delta v=1.0$
intervals. Horizontal dashed line is
$K=K_{planck}$.
\label{fig:7.20}]{\includegraphics[width=0.495\textwidth]{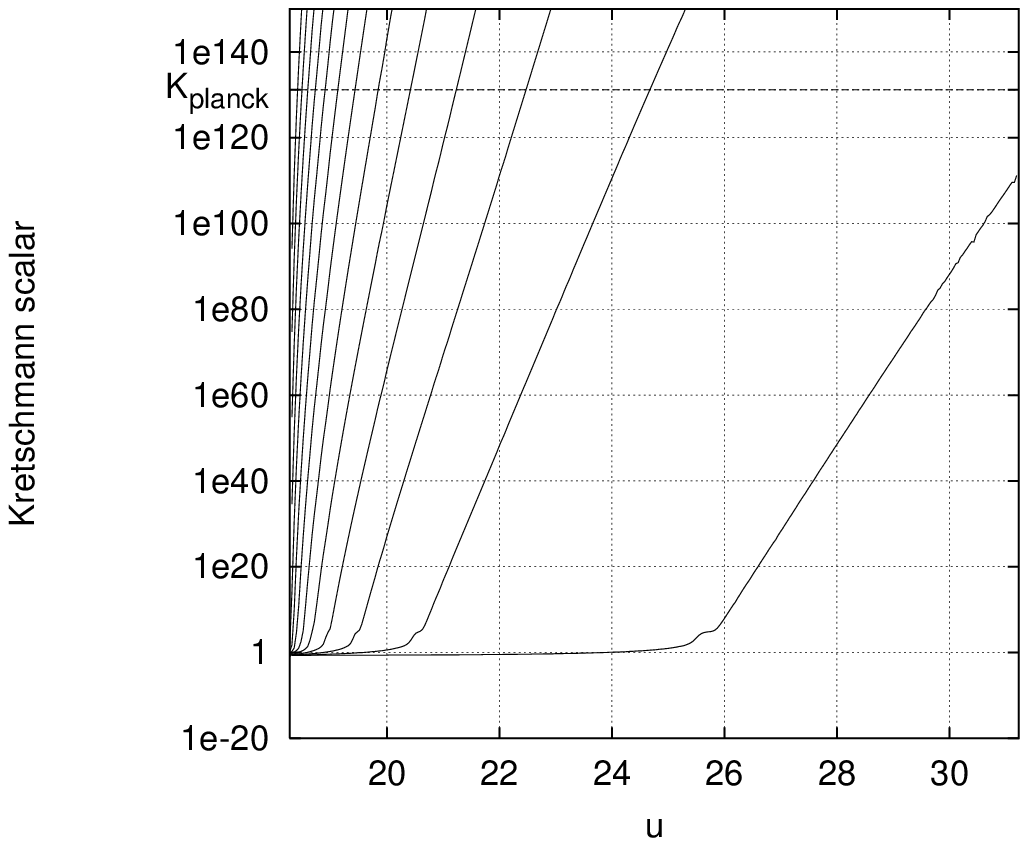}}  
\subfigure[Kretschmann scalar along lines of constant $u$, corresponding to
the symmetrical equivalent of the lines in
fig. \ref{fig:7.20}.
\label{fig:7.21}]{\includegraphics[width=0.495\textwidth]{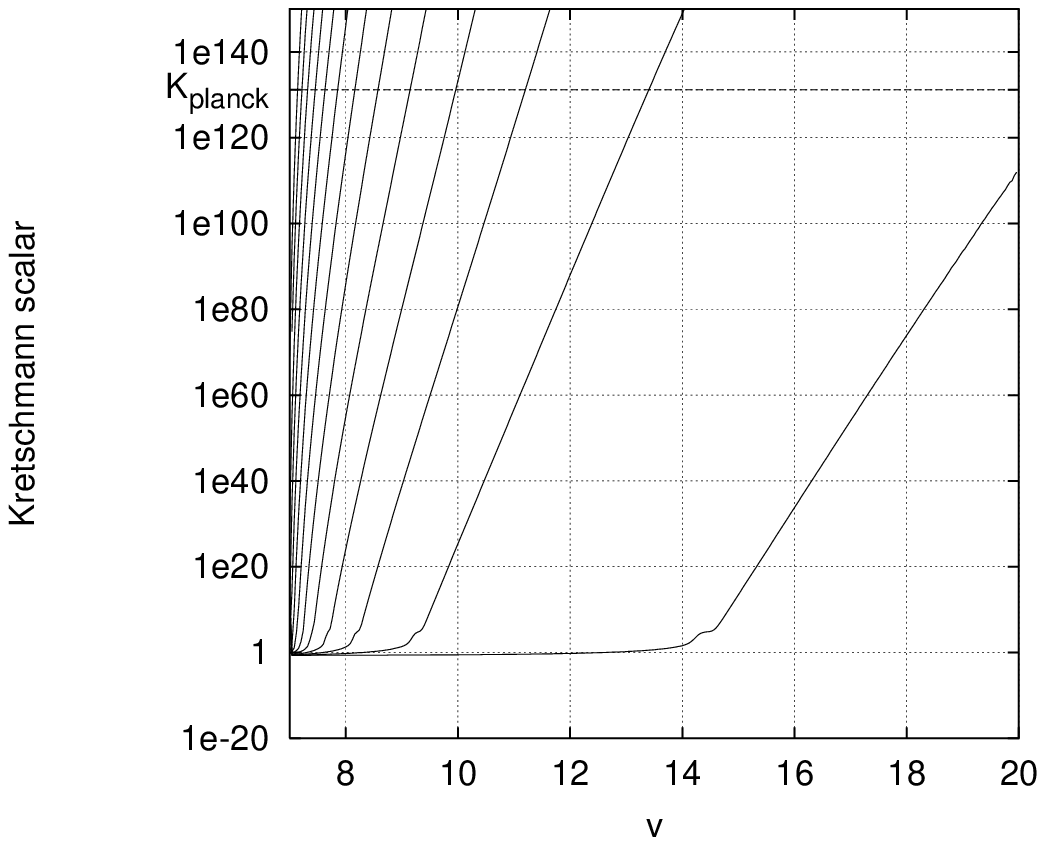}} 
\caption{{\label{fig:7.20-21}Kretschmann scalar for the double-flux case
    double-flux case based on the simple compact pulse: $\Delta =2.0$, $A=0.25$}}
\end{figure*}
\begin{figure*}
\subfigure[$u$ vs. $r$ along lines of constant $v$. Lines are from
$v=7.04$ (rightmost) to $v=19.04$ (bottom left) in $\Delta v=1.0$
intervals
\label{fig:7.22}]{\includegraphics[width=0.495\textwidth]{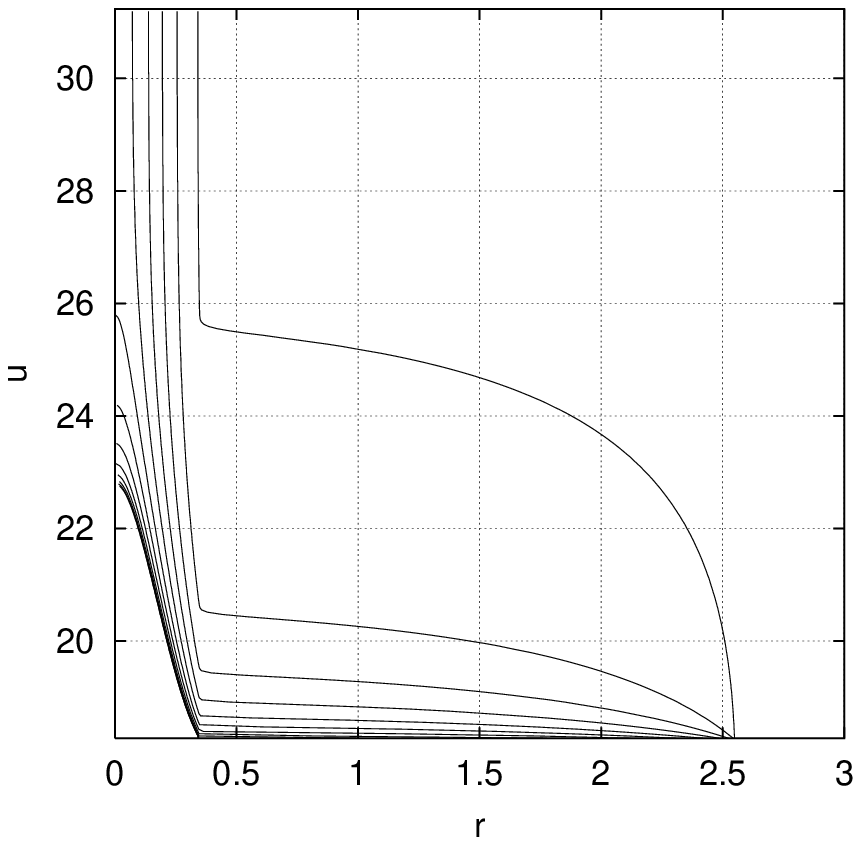}} 
\subfigure[$v$ vs. $r$ along lines of constant $u$, corresponding to
the symmetrical equivalent of the lines in
fig. \ref{fig:7.22}.
\label{fig:7.23}]{\includegraphics[width=0.495\textwidth]{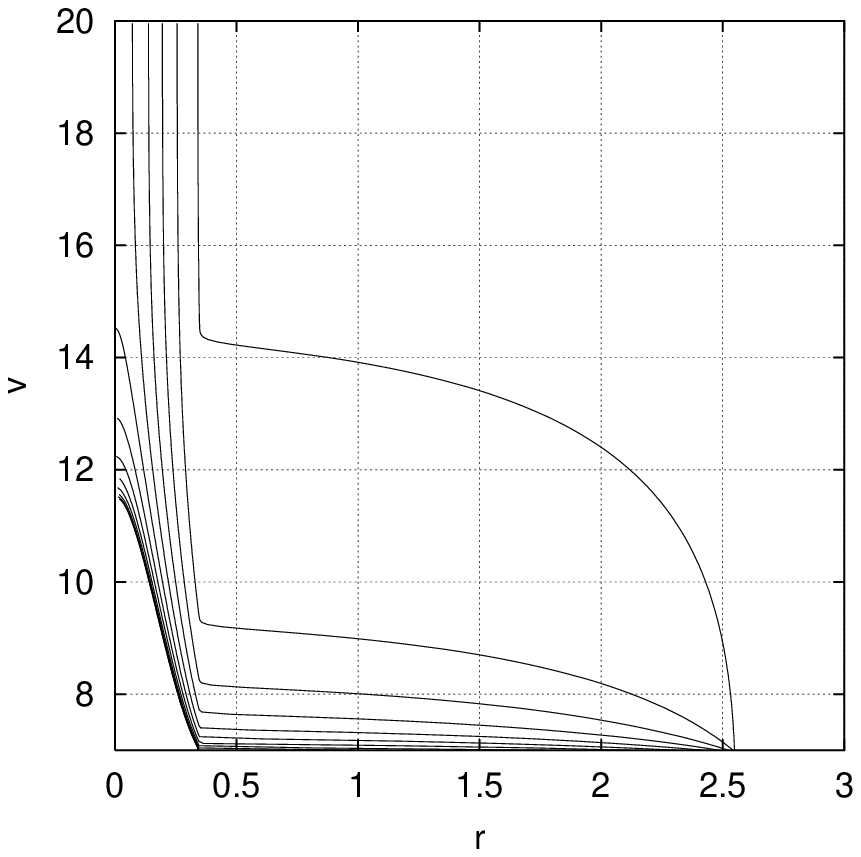}}   
\caption{{\label{fig:7.18-23} $u$ and $v$ versus
    $r$ for the double-flux case double-flux case based on the simple
    compact pulse: $\Delta =2.0$, $A=0.25$}} 
\end{figure*}
So far in all our analyses we have assumed that the flux of the scalar
field $T_{uu}$ through $v=v_0$ was zero and that any $T_{uu}$-flux arised only
as a result of scattering of the $T_{vv}$-flux by the curvature of the
spacetime. Now we would like to consider the influence of a
$T_{uu}$-flux through the surface $v=v_0$. To do this we will consider
an extreme case when $T_{uu}$ through ${v=v_0}$ is equal exactly to
$T_{vv}$ through ${u=const}$ just inside of a black hole. More concretely we
do the following; We specify some
initial $\Phi (u_0,v)$ along $u=u_0$ with a pulse width  
$\Delta = v_1-v_0$ and amplitude $A$ and set all initial conditions equal to
those in subsection \ref{subsec:7.1}, including $\Phi_{,u}(u,v_0)=0$
along $v=v_0$. These initial data are then simulated as usual, however
only in the computational domain of $5<v<20$ and $0<u\le u_{AH}$, where
$u_{AH}$ is the first computational point which is inside of the outer apparent
horizon along $v=v_1$. Because of scattering
of the initial pulse there is now a $T_{vv}$ flux into the black hole
along the line $u=u_{AH}$ for $v>v_1$. We then stop the computation and start a new 
with the following domain of integration: $v_1\le v\le 20$ and
$u_{AH}\le u\le u_{max}$ (where $u_{max}-u_{AH}=(20-v_{1})$). Along
the (new) outgoing initial hypersurface, $u=u_{AH}$, all the variables
are kept as they were in the original simulation, while along the
(new) ingoing initial hypersurface, $v=v_1$, initial data are set
equal to the data along the outgoing hypersurface, hence we have
completely symmetrical initial conditions. Subsequently we performed a
computation for the new computational domain.
It is obvious that all conditions in this domain are
symmetrical with respect to $u$ and $v$ and hence the boundary fluxes
$T_{uu}$ along $v=v_1$ and $T_{vv}$ along $u=u_{AH}$ must be exactly
equal.

We performed our computations for the compact simple pulses
(eq. \eqref{eq:7.1}) for different parameters $\Delta$ and 
$A$. An example of our results can be seen in figs. \ref{fig:7.15-17} -
\ref{fig:7.18-23} for the case $\Delta =2.0, A=0.25$. All pictures are
symmetrical with respect to $u$ and $v$, as they should be.
The outer apparent horizon is naturally outside of our computational
domain.  Figs. \ref{fig:7.15-17} and \ref{fig:7.15} shows the $T_{uu}$
and $T_{vv}$ distributions and $r$ contour lines. On these figures
there is no inner apparent horizon because it coincide with another 
CH singularity at $u\rightarrow \infty$ (see below). But on
fig. \ref{fig:7.15} one can see the mathematical strong singularity
$r=0$ and the physical singularity (with $K=K_{planck}$) which is
reached long before the CH singularity at $u\rightarrow \infty$ and
$v\rightarrow \infty$ in most of the computational domain.

Fig. \ref{fig:7.18-19} shows the huge increase of the mass
function with growing $u$ and $v$, and fig. \ref{fig:7.20-21}
shows the increase of the Kretschmann scalar as
function of $u$ and $v$ up to $K=K_{planck}$. Finally
fig. \ref{fig:7.18-23} shows light signals along
$v=const$ and $u=const$ respectively. This 
figure shows both the existence of null singularities of the CH
singularity-type and $r=0$ singularity. Indeed for example in the case
fig. \ref{fig:7.22} one can see that for some $v=const$ (the
rightmost curves) for $u\rightarrow \infty$, signals goes asymptotically
to practically $r=const$ (asymptotically approach to the null
singularity at $u\rightarrow \infty$). Smaller asymptotic values $r$
are seen for the signals at bigger $v$ corresponding to the focusing
effect. Finally the signals with $v$ big enough, come to the central
singularity $r=0$. The symmetrical picture for the signals with
$u=const$ is shown in fig. \ref{fig:7.23}. Here we see the
asymptotic approach to another null singularity at $v\rightarrow
\infty$.

Note also the different character of the increase of mass function on
fig. \ref{fig:7.18-19} for lines of big and small $v$
and $u$ respectively. For example on
fig. \ref{fig:7.18} the topmost lines (big $v=const$) come to the
central singularity but lines with small $v=const$ (bottom lines) go
to the null singularity at $u\rightarrow \infty$. The different
behaviour of these lines correspond to the different nonlinear processes
influencing the mass function for lines terminating at the $r=0$
singularity and reaching the CH singularity.

Thus in this case there are three different singularities inside the
black hole: two null singularities at $u\rightarrow \infty$ and
$v\rightarrow \infty$ and the physical singularity at $K=K_{planck}$.
In addition there is also a central (mathematical) singularity at $r=0$. 

\FloatBarrier 
\section{\label{sec:8}Conclusions} 
In this paper we investigated the physics of nonlinear processes 
inside of the spherical charged black hole, nonlinearly perturbed by a
selfgravitating, minimally coupled, massless scalar field. For this
purpose we created and tested a numerical code which is stable and
second-order accurate. For our computations we used an adaptive mesh
refinement approach in ingoing $u$-direction.

The following nonlinear physical processes are important inside the
black hole: Scattering of radiation by the curvature of the
spacetime, gravitational focusing effect, mass inflation and squeeze
of matter with pressure. 

At the beginning of our analysis we used a homogeneous approximation to
clarify some physical processes near a spacelike singularity. In
this approximation one supposes that near the singularity temporal
gradients are much higher than the spatial gradients, so one assumes
that all processes depend on the time coordinate only (uniform 
approximation). We used both analytical analyses and a numerical
approach to analyse three different physical matter
contents: dust, a massless scalar field and an ultrarelativistic gas.

For the case $P=0$ (dust) we found that the singularity is at
$r=r_{sing}\neq 0$, $r_{CH}<r_{sing}<r_{EH}$, where $r_{CH}$ and
$r_{EH}$ are the positions of the Cauchy Horizon and the Event Horizon
when there are no dust at all. The value $r_{sing}$ decreases
monotonically with a decrease of the matter contents and tends to
$r_{sing}=r_{CH}$ when the matter contents goes to zero.

In the case of the scalar field, the uniform approximation demonstrates
more a complex behaviour. Here the scalar field can be represented as a
sum of two equal fluxes moving in opposite directions. One can for
this case see the manifestation of both the effect of mass inflation
and the effect of shrinkage of the CH down to $r=0$. 
For very small matter contents
($\epsilon_0 \ll 0.01$) the Kretschmann scalar, $K$, becomes equal to
the Planck value at $r$ close to $r_{CH}$. So in this case the
physical singularity (when $K=K_{planck}$) is at $r\approx
r_{CH}$. For larger values of $\epsilon_0$, (e.g. $0.01 \lesssim
\epsilon_0 \lesssim 0.03$), the CH does not manifest itself and the
model squeezes to $r$ very close to $r=0$ before $K$ reaches
$K_{planck}$. It means that for these values of $\epsilon_0$, the
physical singularity practically coincides with $r=0$. For rather big
$\epsilon_0$, $K$ reaches $K_{planck}$ at rather big $r$ as it was
for the case of dust, $P=0$. 

In the case of matter with isotropic relativistic pressure,
$P=\epsilon /3$, we have the situation intermediate between $P=0$ and
the scalar field. The physical singularity, in this case, is located
at $r$ essentially greater then $r=0$. 

We performed the analysis of the full nonlinear processes inside the
spherical charged black hole with a scalar field using the numerical
approach. This analysis extends the analysis of the earlier works
\cite{Burko97c, Burko02, Burko02b} and reveal new aspects of the  
problem. The detailed description of the results is given in section
\ref{sec:7}. We analysed nonlinear gravitational interaction of the
fluxes of the scalar field, the dependence of the effects on the
boundary conditions, analysed the focusing effects, mass inflation and
squeeze effect and the behaviours of the Kretschmann scalar $K$. We
payed special attention to the analysis of the singularity in 
subsection \ref{subsec:singularity}. We investigated the focusing of the CH
singularity and its dependence on the boundary conditions. We
determined the position of the physical singularity (where
$K=K_{planck}$) inside the black hole and demonstrated that this
position is quite different from the positions of the mathematical
$r=0$ singularity and CH singularity.

The results mentioned above were obtained with the scalar flux into
the black hole in the form of a simple compact sine-pulse with different
amplitudes and widths.

We also analysed the physics in the case of a scalar flux in the form
of a double sine-pulse qualitatively similar to the usage in
\cite{Burko97,Burko98b}. In this case physics is more complicated, but
the main characteristics of the results are the same as for the simple
pulse. 

Finally we investigated the influence of the boundary $T_{uu}$-flux on
the physics of the singularity. We demonstrated that it is possible to
have the existence of three different singularities inside the black
hole: two null singularities at $u\rightarrow\infty$ and
$v\rightarrow\infty$ and the physical singularity $K=K_{planck}$ in
addition to a central mathematical singularity $r=0$.

\begin{acknowledgments}
This work was supported in part by Danmarks Grundforskningsfond through
its support for establishment of the Theoretical Astrophysics Center
and by the Danish SNF Grant 21-03-0336. J. Hansen and I. Novikov
thanks The University of Chicago for hospitality during their visits. 
\end{acknowledgments}

\appendix

\section{\label{sec:3}The numerical code}
In this appendix we describe the structure of the numerical code used
to obtain the numerical results presented in section \ref{sec:7}. 

\subsection{\label{subsec:3.1}The numerical scheme}
Our approach to numerically integrating the field equations is similar to
that of other authors \cite{Oren03, Burko97, Burko97c, Burko02, Burko02b}.
Denoting the three unknowns, $r,\sigma, \Phi$ as $h_j$ with $j=1,3$, the three
evolution equations \eqref{eq:evolve1}-\eqref{eq:evolve3} are each of the form: 
\begin{equation}
  \label{eq:a}
  h_{j,uv}=F_j (h_k, h_{k,v}, h_{k,u}), k=1,3
\end{equation}

\begin{figure}
\includegraphics[width=0.495\textwidth]{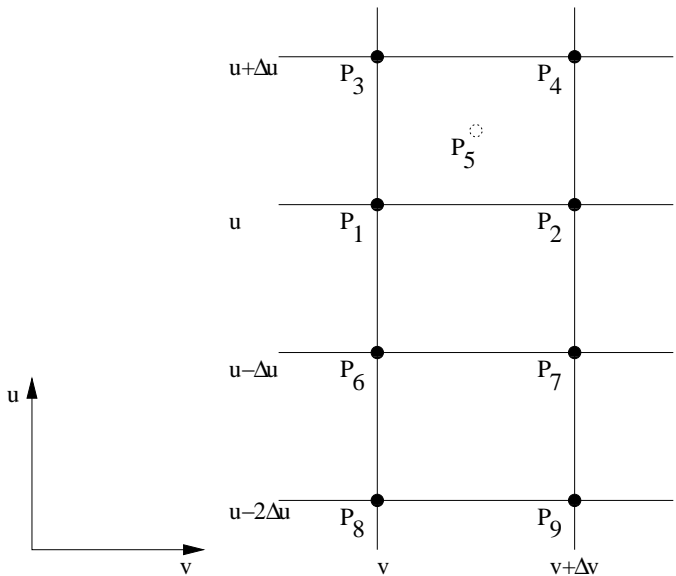}
\caption{Schematic of the points used in the numerical scheme.}
\label{fig:3.1}
\end{figure}

We denote four grid points in the $(u,v)$-domain by $p_1 \equiv
(u, v)$, $p_2 \equiv (u, v+\Delta v)$, $p_3 \equiv (u+ \Delta u,
v)$ and $p_4 \equiv (u+\Delta u, v+\Delta v)$, where
$\Delta v$ and $\Delta u$ are finite increments in the $v$ and $u$
directions (see fig. \ref{fig:3.1}).

By combining a Taylor expansion $h_j$ at these four points (evaluated
around an intermediate point $p_5 \equiv (u+\Delta u/2, v+\Delta v/2
)$) with eq. \eqref{eq:a} we find that:
\begin{eqnarray}
h_j(p_4)&=&h_j(p_3)+h_j(p_2)-h_j(p_1)\nonumber\\
  & &+{\Delta u\Delta
  v}[ F_j(h_k(p_5), h_{k,u}(p_5), h_{k,v}(p_5))\nonumber\\ 
 & & + \frac{{\Delta u}^2}{24}h_{j,vuuu}(p_5) + 
 \frac{{\Delta v}^2}{24}h_{j,uvvv}(p_5)\nonumber\\ 
& & +O(\Delta u^3,\Delta v^3)]  , k=1,3  \label{eq:c}
\end{eqnarray}
i.e. we can evaluate $h_j$ at $p_4$ with second order
accuracy if we know $h_j$ at the four points $p_1, p_2, p_3, p_5$
and $h_{j,v}, h_{j,u}$ at $p_5$. Initially we only know the
values of $h_j$ along in- and outgoing null segments (see section
\ref{sec:2}) i.e. for example at $p_1, p_2$ and $p_3$. However, by
employing eq. \ref{eq:c} twice in a predictor-corrector style we can
evaluate $h_j, h_{j,v}, h_{j,u}$ at $p_5$ with the desired accuracy and
hence subsequently find $h_j$ at $p_4$. 

Equation \eqref{eq:c} constitutes the heart of our
code. As described in section \ref{sec:2}, the values of the unknown
functions along the initial hypersurfaces $u_0, v_0$ are given by the
constraint equations \eqref{eq:constraint1} and
\eqref{eq:constraint2}. We can then use
our numerical scheme to calculate the values of $h_j$ along the 
ingoing ray $v = v_0 + \Delta v$ starting at the point $u = u_0 + \Delta u$,
then solving for $u = u_0 + 2\Delta u$ etc. until we reach the last grid point
at $u = u_{final}$. Using the (now known) values of $h_j$ along $v = v_0 +
\Delta v$, we can then calculate the unknowns along the next
ingoing ray along $v = v_0 + 2\Delta v$ and so on throughout our
computational domain until we reach the end of our computational
domain at $v=v_{final}$. 

\subsection{\label{subsec:3.2}Splitting algorithm}
When outside the event horizon (EH), the scheme described above works
reasonably good as it is, but when
the domain of integration crosses the EH, the unknown functions exhibits
extremely great gradients. This can be seen by considering two
outgoing rays, one just inside and one just outside of the EH. One ray
is destined to be trapped inside the EH, while the 
other must escape to infinity, i.e. regardless of how close the two
rays were initially, their distance will diverge as their advanced
time $v$ grows. Integration across such great gradients using fixed
increments in $u$ and $v$ would cause the numerical errors to explode
unless the initial increments were chosen to be unrealistically small.
 To overcome this difficulty an adaptive mesh refinement
(AMR) approach is crucial for an accurate integration of the
field equations. However, the strong gradients are mainly in the
ingoing $u$-direction, hence AMR is most important in this
direction. While AMR in both $v$ and $u$ directions would be
desirable, implementation of AMR in only one direction, while much less
complicated, still produces good results. This is
supported by other authors who also employ uni-directional AMR
\cite{Oren03, Burko97}.

To determine whether AMR is needed and a cell should be split (or
possibly desplitted), we focus on the truncation error in the
numerical scheme. As can be seen from equation \eqref{eq:c}, the
primary error introduced by the scheme comes mainly from the terms: 
\begin{subequations}
  \label{eq:d}
\begin{equation}  \label{eq:d2}
\epsilon^u_j (p_5)=\frac{{\Delta u}^2}{24}h_{j,vuuu}(p_5)
\end{equation}
\begin{equation}
\epsilon^v_j (p_5)=\frac{{\Delta v}^2}{24}h_{j,uvvv}(p_5)  
\end{equation}
\end{subequations}

Since we use AMR only in the 
ingoing $u$ direction, thus keeping $\Delta v$ constant, we can only
control $\epsilon^u_j(p_5)$ by splitting, hence we will focus on this term. By
introducing four new points: $p_6 \equiv (u-\Delta u,v), p_7 
\equiv (u-\Delta u,
v+\Delta v), p_8 \equiv (u-2\Delta u, v), p_9 \equiv (u-2\Delta u,
v+\Delta v)$, (see fig. \ref{fig:3.1}), we can estimate $h_{j,vuuu}$
(and hence $\epsilon^u_j$) at $p_5$ with the following finite difference
operator: 
\begin{eqnarray}
   h_{j,vuuu}\left(p_5\right)&=&\Delta v \Delta u ^3 \bigg( 3
 h_j\left( p_1\right) - 3 h_j\left(p_3\right) \nonumber \\ 
 & &  - 3 h_j\left(p_6\right)  +  3 h_j\left(p_7\right)- h_j\left(p_2\right)+h_j \left(p_4\right)   \nonumber \\ 
 & &   + h_j \left(p_8\right) -
  h_j \left(p_9\right) \bigg) + O\left(\Delta u\right)   \label{eq:e}
\end{eqnarray}

In practice we proceed as follows; Following a calculation of $h_j$ at
$p_4$ we use eq. \eqref{eq:d2} + \eqref{eq:e} to estimate the  
error $\epsilon^u_j(p_5)$ involved in the calculation and compare it to
$F_j(h_k(p_5), h_{k,u}(p_5), h_{k,v}(p_5))$ in eq. \eqref{eq:c} 
for each of the 
variables, $j=1,3$. We now require that the relative error should be
within some fixed interval: 
\begin{equation}
  \label{eq:f}
S_{desplit}<\frac{\epsilon^u_j(p_5)}{F_j(h_k(p_5),h_{k,u}(p_5), h_{k,v}(p_5))}<S_{split}
\end{equation}
If the relative error is greater then the threshold parameter
$S_{split}$ we perform a split. Conversely, if the
relative error is less than the threshold parameter $S_{desplit}$ a
desplit is made.  Note that because we need knowledge of the points
$p_6$ to $p_9$ to calculate the relative error, we don't use splitting
for the first two cells $u<u_0 + 2\Delta u$ on a given ingoing ray.

If a split is required a new point is introduced at
$p'_3\equiv (u+\Delta u/2,v)$. The values of $h_j$ at $p_3'$ are then found
by a four-point interpolation scheme (using data from the two points
above and the two below $p'_3$). The numerical scheme is then used to
calculate $h_j$ at $p_4'\equiv (u+\Delta u/2,v+\Delta v)$ using the points
$p_1, p_2$ and $p'_3$. If a desplitting is needed the reverse process
takes place, i.e. $p_3$ is deleted and the point above is used in its place.  
\begin{figure*}
\subfigure[Line a) $\psi_N (r)$, b) $\psi_N (\Phi )$ and line
c) $\psi_N (\sigma )$. Line d)
shows a line with a slope of minus two. 
\label{fig:4.1a}]{\includegraphics[width=0.495\textwidth]{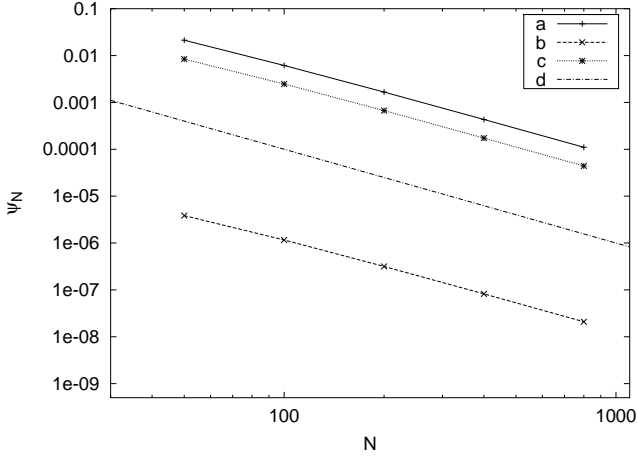}} 
\subfigure[Line a) $\chi (C_1)$ and b) $\chi (C_2)$. Line c)
shows a line with a slope of minus two
\label{fig:4.1b}]{\includegraphics[width=0.495\textwidth]{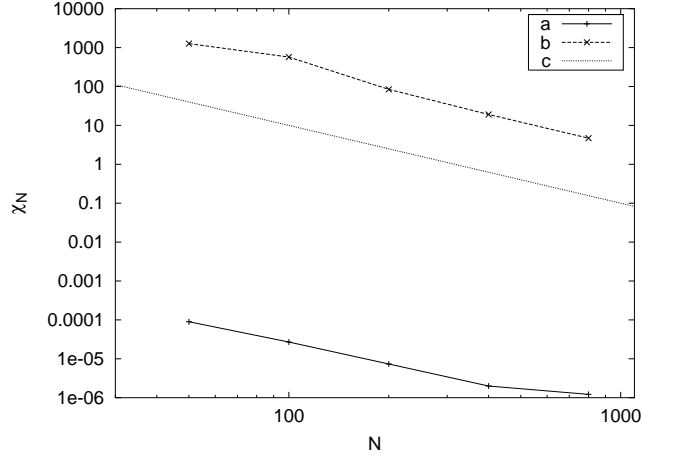}} 
\caption{{Demonstration of the convergence (without AMR) of
    $r$, $\Phi$, $\sigma$, $C_1$ and $C_2$  along an outgoing ray
    $u=23.00$.\label{fig:4.1} }} 
\end{figure*}
\begin{figure*}
\subfigure[\label{fig:4.2a}]{\includegraphics[width=0.495\textwidth]{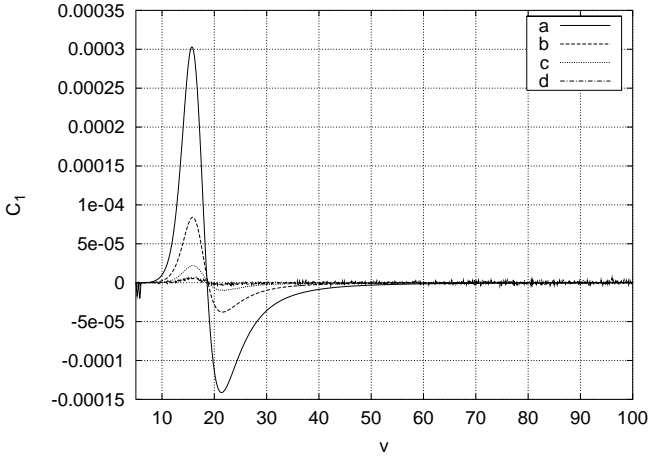}}
\subfigure[\label{fig:4.2b}]{\includegraphics[width=0.495\textwidth]{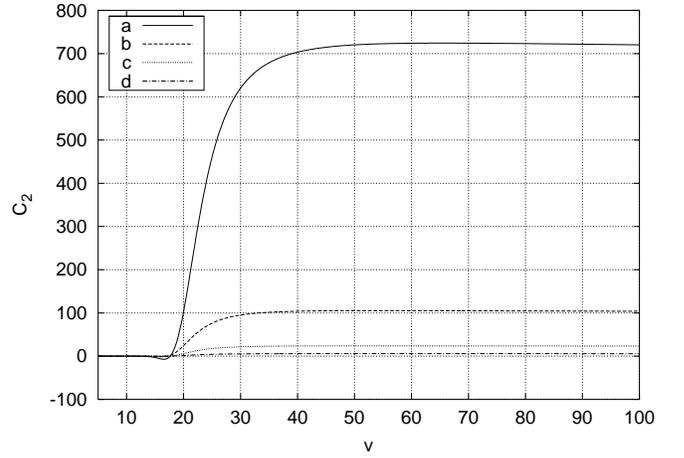}}
\caption{{ $C_1$ (fig. a) and $C_2$ (fig. b)  along an outgoing ray
    $u=23.00$ for resolutions; line a) $N=100$, b) $N=200$, c) $N=400$
    and line d) $N=800$. \label{fig:4.2} }} 
\end{figure*}

To obtain the
results presented in section \ref{sec:7}, the splitting thresholds were set to
$S_{split}=10^{-7}$,  $S_{desplit}=10^{-9}$. 
The resolutions
along the initial outgoing null segment were set to $\Delta u = 1/100$
and $\Delta v = 1/12800$. 
The initial value of $\Delta u$ can be set rather low,
since the splitting algorithm will assure that $\Delta u$ always have
an appropriate value. The value of $\Delta v$ on the other hand, is
constant throughout the computational domain and must initially be
chosen to have a suitable value everywhere. The high value of $\Delta
v$ has been chosen 
such that  $\Delta v$ is as comparable to $\Delta u$ in large parts of
the interior of the black hole. This has been verified by numerical
tests to give the best results.

\section{\label{sec:4}Analysis of the code}
We have tested the code and found it to be stable and second-order
accurate. Here we presents tests to demonstrate this.

\subsection{Basic convergence tests}
We test the basic convergence properties of the code by performing
simulations with the same initial conditions, but with varying
resolutions and with the splitting algorithm disabled. As 
initial conditions we use a Reissner-Nordstr{\" o}m spacetime
(with initial mass $m_0=1$ and charge $q=0.95$) which
is perturbed by a massless scalar field. We set the specific form of the scalar field
along the initial outgoing null segment in the interval $v_0\le v\le v_1$ to be:
\begin{equation}
  \label{eq:4.1}
\Phi (u_0,v)= \frac{A}{4\pi} \left( 2\pi\left( v-v_0\right)
    -\left( v_1-v_0\right)\sin\left(2\pi\frac{v-v_0}{v_1-v_0} \right) \right)
\end{equation}
with $A=0.05$ and
$v_0=5, v_1=6$. For $v>v_1$ we set $\Phi$ constant to $\Phi
(u_0,v)=\Phi(u_0,v_1)$. Along the ingoing null segment $v=v_0$ we
specify $\Phi_{,u} (u, v_0)=0$ (these initial conditions are the same
as those used in section \ref{sec:7}). As domain of integration we
used $5<v<100$ and $0<u< 23.05$ and with the gauge choice described in
section \ref{sec:2}.

We make simulations with varying
resolutions $N=1/\Delta u = 1/\Delta v$ and examine data along an outgoing ray
$u=23.00$ which is just outside of the event horizon. Along this ray we calculate
the average absolute difference at each data point between the unknown 
 functions for two resolutions along the outgoing ray. We define: 
\begin{equation}
  \label{eq:psidef}
  \psi_N (x) \equiv \frac{1}{n}\sum_i^n |x_N^i - x_{2N}^i|
\end{equation}
where $x_N^i$ denotes the value of variable $x$ at the $i$'th data
point (out of a total of $n$ points) from a simulation with 
resolution $N$. I.e. we basically calculate the normalized
$L_1$-norm of the difference between a simulation with resolution $N$
and one with resolution $2N$ at each point for the three
unknowns, which is similar to the convergence tests used in
\cite{Burko97}.
 
In figure \ref{fig:4.1a} are shown $\psi_N (r), \psi_N (\Phi )$ and
$\psi_N (\sigma )$ for $N=50, 100, 200, 400, 800$.
We see that all the lines have a slope of approximately minus two
(compare with line 'd'), indicating that the functions
converge with second order. However,
we also need to show that the converging solution is a
physical solution of our field equations. We have
tested the code against the Reissner-Nordstr{\"
  o}m and Schwarzschild solutions and reproduced all known features
(e.g. location of the event horizon as a function of initial mass
etc.) of these spacetimes. If a scalar field is present there are no
suitable analytic solutions against which we can compare the numerical
results, but we can use the constraint equations
\eqref{eq:constraint1} and \eqref{eq:constraint2} to test that our
code is converging to a physical solution.  

Denoting the left hand side of the constraint equations
\eqref{eq:constraint1} and \eqref{eq:constraint2} by $C_1$ and $C_2$
respectively, we calculate the average absolute value of $C_1$ and
$C_2$ at each point:
\begin{equation}
  \label{eq:psi2def}
  \chi (x) \equiv \frac{1}{n}\sum_i^n |x^i|
\end{equation}
defined similar to eq. \eqref{eq:psidef}, with $x$ being either $C_1$
or $C_2$. In figure \ref{fig:4.1b} are shown $\chi (C_1)$ and $\chi
(C_2)$ along $u=23.00$. It is
noted that $\chi (C_2)$ is much greater then $\chi (C_1)$. This is
mainly because the line $u=23.00$ is very close to the event horizon
and the gradients along ingoing rays 
are here very large. This results in great inaccuracy for coarse resolutions
when calculating the derivatives used to calculate $C_2$. Related to
this is of course the fact that we are not using AMR for these basic
convergence tests and hence the large average value for $C_2$ is also
an indication that splitting would 
be most desirable in this region. However, $\chi (C_2)$ is
decreasing with increasing resolution, which is the important result
in this context. We note also that the typical absolute value of the
biggest term in eq. \eqref{eq:constraint2} is approximately ten times
greater than $C_2$ even for rough resolutions. This means that 
the constraint equation is roughly satisfied for this case even
though $C_2$ has very large values. 
For the highest resolution ($N=800$), $\chi (C_1)$ shows indications
of a slightly decreasing convergence rate. This can also be traced to roundoff
errors when calculating the $r_{vv}$-term of $C_1$.

\begin{figure}
\includegraphics[width=0.495\textwidth]{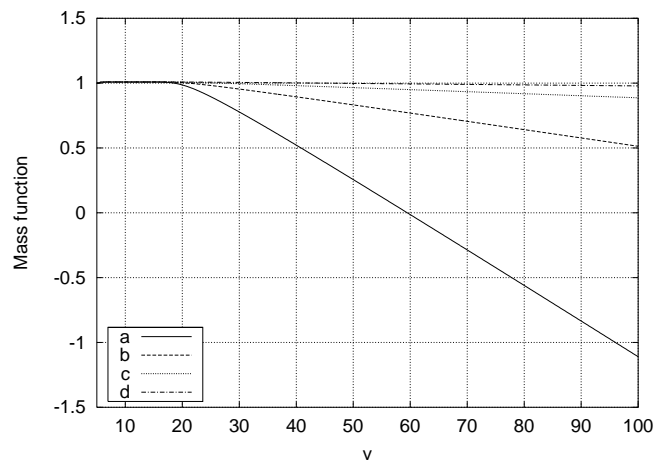}
\caption{Mass function as a function of $v$ for the resolutions; line
  a) $N=200$, b) $N=400$, c) $N=800$ and d) $N=1600$.}
\label{fig:4.3}
\end{figure}
\begin{figure*}
\subfigure[a: $\chi (C_1)$  and b: $\chi (C_2)$.
\label{fig:4.6}]{\includegraphics[width=0.495\textwidth]{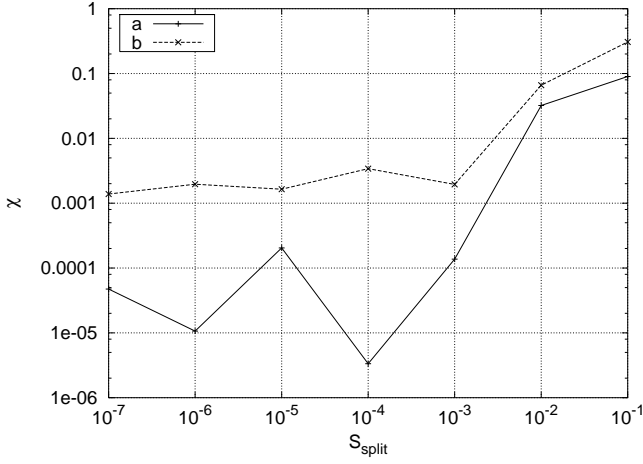}} 
\subfigure[a: $\psi (r)$, b: $\psi (\Phi )$ and c: $\psi (\sigma )$
\label{fig:4.7}]{\includegraphics[width=0.495\textwidth]{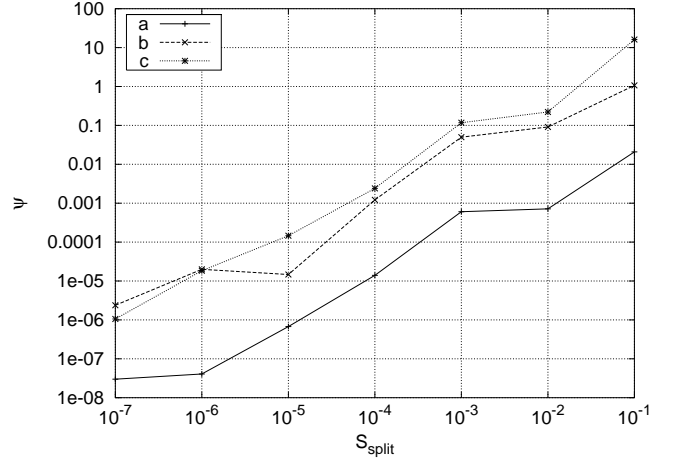}} 
\caption{{Plots illustrating convergence (with AMR) as
    function of splitting criterias along outgoing ray at
    $u=24.60$.\label{fig:4.6-7} }} 
\end{figure*}
\begin{figure*}
\subfigure[$\Phi$ as a function of
$r$.\label{fig:4.8}]{\includegraphics[width=0.495\textwidth]{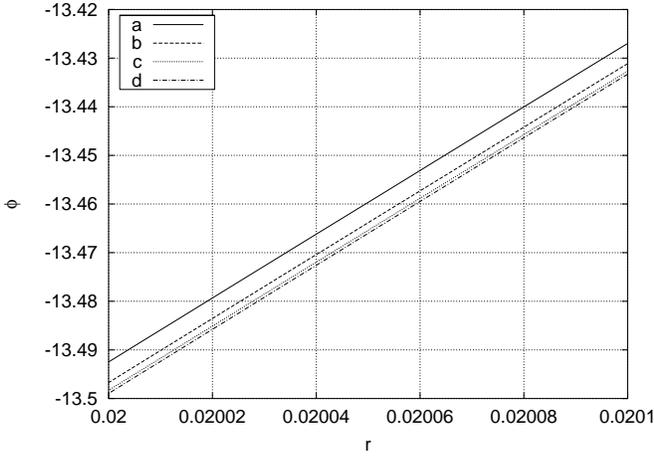}} 
\subfigure[$\sigma$ as a function of
$r$.\label{fig:4.9}]{\includegraphics[width=0.495\textwidth]{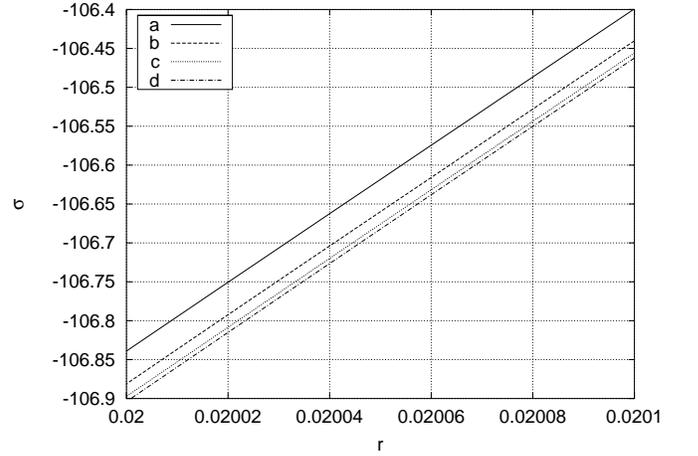}} 
\caption{{$\Phi$ and $\sigma$ as functions of
    $r$ along outgoing ray at
    $u=24.60$. Error tolerance intervals are 
for line a) $S_{split}=10^{-4}$, b) $S_{split}=10^{-5}$, c) $S_{split}=10^{-6}$
and d) $S_{split}=10^{-7}$. In all cases $S_{desplit}=10^{-2}\times
    S_{split}$. \label{fig:4.10} }} 
\end{figure*}

Figure \ref{fig:4.2a} shows $C_1(v)$ for the
resolutions $N=100,200,400,800$. It is seen that $C_1$ is clearly
converging to zero for increasing resolution. It is also clear that
numerical noise play a dominant role for $N=800$, especially at large
$v$, causing the mentioned decreasing convergence rate effect in
fig. \ref{fig:4.1b}. As noted, this numerical noise can be traced to
the numerical calculation of the $r_{vv}$-term 
in eq. \eqref{eq:constraint2}. The reason that this particular term
exhibits large roundoff errors is mainly that $r_{vv}\approx 0$
along the line at which we look. This results in the subtraction of two nearly
equal numbers which results in large roundoff errors.
Figure \ref{fig:4.2b} shows the convergence for $C_2$ again for
$N=100,200,400,800$. Clearly $C_2$ is also converging to a zero value,
but for a given resolution, $C_2$ is much larger then $C_1$ due to the
proximity to the event horizon and the absence of AMR in these
simulations as explained above. 

It is also noted that although $C_1$ and $C_2$ seem
to be converging with second order accuracy, the convergence rates of $C_1$
and $C_2$ are not by themself as important as the convergence rates of
the unknown functions, since $C_1$ and $C_2$ are functions of the
derivatives of the unknown function. The unknown functions, however,
directly measures the convergence rate of the code as these are the
actual variables being evolved.

\subsection{\label{subsec:4.2}Convergence with AMR}
Another test to show that the numerical solution is converging to a
physical solution is to monitor whether the code is mass
conserving. Figure \ref{fig:4.3} shows the mass function
\eqref{eq:2_12} along the outgoing ray $u=23.00$ in the interval
$5<v<100$ for the resolutions $N=200,400,800,1600$. We see that the
mass function is converging to a constant value for high resolutions,
further indicating that the code is converging to a physical solution
of the field equations.  

In the previous subsection we verified that with the fixed increments
$\Delta u = \Delta v$, our code converges to a
physical solution with second order accuracy. In this subsection we
will demonstrate the convergence properties of our code with the
splitting algorithm turned on, as a function of the splitting
criterias used.  

We wish to confirm that our splitting algorithm works as desired and
that the numerical solution converges to a physical 
solution. For this purpose we make a number of simulations with
identical initial conditions, but vary the splitting thresholds. We
use the same pulse shape as in the previous subsection
(eq. \eqref{eq:4.1}), but with a wider and 
stronger pulse with $A=0.20$ and $v_0 = 5.0$ and $v_1=7.0$ which is
strong enough to generate a $r=0$ singularity within out computational
domain. Our computational domain is in the range $5<v<20$ and
$0<u<25$. We examine an outgoing ray ($u=24.60$) which is inside the
event horizon and in fact comes very close to the $r=0$ singularity. 

To measure the convergence rates we use eq. \eqref{eq:psidef} and
\eqref{eq:psi2def} only in this subsection they are functions of
$S_{split}$ (see section \ref{subsec:3.2}) instead of the resolution
$N$.

Figure \ref{fig:4.6} shows $\chi (C_1)$ and $\chi (C_2)$ as 
functions of the splitting threshold
$S_{split}$ (see section \ref{subsec:3.2}) that we require
the relative numerical error to be below for all computational cells,
i.e. the left hand side of the figure denotes 
a more strict splitting policy. The threshold
for desplitting is here and throughout the paper set to two orders of
magnitude lower then the splitting threshold, 
$S_{deplit}=10^{-2}\cdot S_{split}$. We see that as the splitting threshold
$S_{split}$ is decreased, so is the average absolute of
$C_1$ and $C_2$, indicating that the numerical solution converges to a
physical solution with decreasing error tolerance.  

For the lowest splitting threshold at the left side of the figure we
see that lower splitting thresholds does 
not result in significant smaller values of $C_1$ and $C_2$. This
arises because $C_1$ and $C_2$ becomes so small that numerical noise
(which is little affected by resolution) begins to dominate the
calculations of $C_1$ and $C_2$.

Figure \ref{fig:4.7} illustrates the
convergence of the functions $r$, $\Phi$ and $\sigma$ along the same
outgoing ray $u=24.60$. The figure, which is qualitatively similar to figure
\ref{fig:4.1a}, shows  the difference between simulations ($\psi$)
with different the splittings thresholds $S_{split}=10^{-7}$,
$10^{-6}$, $10^{-5}$, $10^{-4}$, $10^{-3}$, $10^{-2}$, $10^{-1}$. Since
$\psi$ in this subsection is a function of $S_{split}$ instead of
$N$, a point in fig.  \ref{fig:4.7} illustrates the average absolute
difference between the simulation a $S_{split}$
and the simulation with the next higher splittings threshold. For
example the leftmost points in fig. \ref{fig:4.7} (at
$S_{split}=10^{-7}$) illustrates the difference between simulations
with $S_{split}=10^{-7}$ and $S_{split}=10^{-6}$.    

As the splitting thresholds decreases we expect to see $\psi$
convergece to zero. We see that for smaller $S_{split}$, the
$\psi$ decrease indication a convergence of the unknown
functions. This convergence is further demonstrated in figure
\ref{fig:4.10} which shows $\Phi$ and 
$\sigma$ as functions of $r$ close to the $r=0$ singularity (in the
interval $0.02<r<0.0201$) along the same outgoing ray that was used
for previous plots. From these figures it is clear that even in the
vicinity of the $r=0$ singularity, the unknown
functions are converging. It is also important to notice that even
though $C_1$ and $C_2$ cease to decrease at rather high splitting
thresholds, the unknown functions themselves continue to benefit from
lower splitting thresholds.

\bibliography{paper.bbl}

\begin{thebibliography}{34}
\expandafter\ifx\csname natexlab\endcsname\relax\def\natexlab#1{#1}\fi
\expandafter\ifx\csname bibnamefont\endcsname\relax
  \def\bibnamefont#1{#1}\fi
\expandafter\ifx\csname bibfnamefont\endcsname\relax
  \def\bibfnamefont#1{#1}\fi
\expandafter\ifx\csname citenamefont\endcsname\relax
  \def\citenamefont#1{#1}\fi
\expandafter\ifx\csname url\endcsname\relax
  \def\url#1{\texttt{#1}}\fi
\expandafter\ifx\csname urlprefix\endcsname\relax\def\urlprefix{URL }\fi
\providecommand{\bibinfo}[2]{#2}
\providecommand{\eprint}[2][]{\url{#2}}

\bibitem[{\citenamefont{Goldwirth and Piran}(1987)}]{Goldwirth87}
\bibinfo{author}{\bibfnamefont{D.}~\bibnamefont{Goldwirth}} \bibnamefont{and}
  \bibinfo{author}{\bibfnamefont{T.}~\bibnamefont{Piran}},
  \bibinfo{journal}{Phys. Rev. D} \textbf{\bibinfo{volume}{36}},
  \bibinfo{pages}{3575} (\bibinfo{year}{1987}).

\bibitem[{\citenamefont{Poisson and Israel}(1990)}]{Poisson90}
\bibinfo{author}{\bibfnamefont{E.}~\bibnamefont{Poisson}} \bibnamefont{and}
  \bibinfo{author}{\bibfnamefont{W.}~\bibnamefont{Israel}},
  \bibinfo{journal}{Phys. Rev. D} \textbf{\bibinfo{volume}{41}},
  \bibinfo{pages}{1796} (\bibinfo{year}{1990}).

\bibitem[{\citenamefont{Ori}(1991)}]{Ori91}
\bibinfo{author}{\bibfnamefont{A.}~\bibnamefont{Ori}}, \bibinfo{journal}{Phys.
  Rev. Lett.} \textbf{\bibinfo{volume}{67}}, \bibinfo{pages}{789}
  (\bibinfo{year}{1991}).

\bibitem[{\citenamefont{Ori}(1992)}]{Ori92}
\bibinfo{author}{\bibfnamefont{A.}~\bibnamefont{Ori}}, \bibinfo{journal}{Phys.
  Rev. Lett.} \textbf{\bibinfo{volume}{68}}, \bibinfo{pages}{2117}
  (\bibinfo{year}{1992}).

\bibitem[{\citenamefont{Gnedin and Gnedin}(1993)}]{Gnedin93}
\bibinfo{author}{\bibfnamefont{M.}~\bibnamefont{Gnedin}} \bibnamefont{and}
  \bibinfo{author}{\bibfnamefont{N.}~\bibnamefont{Gnedin}},
  \bibinfo{journal}{Class. Quant. Grav.} \textbf{\bibinfo{volume}{10}},
  \bibinfo{pages}{1083} (\bibinfo{year}{1993}).

\bibitem[{\citenamefont{Bonanno et~al.}(1994)\citenamefont{Bonanno, Droz,
  Israel, and Morsink}}]{Bonanno94}
\bibinfo{author}{\bibfnamefont{A.}~\bibnamefont{Bonanno}},
  \bibinfo{author}{\bibfnamefont{S.}~\bibnamefont{Droz}},
  \bibinfo{author}{\bibfnamefont{W.}~\bibnamefont{Israel}}, \bibnamefont{and}
  \bibinfo{author}{\bibfnamefont{S.}~\bibnamefont{Morsink}},
  \bibinfo{journal}{Proc. Roy. Soc. London A} \textbf{\bibinfo{volume}{450}},
  \bibinfo{pages}{553} (\bibinfo{year}{1994}).

\bibitem[{\citenamefont{Brady and Smith}(1995)}]{Brady95}
\bibinfo{author}{\bibfnamefont{P.}~\bibnamefont{Brady}} \bibnamefont{and}
  \bibinfo{author}{\bibfnamefont{J.}~\bibnamefont{Smith}},
  \bibinfo{journal}{Phys. Rev. Lett.} \textbf{\bibinfo{volume}{75}},
  \bibinfo{pages}{1256} (\bibinfo{year}{1995}).

\bibitem[{\citenamefont{Droz}(1996)}]{Droz96}
\bibinfo{author}{\bibfnamefont{S.}~\bibnamefont{Droz}},
  \bibinfo{journal}{Helv.Phys.Acta} \textbf{\bibinfo{volume}{69}},
  \bibinfo{pages}{257} (\bibinfo{year}{1996}).

\bibitem[{\citenamefont{Burko and Ori}(1997{\natexlab{a}})}]{Burko97b}
\bibinfo{editor}{\bibfnamefont{L.~M.} \bibnamefont{Burko}} \bibnamefont{and}
  \bibinfo{editor}{\bibfnamefont{A.}~\bibnamefont{Ori}}, eds.,
  \emph{\bibinfo{title}{Internal structure of black holes and spacetime
  singularities}}, vol. \bibinfo{volume}{13 of the Annals of the Israel
  Physical Society} (\bibinfo{publisher}{Israel Physical Society},
  \bibinfo{address}{Jerusalem}, \bibinfo{year}{1997}{\natexlab{a}}), ISBN
  \bibinfo{isbn}{0-7503-05487}.

\bibitem[{\citenamefont{Burko}(1997)}]{Burko97c}
\bibinfo{author}{\bibfnamefont{L.~M.} \bibnamefont{Burko}},
  \bibinfo{journal}{Phys. Rev. Lett.} \textbf{\bibinfo{volume}{79}},
  \bibinfo{pages}{4958} (\bibinfo{year}{1997}).

\bibitem[{\citenamefont{Burko}(1999{\natexlab{a}})}]{Burko98}
\bibinfo{author}{\bibfnamefont{L.~M.} \bibnamefont{Burko}},
  \bibinfo{journal}{Phys. Rev. D.} \textbf{\bibinfo{volume}{59}},
  \bibinfo{pages}{020411} (\bibinfo{year}{1999}{\natexlab{a}}).

\bibitem[{\citenamefont{Burko and Ori}(1998)}]{Burko98c}
\bibinfo{author}{\bibfnamefont{L.~M.} \bibnamefont{Burko}} \bibnamefont{and}
  \bibinfo{author}{\bibfnamefont{A.}~\bibnamefont{Ori}},
  \bibinfo{journal}{Phys. Rev. D} \textbf{\bibinfo{volume}{57}},
  \bibinfo{pages}{7084} (\bibinfo{year}{1998}).

\bibitem[{\citenamefont{Burko}(1999{\natexlab{b}})}]{Burko99}
\bibinfo{author}{\bibfnamefont{L.~M.} \bibnamefont{Burko}},
  \bibinfo{journal}{Phys. Rev. D} \textbf{\bibinfo{volume}{60}},
  \bibinfo{pages}{104033} (\bibinfo{year}{1999}{\natexlab{b}}).

\bibitem[{\citenamefont{Burko}(2002)}]{Burko02}
\bibinfo{author}{\bibfnamefont{L.~M.} \bibnamefont{Burko}},
  \bibinfo{journal}{Phys. Rev. D} \textbf{\bibinfo{volume}{66}},
  \bibinfo{pages}{024046} (\bibinfo{year}{2002}).

\bibitem[{\citenamefont{Burko}(2003)}]{Burko02b}
\bibinfo{author}{\bibfnamefont{L.~M.} \bibnamefont{Burko}},
  \bibinfo{journal}{Phys. Rev. Lett.} \textbf{\bibinfo{volume}{90}},
  \bibinfo{pages}{121101} (\bibinfo{year}{2003}), \bibinfo{note}{erratum in
  Phys. Rev. Lett. 90, 249902 (E)}.

\bibitem[{\citenamefont{Oren and Piran}(2003)}]{Oren03}
\bibinfo{author}{\bibfnamefont{Y.}~\bibnamefont{Oren}} \bibnamefont{and}
  \bibinfo{author}{\bibfnamefont{T.}~\bibnamefont{Piran}},
  \bibinfo{journal}{Phys. Rev. D} \textbf{\bibinfo{volume}{68}},
  \bibinfo{pages}{044013} (\bibinfo{year}{2003}).

\bibitem[{\citenamefont{Hod and Piran}(1997)}]{Hod97}
\bibinfo{author}{\bibfnamefont{S.}~\bibnamefont{Hod}} \bibnamefont{and}
  \bibinfo{author}{\bibfnamefont{T.}~\bibnamefont{Piran}},
  \bibinfo{journal}{Phys. Rev. D} \textbf{\bibinfo{volume}{55}},
  \bibinfo{pages}{3485} (\bibinfo{year}{1997}).

\bibitem[{\citenamefont{Hod and Piran}(1998{\natexlab{a}})}]{Hod98}
\bibinfo{author}{\bibfnamefont{S.}~\bibnamefont{Hod}} \bibnamefont{and}
  \bibinfo{author}{\bibfnamefont{T.}~\bibnamefont{Piran}},
  \bibinfo{journal}{Gen. Rel. Grav.} \textbf{\bibinfo{volume}{30}},
  \bibinfo{pages}{1555} (\bibinfo{year}{1998}{\natexlab{a}}).

\bibitem[{\citenamefont{Hod and Piran}(1998{\natexlab{b}})}]{Hod98b}
\bibinfo{author}{\bibfnamefont{S.}~\bibnamefont{Hod}} \bibnamefont{and}
  \bibinfo{author}{\bibfnamefont{T.}~\bibnamefont{Piran}},
  \bibinfo{journal}{Phys. Rev. Lett.} \textbf{\bibinfo{volume}{81}},
  \bibinfo{pages}{1554} (\bibinfo{year}{1998}{\natexlab{b}}).

\bibitem[{\citenamefont{Ori}(1999)}]{Ori99}
\bibinfo{author}{\bibfnamefont{A.}~\bibnamefont{Ori}}, \bibinfo{journal}{Phys.
  Rev. Lett.} \textbf{\bibinfo{volume}{83}}, \bibinfo{pages}{5423}
  (\bibinfo{year}{1999}).

\bibitem[{\citenamefont{Ori}(2000)}]{Ori99b}
\bibinfo{author}{\bibfnamefont{A.}~\bibnamefont{Ori}}, \bibinfo{journal}{Phys.
  Rev. D} \textbf{\bibinfo{volume}{61}}, \bibinfo{pages}{024001}
  (\bibinfo{year}{2000}).

\bibitem[{\citenamefont{Berger}(2002)}]{Berger02}
\bibinfo{author}{\bibfnamefont{B.~K.} \bibnamefont{Berger}},
  \bibinfo{journal}{Living Reviews Relativity}  (\bibinfo{year}{2002}),
  \bibinfo{note}{http://www.livingreviews.org/ - gr-qc/0201056}.

\bibitem[{\citenamefont{Hamilton and
  Pollack}(2004{\natexlab{a}})}]{Hamilton04a}
\bibinfo{author}{\bibfnamefont{A.}~\bibnamefont{Hamilton}} \bibnamefont{and}
  \bibinfo{author}{\bibfnamefont{S.}~\bibnamefont{Pollack}},
  \bibinfo{journal}{gr-qc/0411061}  (\bibinfo{year}{2004}{\natexlab{a}}).

\bibitem[{\citenamefont{Hamilton and
  Pollack}(2004{\natexlab{b}})}]{Hamilton04b}
\bibinfo{author}{\bibfnamefont{A.}~\bibnamefont{Hamilton}} \bibnamefont{and}
  \bibinfo{author}{\bibfnamefont{S.}~\bibnamefont{Pollack}},
  \bibinfo{journal}{gr-qc/0411062}  (\bibinfo{year}{2004}{\natexlab{b}}).

\bibitem[{\citenamefont{Dafermos}(2004)}]{Dafermos04}
\bibinfo{author}{\bibfnamefont{M.}~\bibnamefont{Dafermos}},
  \bibinfo{journal}{Proceedings of the Seventh Hungarian Relativity Workshop}
  (\bibinfo{year}{2004}), \bibinfo{note}{to appear, gr-qc/0401121}.

\bibitem[{\citenamefont{Frolov and Novikov}(1998)}]{Frolov98}
\bibinfo{author}{\bibfnamefont{V.~P.} \bibnamefont{Frolov}} \bibnamefont{and}
  \bibinfo{author}{\bibfnamefont{I.~D.} \bibnamefont{Novikov}},
  \emph{\bibinfo{title}{Black Hole Physics}} (\bibinfo{publisher}{Kluwer
  Academic Publishers}, \bibinfo{year}{1998}).

\bibitem[{\citenamefont{Gurtug and Halilsoy}(2002)}]{Gurtug02}
\bibinfo{author}{\bibfnamefont{O.}~\bibnamefont{Gurtug}} \bibnamefont{and}
  \bibinfo{author}{\bibfnamefont{M.}~\bibnamefont{Halilsoy}},
  \bibinfo{journal}{gr-qc/0203019}  (\bibinfo{year}{2002}).

\bibitem[{\citenamefont{Burko and Ori}(1997{\natexlab{b}})}]{Burko97}
\bibinfo{author}{\bibfnamefont{L.~M.} \bibnamefont{Burko}} \bibnamefont{and}
  \bibinfo{author}{\bibfnamefont{A.}~\bibnamefont{Ori}},
  \bibinfo{journal}{Phys. Rev. D} \textbf{\bibinfo{volume}{56}},
  \bibinfo{pages}{7820} (\bibinfo{year}{1997}{\natexlab{b}}).

\bibitem[{\citenamefont{Burko}(1998)}]{Burko98b}
\bibinfo{author}{\bibfnamefont{L.~M.} \bibnamefont{Burko}},
  \bibinfo{journal}{Phys. Rev. D.} \textbf{\bibinfo{volume}{58}},
  \bibinfo{pages}{084013} (\bibinfo{year}{1998}).

\bibitem[{\citenamefont{Misner et~al.}(1973)\citenamefont{Misner, Thorne, and
  Wheeler}}]{MTW73}
\bibinfo{author}{\bibfnamefont{C.~W.} \bibnamefont{Misner}},
  \bibinfo{author}{\bibfnamefont{K.~S.} \bibnamefont{Thorne}},
  \bibnamefont{and} \bibinfo{author}{\bibfnamefont{J.~A.}
  \bibnamefont{Wheeler}}, \emph{\bibinfo{title}{Gravitation}}
  (\bibinfo{publisher}{W. H. Freeman}, \bibinfo{address}{San Francisco},
  \bibinfo{year}{1973}).

\bibitem[{\citenamefont{Poisson and Israel}(1989)}]{Poisson89a}
\bibinfo{author}{\bibfnamefont{E.}~\bibnamefont{Poisson}} \bibnamefont{and}
  \bibinfo{author}{\bibfnamefont{W.}~\bibnamefont{Israel}},
  \bibinfo{journal}{Phys. Rev. Lett.} \textbf{\bibinfo{volume}{63}},
  \bibinfo{pages}{1663} (\bibinfo{year}{1989}).

\bibitem[{\citenamefont{Zeldovich and Novikov}(1983)}]{Zeldovich83a}
\bibinfo{author}{\bibfnamefont{Y.~B.} \bibnamefont{Zeldovich}}
  \bibnamefont{and} \bibinfo{author}{\bibfnamefont{I.~D.}
  \bibnamefont{Novikov}}, \emph{\bibinfo{title}{Relativistic Astrophysics}},
  vol. \bibinfo{volume}{{II, The Structure and Evolution of the Universe}}
  (\bibinfo{publisher}{The University of Chicago press}, \bibinfo{year}{1983}).

\bibitem[{\citenamefont{Bonnor and Vaidya}(1970)}]{Bonnor70}
\bibinfo{author}{\bibfnamefont{W.~B.} \bibnamefont{Bonnor}} \bibnamefont{and}
  \bibinfo{author}{\bibfnamefont{P.~C.} \bibnamefont{Vaidya}},
  \bibinfo{journal}{Gen. Relativ. Gravit.} \textbf{\bibinfo{volume}{1}}
  (\bibinfo{year}{1970}).

\bibitem[{\citenamefont{Grishchuk}(1970)}]{Grishchuk70}
\bibinfo{author}{\bibfnamefont{L.~P.} \bibnamefont{Grishchuk}},
  \bibinfo{journal}{Sov. Phys. Doklady} \textbf{\bibinfo{volume}{190}},
  \bibinfo{pages}{1066} (\bibinfo{year}{1970}).

\end{thebibliography}

\end{document}